\documentclass[aps,preprint,prl]{revtex4}

\pdfoutput=1
\usepackage{amsmath,amssymb}
\usepackage{graphicx}

\begin{document}

\title{\bf High -- Temperature Superconductivity in FeSe Monolayers}

\author{M.V.~Sadovskii
\\
{Institute for Electrophysics, Russian Academy of Sciences, Ural Branch}\\ 
{\sl Amundsen str. 106, Ekaterinburg 620016, Russia}\\
{M.N. Mikheev Institute for Metal Physics, Russian Academy of Sciences, Ural Branch,}\\ 
{\sl S. Kovalevsky str. 18, Ekaterinburg 620290, Russia}}


\begin{abstract}

This review discusses the main experiments and theoretical views related to
observation of high -- temperature superconductivity in intercalated FeSe
compounds and single layer films of FeSe on substrates like SrTiO$_3$.
We consider in detail the electronic structure of these systems, both theoretical
calculations of this structure at hand and their correspondence with ARPES experiments.
It is stressed that electronic spectrum of these systems is qualitatively different
from typical picture of the spectrum in well studied FeAs superconductors and
the related problems of theoretical description of spectrum formation are 
also discussed.

We also discuss the possible mechanisms of Cooper pairing in monolayers of FeSe
and problems appearing here. As single layer films of FeSe on SrTiO$_3$ 
can be represented as typical Ginzburg ``sandwiches'', we analyze the possibility
of rising the critical temperature of superconducting transition $T_c$ due to
different variants of  ``excitonic'' mechanism of superconductivity. It is shown,
that in its classic variant (as proposed for such systems by Allender, Bray and
Bardeen) this mechanism is unable to explain the observed values of $T_c$, 
but situation is different when we consider instead of ``excitons'' the optical
phonons in SrTiO$_3$ (with energy of the order of 100 meV). We consider both the
simplest model of $T_c$ enhancement due to interaction with such phonons and
more specific models with dominant ``forward'' scattering, which allow to
understand the growth of $T_c$ as compared with the case of bulk FeSe and
intercalated FeSe systems. We also discuss the problems connected with
antiadiabatic nature of superconductivity due to such mechanism.

PACS: 74.20.-z, 74.20.Fg, 74.20.Mn, 74.20.Rp, 74.25.Jb, 74.62.-c, 74.70.-b

\end{abstract}

\maketitle


\tableofcontents


\section{Introduction}

Discovery of the new class of superconductors based upon iron pnictides has
opened the new perspectives in the studies of high -- temperature 
superconductivity. While possessing the main superconducting characteristics
somehow inferior to those of copper oxides (cuprates), these systems attracted 
much attention of researchers, as the nature of superconductivity and  other 
physical properties here are in many respects different from those of cuprates,
while preserving many common features, which leads to the hopes for more deep
understanding of the problem of high -- temperature superconductivity as a whole.  
And this problem, which was put into the agenda mainly due to the enthusiasm of
V.L. Ginzburg \cite{VLG,HTSC77}, still remains among the central problems of the
modern physics of condensed matter.

At present the properties iron pnictide superconductors are rather well studied
experimentally, there is also almost overwhelmingly accepted theoretical
picture of superconductivity in these systems, which is based on the idea of
leading role of pairing interaction due to exchange of (antiferro)magnetic
fluctuations, which in most cases lead to  $s^{\pm}$ pairing on different sheets
of the Fermi surface, which appear in these {\em multiple bands} systems. 
There is a number of review papers with detailed presentation of modern
experimental situation and basic theoretical concepts, used to describe these
systems \cite{Sad_08,Hoso_09,John,MazKor,Stew,Kord_12}.

Soon after the discovery of superconductivity in iron pnictides, it was
followed by its discovery in iron {\em chalcogenide} FeSe, which attracted
attention probably only due to the unusual simplicity of this compound, while
its superconducting characteristics (under normal conditions) were rather modest
($T_c\sim$8K), and its electronic structure was quite similar to that in iron
pnictides. However, this system was also thoroughly studied 
(cf. review in \cite{FeSe}).

Situation with iron chalcogenides undergone the major change with the
appearance of {\em intercalated} FeSe systems, where the values of
$T_c\sim$ 30-40K were obtained, and which attracted much attention because of
their unusual electronic structure \cite{JMMM,JTLRev}. At present a number of 
such compounds are known with properties significantly different from 
traditional iron pnictides and which require the development of new
theoretical understanding of mechanisms of superconductivity, as the traditional
for pnictides picture of $s^{\pm}$-pairing is apparently not working here.

All these problems rather sharpened after the experimental observation of
superconductivity with $T_c\sim$ 80-100K in monolayers of FeSe (epitaxial films),
grown on SrTiO$_3$ substrate (and the number of similar compounds).
At present we can speak of the  ``new frontier'' in the studies of 
high -- temperature superconductivity \cite{Boz2014}.

This small review is devoted to the description of the main experimental results
on superconductivity in intercalated FeSe monolayers and single layer films of
FeSe on substrates like SrTiO$_3$, and to discussion of a number of related
theoretical problems, including the possible mechanisms leading to significant
enhancement of $T_c$. It should be said, that here we remain with more
questions, than answers, but this is what attracts most of the researchers 
to the studies of systems discussed in this review. This field develops very
fast and we can not claim for the overwhelming discussion of all available
literature. Our presentation will be necessarily on rather elementary
(general physics) level, with the hope to make it understandable for 
nonspecialists. The references to many important works can be found in 
papers quoted below, many papers are not mentioned simple because of the
limited space for the review. However, the author hopes that this review
will be of interest to a wide community of {\em Physics Uspekhi} readers as a kind
of introduction to this new field of research, especially in connection with
centenary of great physicist --- V.L. Ginzburg, whose ideas and views on
the problem of high -- temperature superconductivity had so much influence
on everybody who is involved in this field.

\section{Main systems and experiments}

\subsection{Intercalated FeSe systems}

\begin{figure}
\includegraphics[clip=true,width=0.8\textwidth]{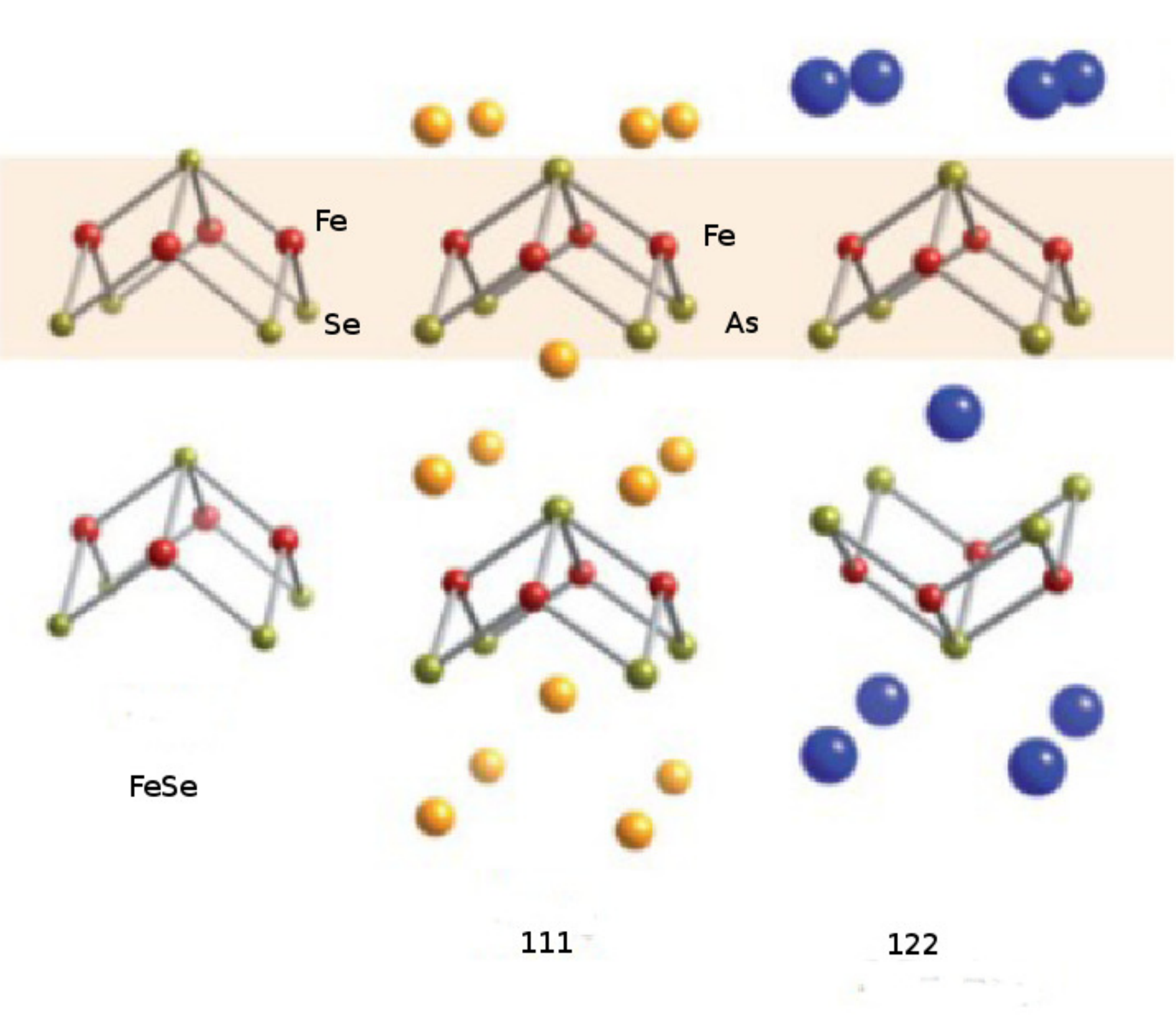}
\includegraphics[clip=true,width=0.8\textwidth]{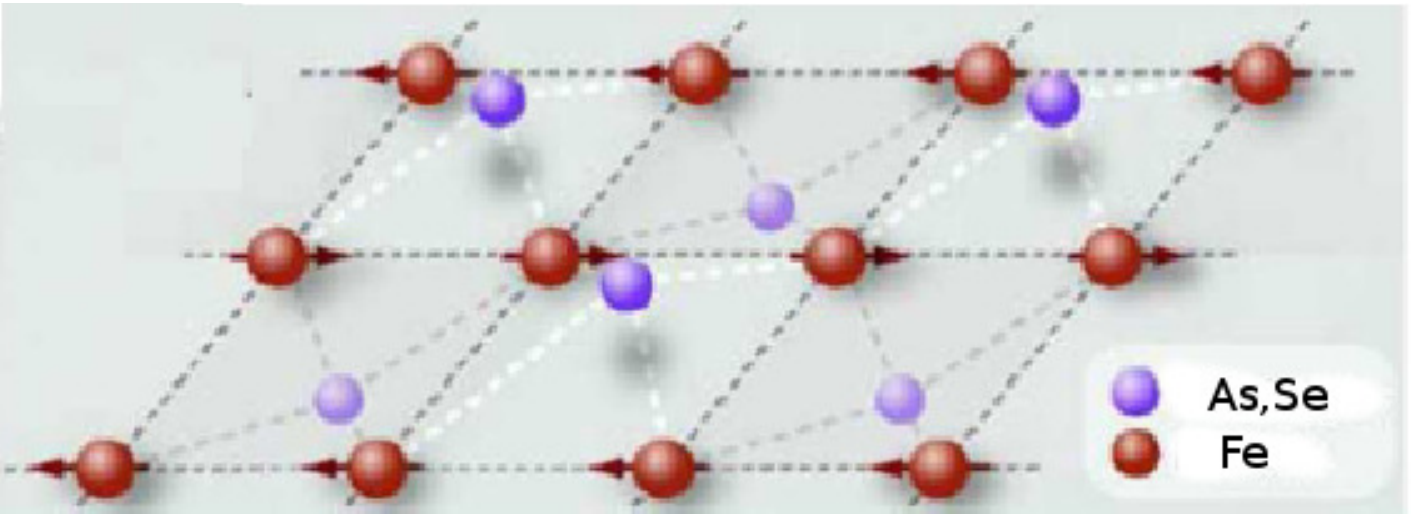}
\caption{(a) --  crystal structure of simplest iron based superconductors,
(b) -- structure of highly conducting plane (layer) of iron ions and pnictogens
(chalcogens). Arrows show direction of spins for typical ordering in
antiferromagnetic phase.}
\label{StrucFePn}
\end{figure}

In Fig. \ref{StrucFePn} (a) we show schematically the simplest crystal structures
of iron based superconductors \cite{Sad_08,Hoso_09,John,MazKor,Stew,Kord_12,FeSe}.
The common element here is the presence of FeAs or FeSe plane (layer), where
ions of Fe form the simple square lattice, while ions of pnictogens (Pn -- As)
or chalcogens (Ch -- Se) are placed in the centers of these squares, above and
below the Fe plane in chess -- board order. In Fig. \ref{StrucFePn} (b) the
structure of this layer is shown in more details. Actually the electronic states
of Fe ions in FePn(Ch) plane play decisive role in the formation of electronic
properties of these systems and among them --- superconductivity. In this sense
these layers are quite similar to CuO$_2$ planes in cuprates (copper oxides)
and these systems can be considered, in the first approximation, as quasi --
two -- dimensional, though the anisotropy in most of them may be not so strong.
Below we shall mainly limit ourselves to such oversimplified picture and speak
about the physics of FeSe planes (monolayers).

In Fig.  \ref{StrucFePn} (b) arrows show direction of spins on Fe in
antiferromagnetic structure, which is typically realized in stoichiometric
state of FeAs based systems \cite{Sad_08,Hoso_09,John,MazKor,Stew,Kord_12},
which are (in their ground state) antiferromagnetic metals. Antiferromagnetic
ordering is destroyed under electron or hole doping, when superconducting phase
just appear. In this sense the phase diagrams of systems under consideration are
quite similar to phase diagrams of cuprates
\cite{Sad_08,Hoso_09,John,MazKor,Stew,Kord_12}. These phase diagrams at present
are rather well studied. In FeSe systems, which will be considered below, the
character of magnetic ordering is known not so well. Because of this, as well as
due to the lack of space, we practically shall not discuss magnetic properties
of FeSe systems.

Note that all FeAs structures shown in Fig. \ref{StrucFePn} (a) are simple
ionic -- covalent crystals. The chemical formula say for typical 122 -- system
can be written, example, as Ba$^{+2}$(Fe$^{+2}$)$_2$(As$^{-3}$)$_2$.
The charged FeAs layers are hold together by Coulomb forces from
surrounding ions. In the bulk FeSe electroneutral FeSe layers are hold by
much weaker van der Waals interactions. This makes this system convenient for
intercalation by different atoms or molecules, which can easily enough
penetrate between FeSe layers. The chemistry of intercalation of iron selenide
superconductors is discussed in detail in a recent review \cite{VivRod}.

As we already noted, superconductivity in bulk FeSe, discovered immediately
after high -- temperature superconductivity was observed in iron pnictides,
was studied more or less in detail \cite{FeSe}, but initially has not attracted
much interest because of its similarity to superconductivity in
iron pnictides and low enough superconducting characteristics. 
This situation changed drastically after the discovery of high -- temperature 
superconductivity in intercalated FeSe compounds and
especially after the achievement of record breaking values of $T_c$ in single
layer films of FeSe on SrTiO$_3$.

\begin{figure}
\includegraphics[clip=true,width=0.4\textwidth]{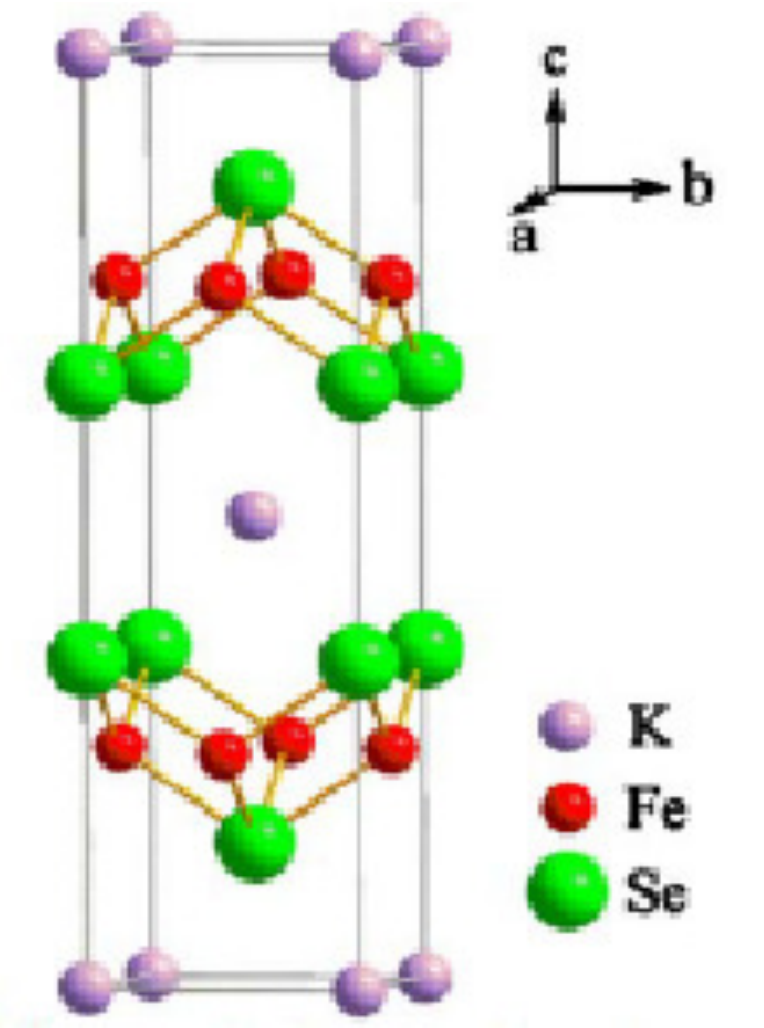}
\includegraphics[clip=true,width=0.35\textwidth]{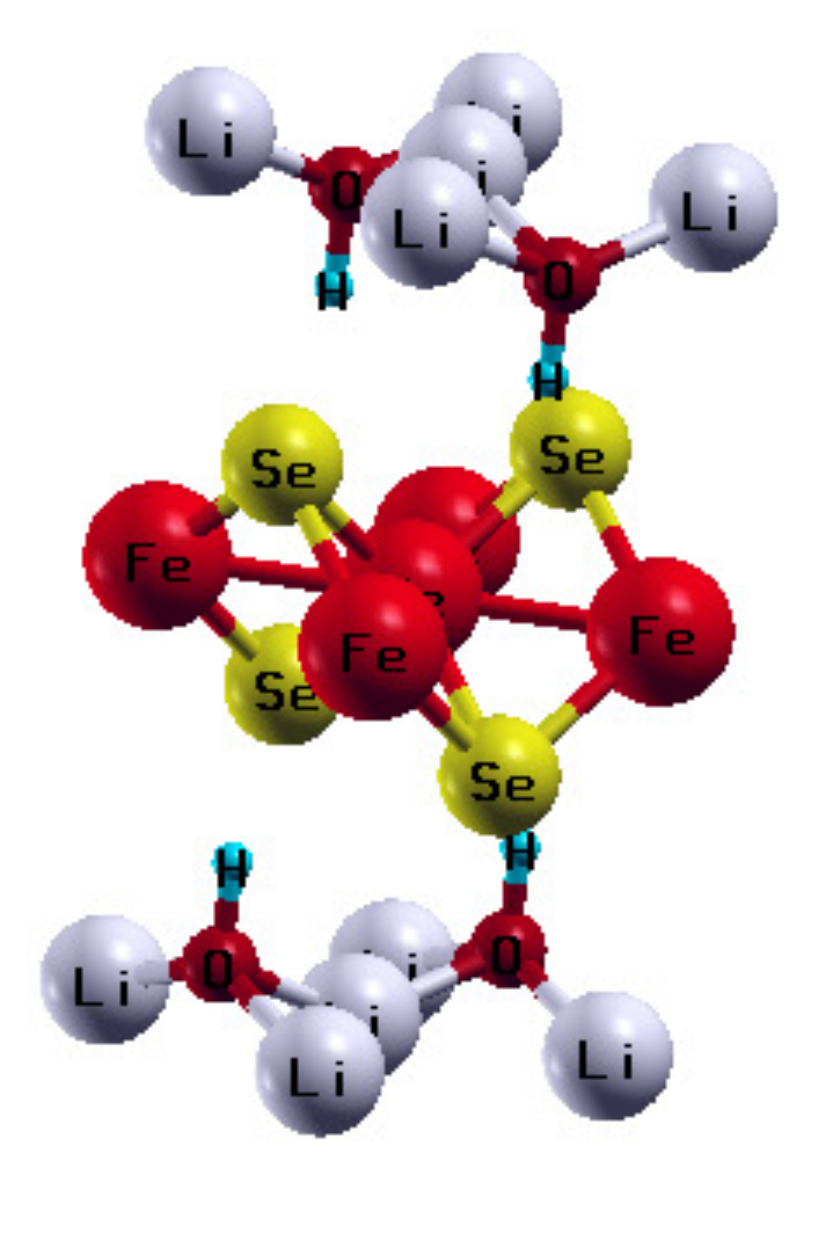}
\caption{(a) --  ideal ($x$=1) crystal structure
(122 -- type) of K$_x$Fe$_2$Se$_2$ compound, (b) -- ideal ($x$=0)
crystal structure of  [Li$_{1-x}$F$_x$OH]FeSe compound.}
\label{interFeCh}
\end{figure}

First systems of this kind were A$_x$Fe$_{2-y}$Se$_2$ (A=K,Rb,Cs) compounds,
with the values of $T_c\sim$ 30K \cite{AFeSe,AFeSe2}. It is commonly assumed
that superconductivity here is realized in 122 -- like  structure shown in
Fig. \ref{interFeCh} (a), while real samples, studied up to now, were always
multiphased, consisting of mesoscopic mixture of superconducting and insulating
(antiferromagnetic) structures like K$_2$Fe$_4$Se$_5$, which naturally complicates
the general picture. Significant further increase of $T_c$ up to the values of
the order of 45K was achieved by intercalating the FeSe layers by large enough
molecules in compounds like Li$_x$(C$_2$H$_8$N$_2$)Fe$_{2-y}$Se$_2$ \cite{intCH}
and Li$_x$(NH$_2$)$_y$(NH$_3$)$_{1-y}$Fe$_2$Se$_2$ \cite{intNH}. The increase of
$T_c$ in these systems can be supposed to be related with the growth of
spacing between FeSe layers from 5.5\AA\ in bulk FeSe to $\sim$7\AA\ in
A$_x$Fe$_{2-y}$Se$_2$ and to 8-11\AA\ in systems intercalated by large molecules.
i.e. with the growth of their two -- dimensional nature.

Recently the active studies has begun of [Li$_{1-x}$Fe$_x$OH]FeSe system, where
the values of $T_c\sim$ 43K were reached \cite{LiOH1,LiOH2} and it was possible
to obtain rather good single -- phase samples and single crystals.
Crystal structure of this system is shown in Fig. \ref{interFeCh} (b).
An interesting discussion has developed on the nature of possible magnetic
ordering on Fe ions replacing Li in intercalating layers of LiOH.
In Ref. \cite{LiOH1} it was claimed that this ordering corresponds to a canted
antiferromagnet. However, magnetic measurements of Ref. \cite{LiOH2} has lead to
unexpected conclusion on {\em ferromagnetic} character of this ordering with
Curie temperature $T_C\sim$ 10K, i.e. much lower than superconducting transition
temperature. This conclusion was indirectly confirmed in Ref. \cite{LiOH3} by
the observation of neutron scattering on the lattice of Abrikosov's vortices,
supposedly induced in FeSe layers by ferromagnetic ordering of spins of Fe in
Li$_{1-x}$Fe$_x$OH layers. At the same time, it was claimed in Ref. \cite{LiOH4}
that M\"ossbauer measurements on this system indicate the absence of any kind of
magnetic ordering on Fe ions.

\subsection{Superconductivity in FeSe monolayer on SrTiO$_3$}

\begin{figure}
\includegraphics[clip=true,width=0.8\textwidth]{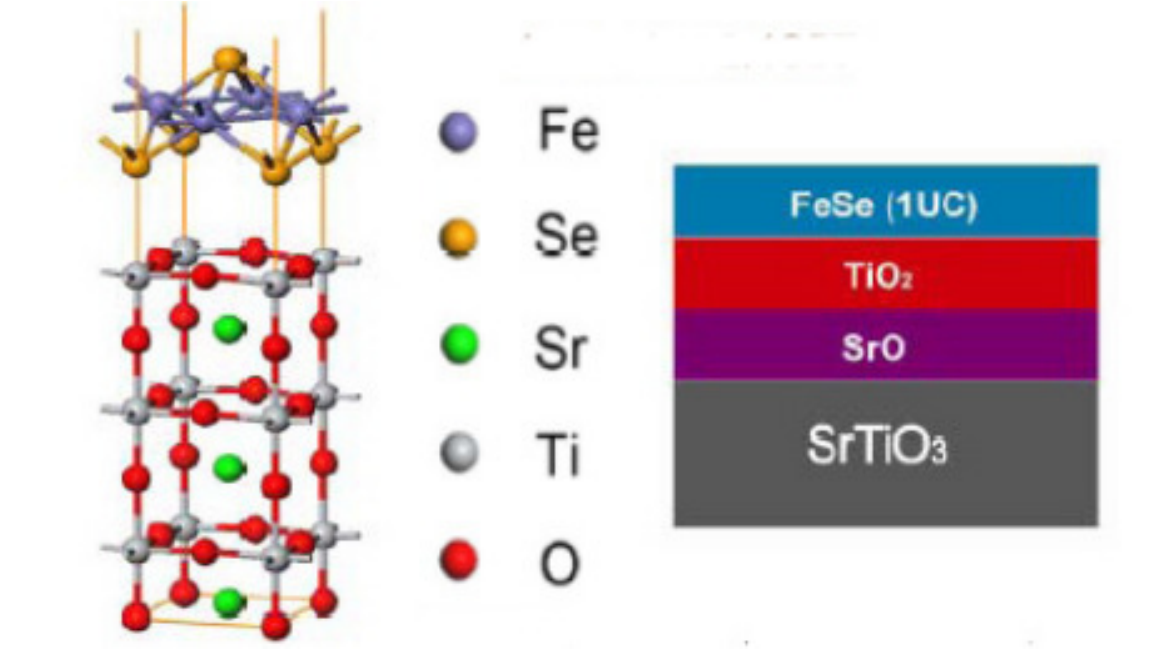}
\caption{Structure of FeSe monolayer on SrTiO$_3$ substrate (001).}
\label{FeSeSTO}
\end{figure}

\begin{figure}
\includegraphics[clip=true,width=\textwidth]{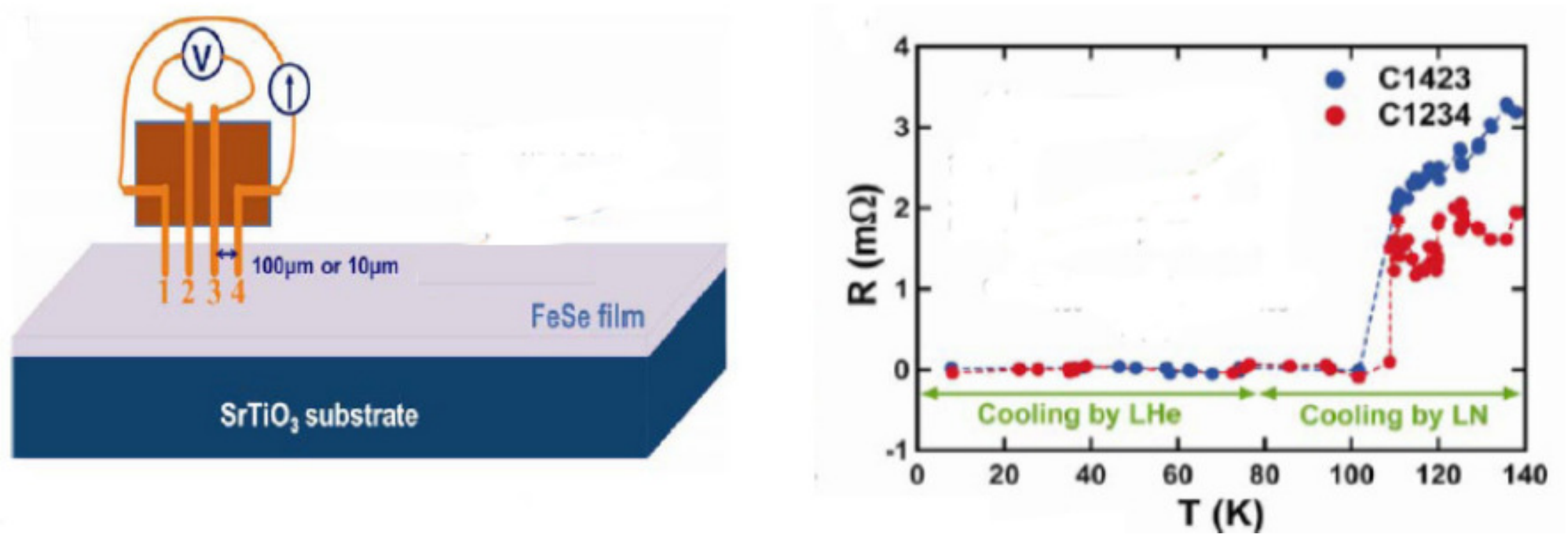}
\caption{Experimental setup to measure resistance of single layer FeSe film
on SrTiO$_3$ substrate and temperature dependence of resistivity obtained on two
samples \cite{FeSe_STO2}.}
\label{FeSeSTOres}
\end{figure}

The major breakthrough in the studies of superconductivity in FeSe systems,
as already was noted above, is connected with the observation of record breaking
values of $T_c$ in epitaxial films of monolayer of FeSe on SrTiO$_3$ (STO)
substrate \cite{FeSe_STO1}. These films were grown in Ref. \cite{FeSe_STO1} and
in most of the papers to follow on 001 plane of STO. The structure of these films
is shown in Fig. \ref{FeSeSTO}, where we can see, in particular, that the FeSe
layer is adjacent to TiO$_2$ layer on the surface of STO. Note that the lattice
constant in FeSe layer of bulk samples is 3.77 \AA, while in STO it is
significantly larger being equal to 3.905 \AA,  so that the single layerтак FeSe
films are  noticeably stretched, as compared to the bulk FeSe and are in a
stressed state, which disappears fast with addition of the next layers.
Tunneling measurements of Ref. \cite{FeSe_STO1} has demonstrated the record
values of the energy gap, while in resistance measurements the temperature of the
onset superconducting transition essentially exceeded 50K. It should be stressed
that films under study were quite unstable on the air, so that in most of the
works resistive transitions were usually studied on films covered by amorphous
Si or a number of layers of FeTe, which significantly reduced the observed
values of $T_c$. The unique {\em in situ} measurements of FeSe films on STO,
made in Ref. \cite{FeSe_STO2}, has given the record breaking values of $T_c>$ 100K,
which can be seen from the data shown in Fig. \ref{FeSeSTOres}. Up to now these
results are not confirmed by other authors, but ARPES measurements of
temperature behavior of energy gap in such films {\em in situ} at present
routinely demonstrate the values of $T_c$ in the interval of 65-75 K.

In films consisting of several layers of FeSe the observed values of $T_c$ are
significantly lower than record values of single layer films \cite{FeSe13UCK}.
Recently the monolayer FeSe films were grown also on 110 plane of STO
\cite{FeSe_STO_FeTe}, with cover up by several FeTe layers. Resistive
measurements on these films (including the measurements of the upper critical
magnetic field $H_{c2}$) has given the values of $T_c\sim$ 30 K. At the same time,
FeSe films grown on BaTiO$_3$ (BTO), doped with Nb (with even larger values
of the lattice constant $\sim$ 3.99\AA), have shown (in ARPES measurementsи)
the values of $T_c\sim$70 K \cite{FeSe_BTO}. A recent paper \cite{FeSe_Anatas}
has reported the observation of record (for FeSe systems) values of
superconducting gap (from tunneling) in FeSe monolayers on 001 plane of
TiO$_2$ (anatase), grown on 001 plane of SrTiO$_3$. It was noted that the
lattice constants of anatase are quite close to those of the bulk FeSe, so that
FeSe films is practically non stretched.

Single FeSe layer films were also grown on the graphene substrate
\cite{FeSeGraph}, but the values of $T_c$ of these films have not exceeded 8-10 K,
characteristic of the bulk FeSe, which stresses the role of the substrates like
Sr(Ba)TiO$_3$, with the unique properties, which may be determining for the
strong enhancement of $T_c$.

We shall limit ourselves with this short review of experimental situation with
observation of superconductivity in FeSe monolayers to concentrate below on
the discussion of electronic structure  and possible mechanisms, explaining  the
record (for iron based superconductors) values of $T_c$. More detailed
information on experiments on this system can be found in a recent review
\cite{FeSe1UC_rev}.

\section{Electronic structure of iron -- selenium systems}

\begin{figure}
\includegraphics[clip=true,width=0.9\textwidth]{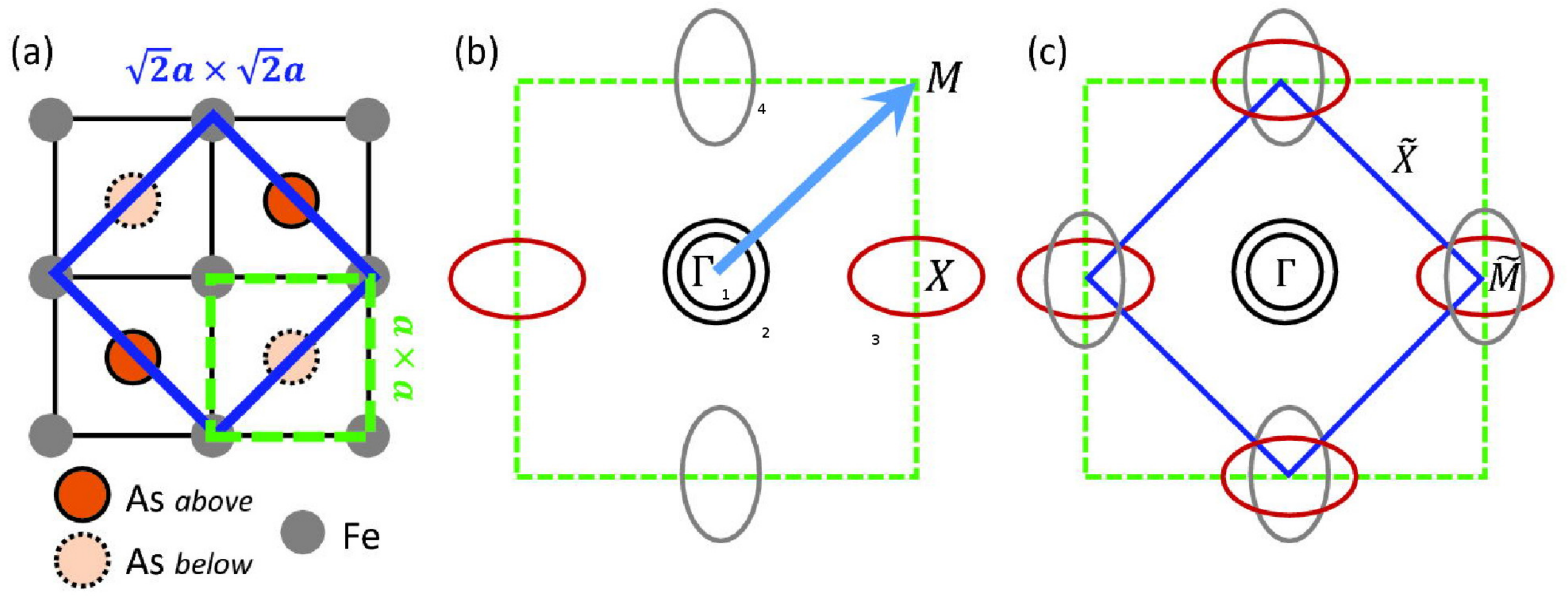}
\caption{(a) -- different choices of elementary cell in FeAs(Se) plane,
(b) -- Brillouin zone and Fermi surfaces for the case of one Fe ion in elementary
cell, (c) -- folded Brillouin zone and Fermi surfaces for the case of two Fe
ions in elementary cell.}
\label{BZFS}
\end{figure}

Electronic spectrum of iron pnictides is now well studied, both with theoretical
calculations based on modern energy band theory and also experimentally, where
the decisive role was played by ARPES experiments
\cite{Sad_08,Hoso_09,John,MazKor,Stew,Kord_12}. As we already noted above,
almost all effects of interest to us are determined by electronic states of
FeAs plane (layer), shown in Fig. \ref{StrucFePn} (b). The spectrum of carriers
in the vicinity of the Fermi level (with the width $\sim$ 0.5 eV, where
everything concerning superconductivity obviously takes place) is practically
formed only by $d$-states of Fe. Hybridization of Fe and As states according to 
all band structure calculations is very small.  
Accordingly, up to five bands (two or three
hole -- like and two electron -- like) cross the Fermi level, forming the
spectrum typical for a  semi -- metal. The schematic picture of Brillouin zones
and Fermi surfaces is shown in Fig. \ref{BZFS}, and it is essentially rather
simple.

In first approximation, assuming that all As ions belong to the same plane as
Fe ions, we have an elementary cell with one Fe and (square) lattice constant
$a$ (cf. Fig. \ref{BZFS} (a)). Corresponding Brillouin zone is shown in
Fig. \ref{BZFS} (b). If we take into account that As ions are in fact placed
above and below Fe plane, as shown in Fig. \ref{StrucFePn} (b), the elementary
cell will contain two Fe ions and Brillouin zone is reduced by a factor of two as
shown in Fig. \ref{BZFS} (c). Two -- dimensional Fermi surfaces for the case of
four bands (two hole -- like in the center and two electron -- like at the edges
or in the corners of appropriate Brillouin zones) are also schematically shown
in Fig. \ref{BZFS} (b,c).

In the energy interval around the Fermi level, which is of interest to us,
energy bands can be considered parabolic, so that the Hamiltonian of free
carriers can be written as \cite{MazKor}:
\begin{equation}
H = \sum_{{\bf k}, \sigma, i=\alpha_1,\alpha_2,\beta_1,\beta_2} 
\varepsilon^{i}_{\bf k}
c_{i {\bf k} \sigma}^\dag c_{i {\bf k} \sigma}.
\label{Hparabolic}
\end{equation}
where $c_{i {\bf k} \sigma}$ is annihilation operator of an electron with
momentum and spin ${\bf k}$,  $\sigma$ and band index $i$, and the hole bands
$\alpha_i$ dispersions take the form:
\begin{equation}
\varepsilon^{\alpha_{1,2}}_{\bf k} =
- \frac{k^2}{2 m_{1,2}} +\mu  
\label{holes}
\end{equation}
while the electron bands $\beta_i$ dispersions are written (in the coordinates
of Brillouin zone of Fig. \ref{BZFS} (b)) as:
\begin{eqnarray}
\varepsilon^{\beta_1}_{\bf k} =
\frac{(k_x-\pi/a)^2}{2 m_x} + \frac{k_y^2}{2 m_y} - \mu\nonumber\\
\varepsilon^{\beta_2}_{\bf k} = \frac{k_x^2}{2 m_y} + 
\frac{(k_y-\pi/a)^2}{2 m_x}-\mu
\label{electrons}
\end{eqnarray}

More complicated band structure models valid in the vicinity of the Fermi level
and being in direct correspondence with LDA calculations can also be proposed
(see e.g.  \cite{KuchSad10}), but the general, rather simple, picture of this
``standard model'' of iron pnictides spectrum remains the same. LDA+DMFT
calculations \cite{DMFT1,DMFT2}, taking into account the role of electron
correlations, show that in iron pnictides, in contrast to cuprates, this role is
rather irrelevant and reduced to (actually noticeable) renormalization of
the effective masses of electron and hole dispersions, as well to the general
``compression'' (reduced  width) of the bands.

The presence of electron and hole Fermi surfaces with close sizes, satisfying
(approximate!) ``nesting'' conditions, is very significant for the
theories of superconducting pairing in iron pnictides based on the decisive role
of antiferromagnetic spin fluctuations \cite{MazKor}. Below we shall see that
electronic spectrum and Fermi surfaces in Fe chalcogenides are significantly
different from the qualitative picture presented above, which poses new
(and far from being solved) problems of explaining the microscopic mechanism
of superconductivity in these systems.

\subsection{A$_x$Fe$_{2-y}$Se$_2$ system}

\begin{figure}
\includegraphics[clip=true,width=0.45\textwidth]{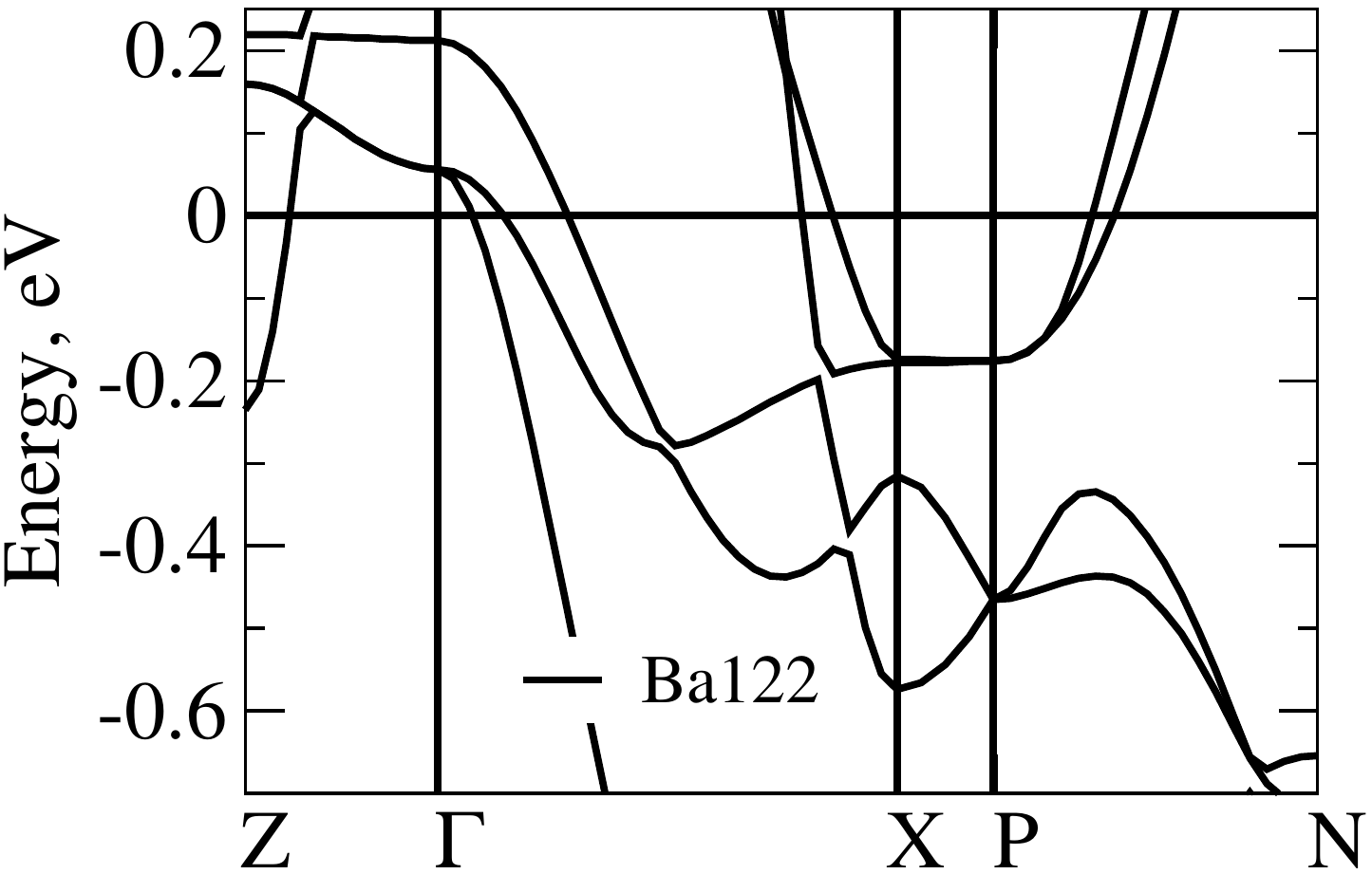}
\includegraphics[clip=true,width=0.45\textwidth]{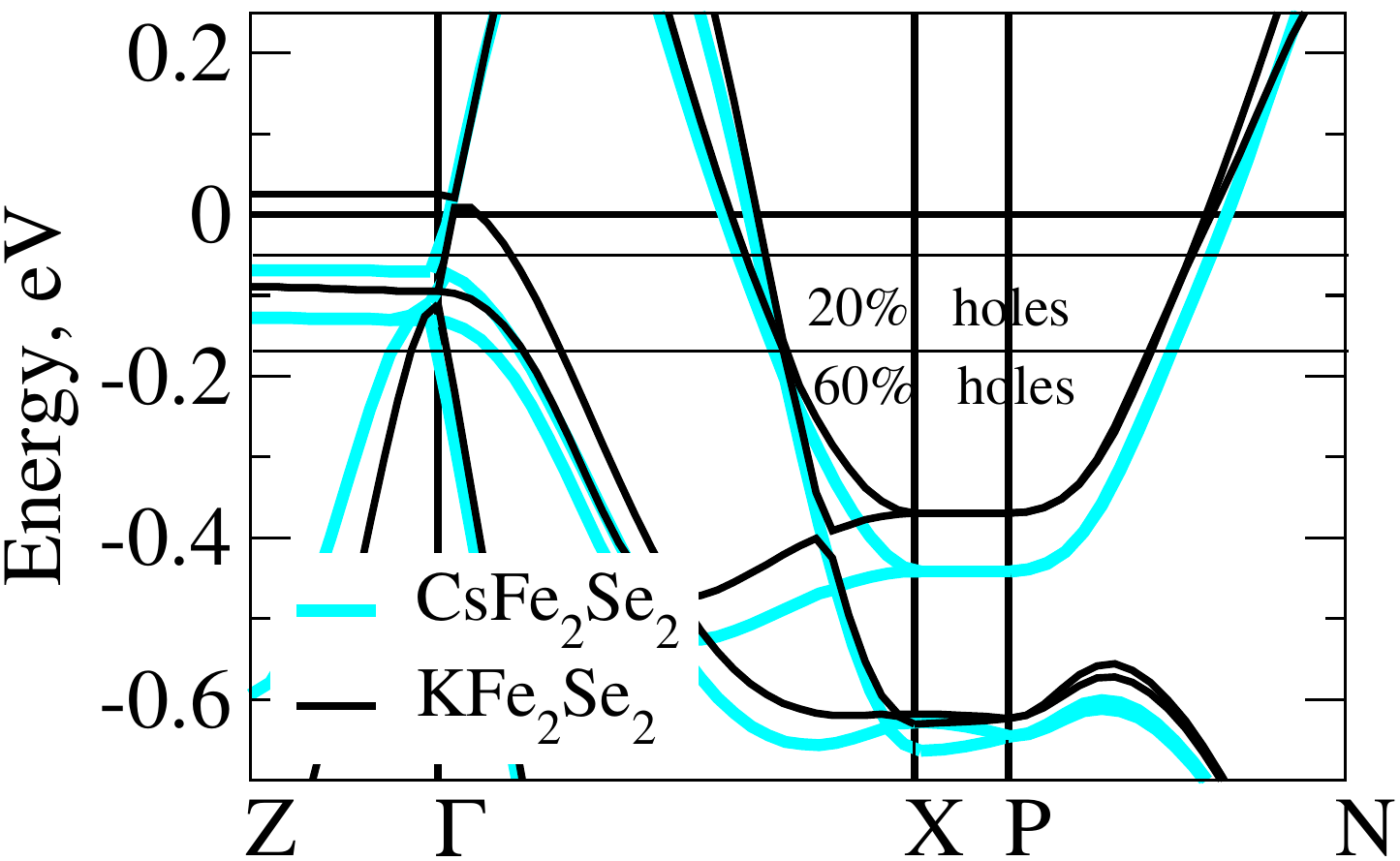}
\caption{(a) -- LDA bands of Ba122 close to the Fermi level ($E=$0) \cite{Ba122},
(b) -- LDA bands of K$_x$Fe$_2$Se$_2$ (black lines) and Cs$_x$Fe$_2$Se$_2$
(blue lines). Additional horizontal lines correspond to Fermi level at
20\% and 60\% hole doping \cite{KFe2Se2}.}
\label{122comp}
\end{figure}

LDA calculations of electronic spectrum of A$_x$Fe$_{2-y}$Se$_2$ (A=K,Cs)
system were performed immediately after its experimental discovery
\cite{KFe2Se2,KFe2Se2SI}. Rather unexpectedly this spectrum was found to be
qualitatively different from the spectrum of bulk FeSe and the spectra of all
the known FeAs systems. In Fig. \ref{122comp} we compare the spectrum of
BaFe$_2$As$_2$ (Ba122) \cite{Ba122}, which is typical for all FeSe based
systems, and the spectrum of A$_x$Fe$_{2-y}$Se$_2$ (A=K,Cs), obtained in
Ref. \cite{KFe2Se2}. We can see the clear difference of these spectra in the
vicinity of the Fermi level.

\begin{figure}
\includegraphics[clip=true,width=0.9\textwidth]{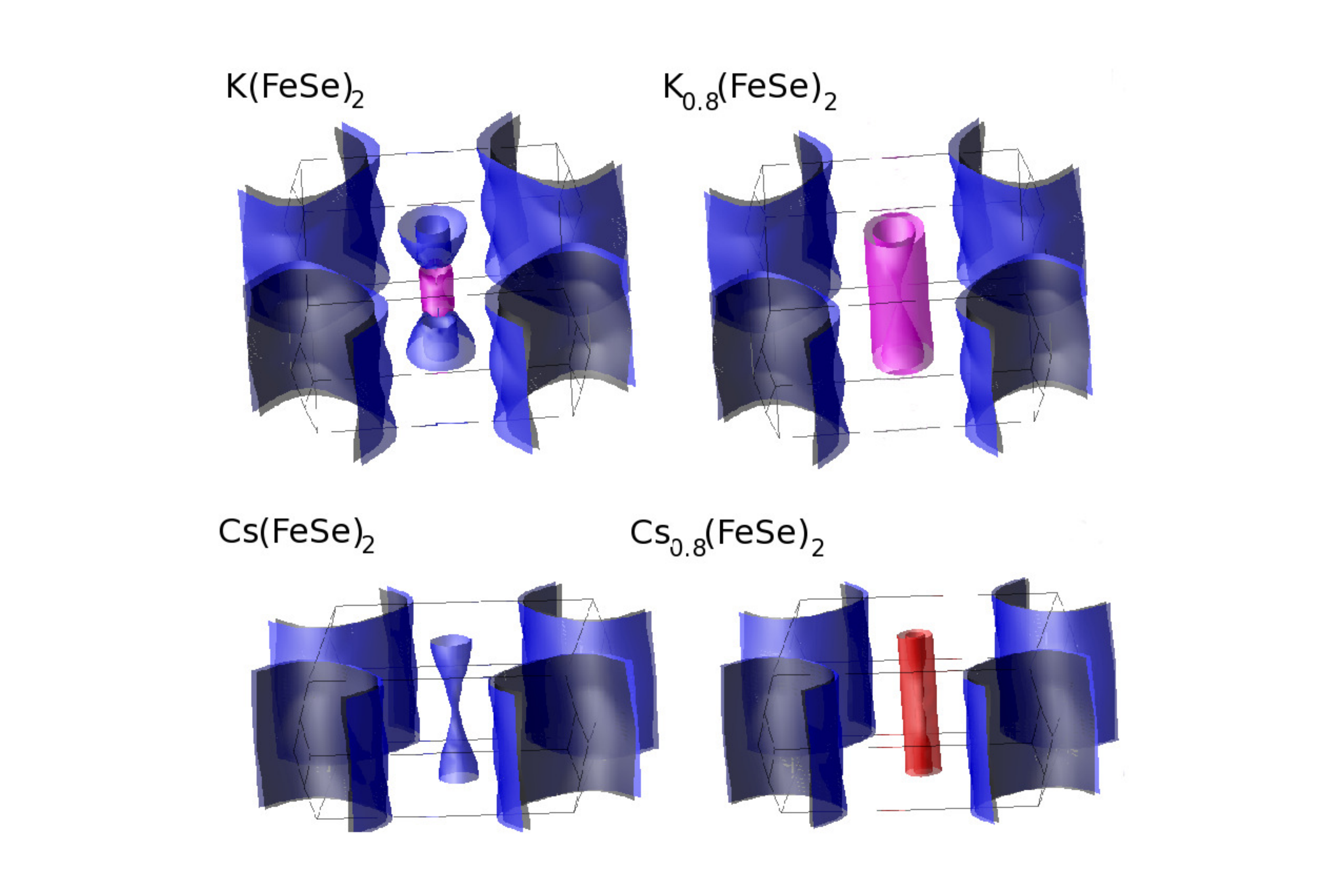}
\caption{Fermi surfaces of A$_x$Fe$_2$Se$_2$ (A=K,Cs) for stoichiometric
composition and for the case of 20\% hole doping \cite{KFe2Se2}. }
\label{FSAFeSe}
\end{figure}

In Fig. \ref{FSAFeSe} we show the Fermi surfaces calculated in Ref. \cite{KFe2Se2}
for two typical compositions of A$_x$Fe$_{2-y}$Se$_2$ (A=K,Cs). We can see that
these are quite different from the Fermi surfaces of FeAs systems --- in the
center of Brillouin zone there are only small (electron -- like!) Fermi surfaces,
while electron -- like cylinders at the corners of Brillouin zone are much larger.
The shape of Fermi surfaces typical for the bulk FeSe and FeAs based systems is
reproduced only for much larger (unreachable) hole doping levels \cite{KFe2Se2}.

\begin{figure}
\includegraphics[clip=true,width=0.85\textwidth]{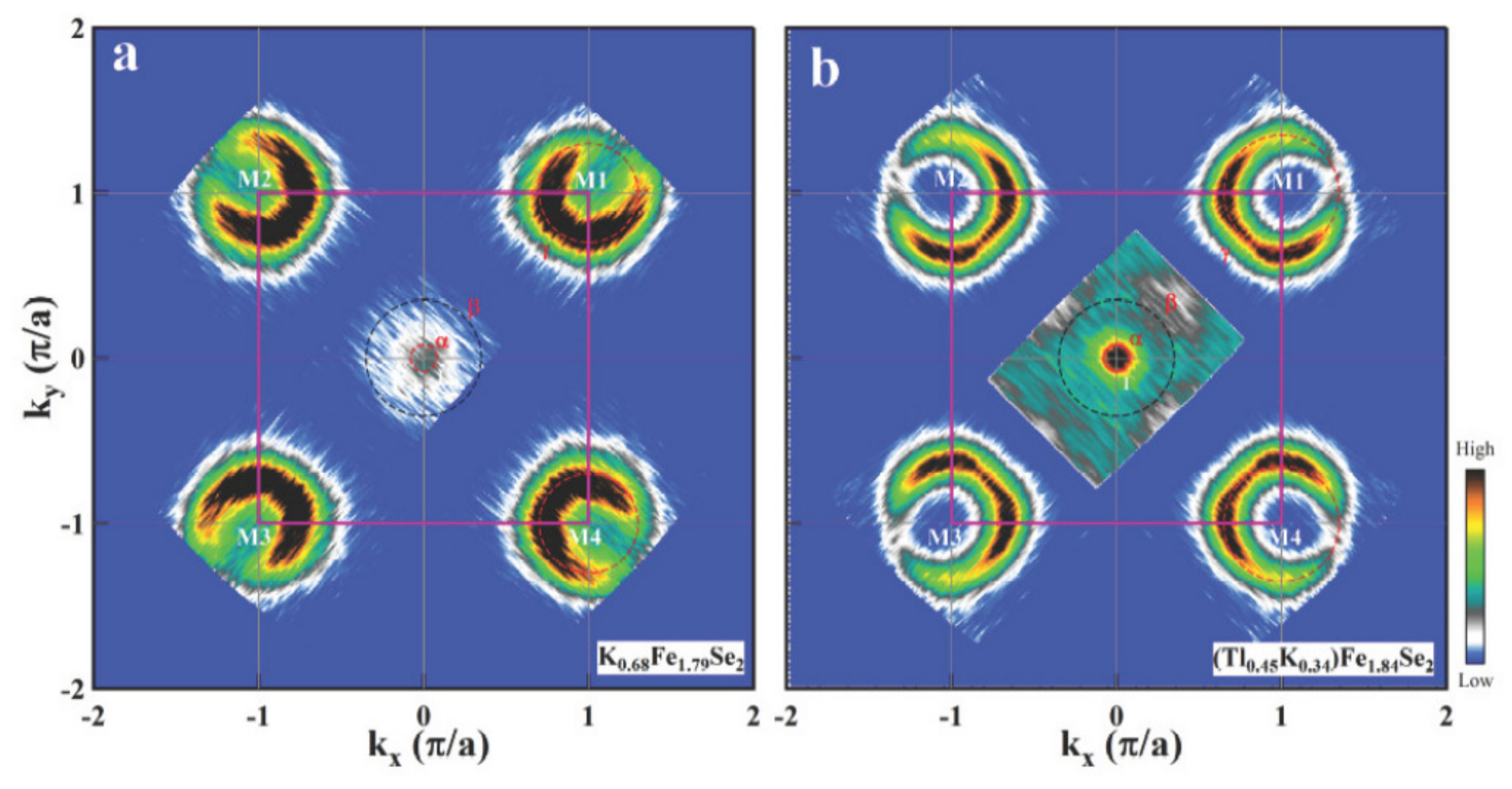}
\caption{ARPES Fermi surfaces of K$_{0.68}$Fe$_{1.79}$Se$_2$
($T_c$=32K) and  Tl$_{0.45}$K$_{0.34}$Fe$_{1.84}$Se$_2$ ($T_c$=28K)
\cite{KFe2Se2ARPES}.}   
\label{FSFeSeARP}
\end{figure}

This form of the Fermi surfaces in A$_x$Fe$_{2-y}$Se$_2$ was soon confirmed by
ARPES experiments. As an example, in Fig. \ref{FSFeSeARP} we show ARPES data
of Ref. \cite{KFe2Se2ARPES}, which are in obvious qualitative correspondence
with LDA calculations \cite{KFe2Se2,KFe2Se2SI}.

It is seen that in this system we can not speak of any, even approximate,
``nesting'' properties of electron -- like and hole -- like Fermi surfaces,
while it is precisely these properties that form the basis of the most of
theoretical approaches to microscopic description of FeAs based systems
\cite{MazKor}, where ``nesting'' of electron -- like and hole -- like Fermi
surfaces leads to the picture of well developed spin fluctuations, which
are considered as the main mechanism of pairing interaction.

LDA+DMFT calculations of K$_{1-x}$Fe$_{2-y}$Se$_2$ for different doping levels
were performed in Refs. \cite{KFeSeLDADMFT1,KFeSeLDADMFT2}. There, besides the
standard LDA+DMFT approach we have used also the modified LDA$'$+DMFT
developed by us in Refs. \cite{CLDA,CLDA_long}, which allows, in our opinion,
the more consistent solution of the ``double -- counting'' problem of
Coulomb interactions in LDA+DMFT.  For DMFT calculations we have chosen the
$U=3.75$~eV and  $J=0.56$~eV as the values of Coulomb and exchange interactions
in 3$d$ shell of Fe. As {\em impurity solver} we have used the Quantum Monte --
Carlo (QMC). The results of these calculations were directly compared with
ARPES data of Refs. \cite{KFe2Se2_ARPES,KFe2Se2_ARPES_2}.

It can be seen that for K$_{1-x}$Fe$_{2-y}$Se$_2$ system the correlation effects
play rather significant role. They lead to noticeable change of LDA dispersions.
In contrast to iron arsenides, where the quasi -- particle bands close to the
Fermi level remain well  defined, in K$_{1-x}$Fe$_{2-y}$Se$_2$ compounds, in
the vicinity of the Fermi level we observe rather strong suppression of
quasi -- particle bands. This reflects the fact, that correlation effects in
this system are more strong, than in iron arsenides. The value of correlation
renormalization (correlation narrowing) of the bands close to the Fermi level
is given by the factor of 4 or 5, while in iron arsenides this factor is usually
of the order of 2 or 3, for the same values of interaction parameters.

Results of these calculations are in general qualitative agreement with ARPES
data of Refs. \cite{KFe2Se2_ARPES,KFe2Se2_ARPES_2}, which also demonstrate the
strong damping of quasi -- particles in the immediate vicinity of the Fermi
level and stronger renormalization of effective masses in comparison to FeAs
systems. At the same time, our calculations do not reveal the formation of
unusually ``shallow'' ($\sim$ 0.05 eV deep below the Fermi level) electron --
like band at the $X$ point in Brillouin zone, which was observed in ARPES
experiments.

\subsection{[Li$_{1-x}$Fe$_x$OH]FeSe system}

\begin{figure}
\includegraphics[clip=true,width=0.35\textwidth]{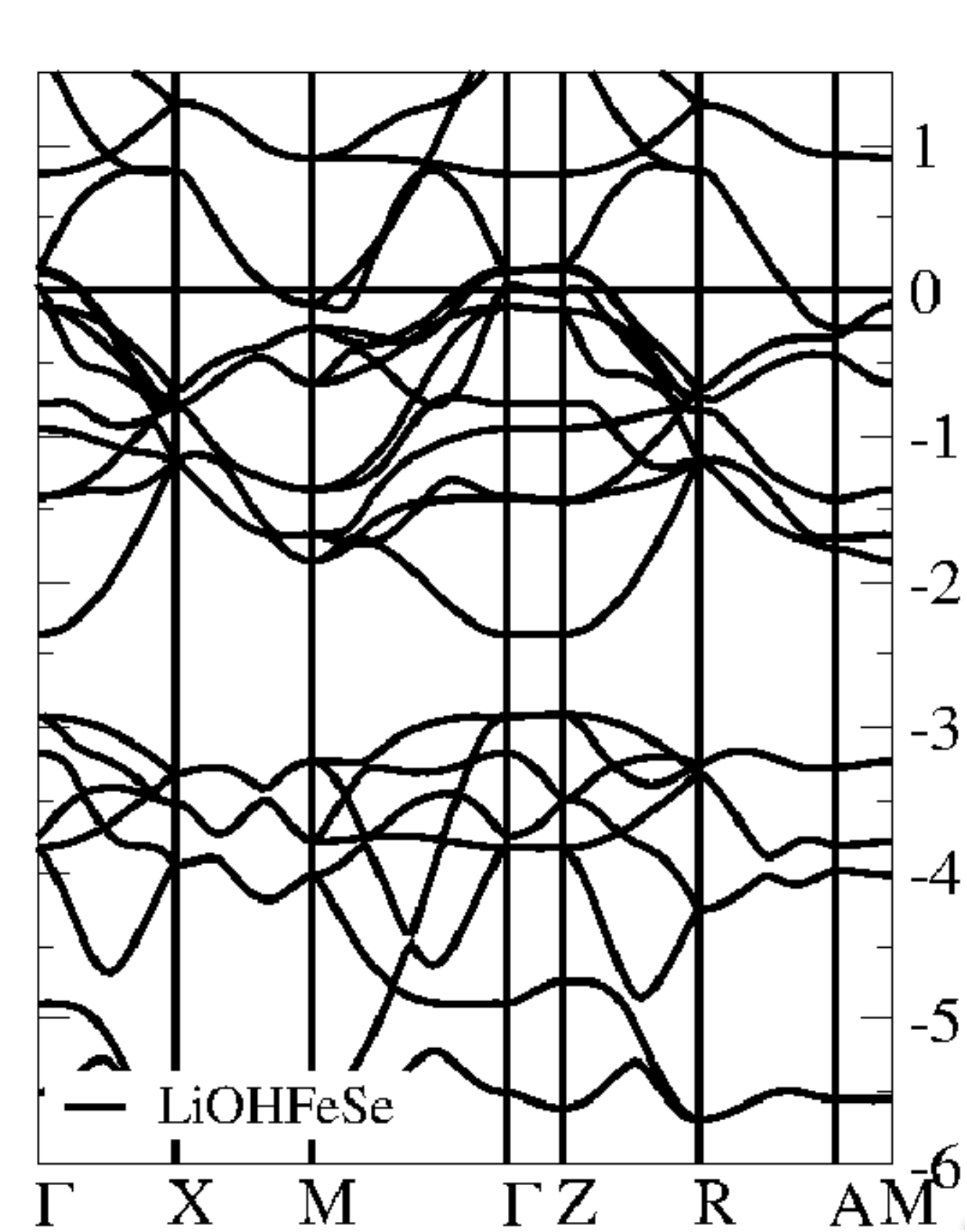}
\includegraphics[clip=true,width=0.42\textwidth]{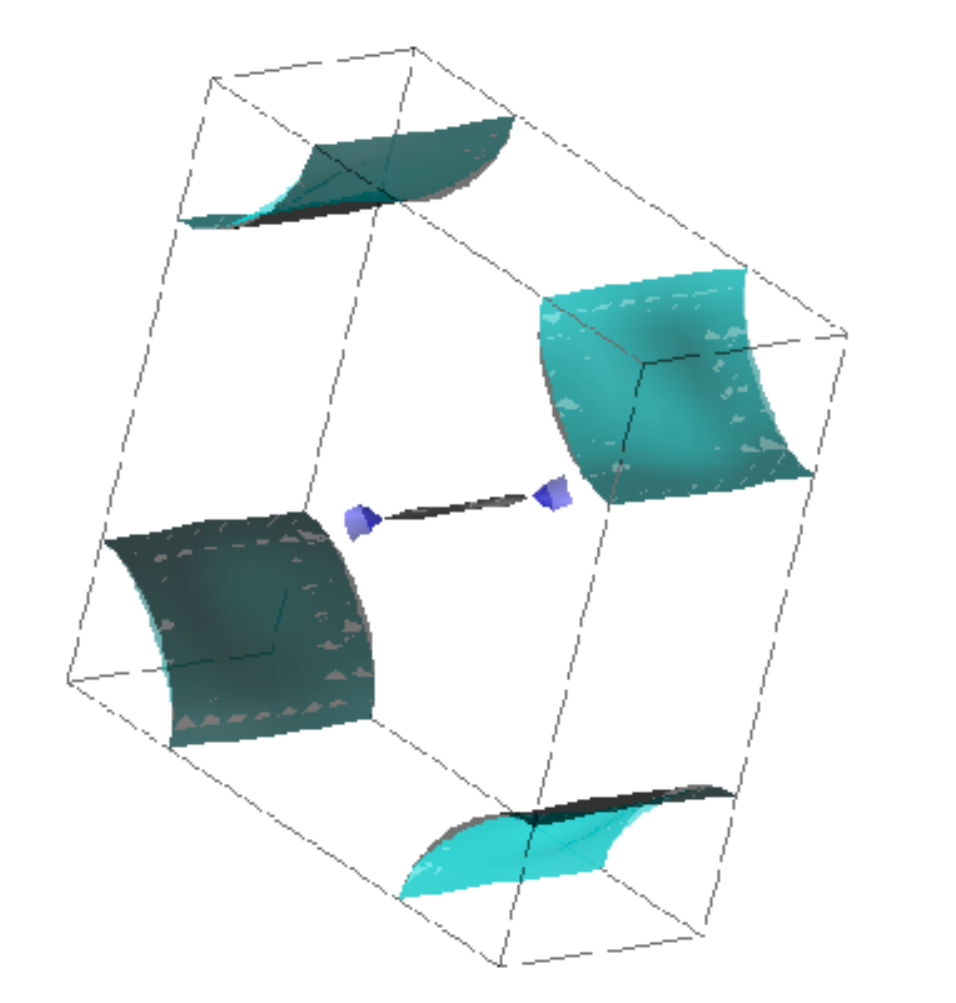}
\includegraphics[clip=true,width=0.45\textwidth]{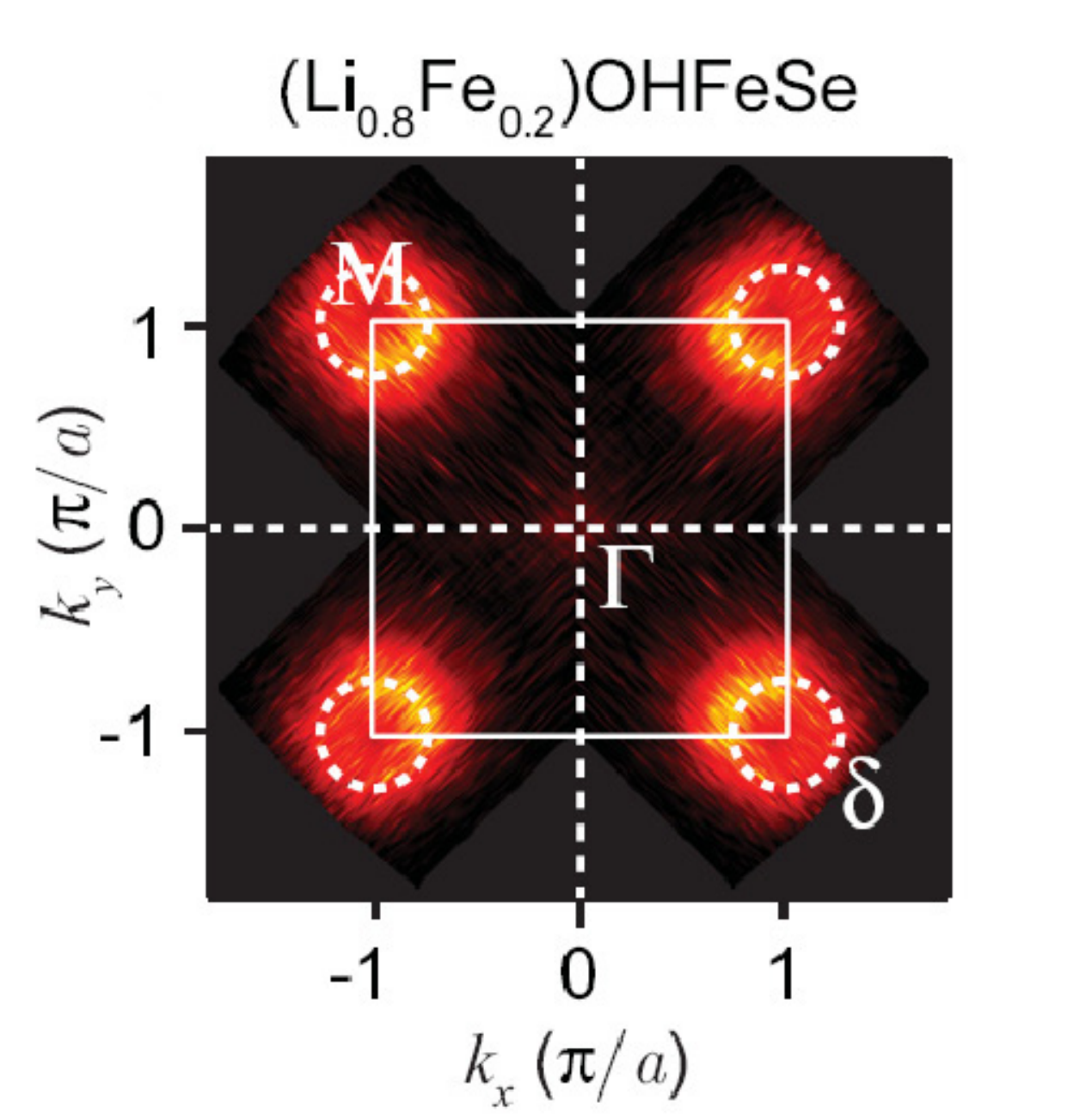}
\includegraphics[clip=true,width=0.45\textwidth]{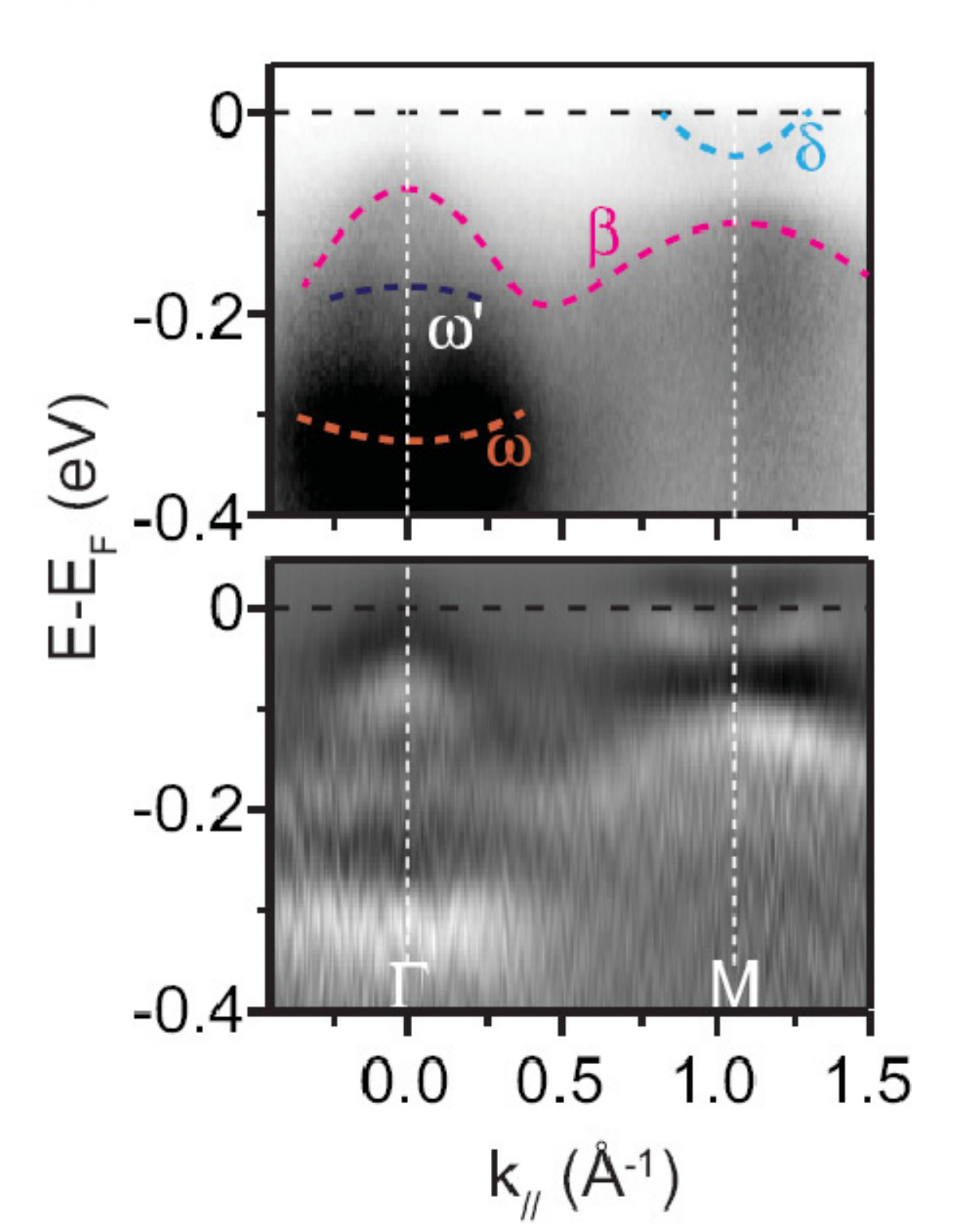}
\caption{(a) -- LDA bands of LiOHFeSe
(Fermi level at Ферми $E=$0) \cite{LiOHFeSe_NS},
(b) -- Fermi surface of LiOHFeSe corresponding doping level of 0.3
electrons per unit cell,
(c) -- ARPES Fermi surfaces of [Li$_{0.8}$Fe$_{0.2}$OH]FeSe \cite{LiOHFeSe_ARP},
(d) -- ARPES bands close to the Fermi level in [Li$_{0.8}$Fe$_{0.2}$OH]FeSe
\cite{LiOHFeSe_ARP}.}
\label{LiOHFeSe_spectr_FS}
\end{figure}

In Ref. \cite{LiOHFeSe_NS} we have performed LDA calculations of stoichiometric
LiOHFeSe compound, the appropriate results for energy dispersions are shown in
Fig. \ref{LiOHFeSe_spectr_FS} (a). On a first sight  the energy spectrum of this
system is quite analogous to the spectra of the majority of FeAs systems and
that of the bulk FeSe. In particular, the main contribution to the density of
states in rather wide energy region around Fermi level comes from $d$-states
of Fe, while the Fermi surfaces qualitatively have the same form as in the
majority of Fe based superconductors. However, this impression is wrong ---
in real [Li$_{0.8}$Fe$_{0.2}$OH]FeSe superconductor, the partial replacement of
Li by Fe in intercalating LiOH layers leads to significant {\em electron}
doping, so that the Fermi level goes upward in energy (as compared to
stoichiometric case) by 0.15 -- 0.2 eV. Then, as it is clear from Fig.
\ref{LiOHFeSe_spectr_FS} (a) hole -- like bands in the vicinity of $\Gamma$
point move below the Fermi level, so that hole -- like cylinders of the Fermi
surface just vanish. The general form of the Fermi surfaces for such electron
doping level following from LDA calculations is shown in Fig.
\ref{LiOHFeSe_spectr_FS} (b) and it has much in common with similar results for
A$_x$Fe$_{2-y}$Se$_2$ system (cf. Fig. \ref{FSAFeSe}). This conclusion is
confirmed by direct ARPES experiments \cite{LiOHFeSe_ARP}, the results of
these are shown in Fig. \ref{LiOHFeSe_spectr_FS} (c).

In particular, we can see from Fig. \ref{LiOHFeSe_spectr_FS} (b) that Fermi
surfaces consist mainly of electron -- like cylinders around points $M$,
while in the vicinity of $\Gamma$ point Fermi surface is either absent or is
quite small. In any case for this system there are no ``nesting'' properties
between electron and hole surfaces in any sense. Electronic dispersions
determined from ARPES are quite similar to corresponding dispersions measured
in Refs. \cite{KFe2Se2_ARPES,KFe2Se2_ARPES_2} for K$_{1-x}$Fe$_{2-y}$Se$_2$
system. These are qualitatively similar to dispersions  obtained in LDA
calculations, taking into account strong enough correlation narrowing of bands
(compression factor is actually different for different bands, as was shown by
LDA+DMFT calculations of Refs. \cite{KFeSeLDADMFT1,KFeSeLDADMFT2}).
At the same time the origin of unusually ``shallow'' electronic band $\sim$ 0.05
eV deep eV close to $M$ point remains unclear. To explain this we need unusually
strong correlation narrowing (while conserving the diameter of electron -- like
cylinders around the $M$ point, which practically coincides with the results of
LDA calculations), which is difficult to obtain from LDA+DMFT calculations.

In Ref. \cite{LiOHFeSe_NS} LSDA calculations of exchange parameters were
performed for different configurations of Fe ions, replacing Li in LiOH layers.
For most probable configuration, leading to magnetic ordering, the positive
(ferromagnetic) sign of exchange interaction was obtained and the simplest
estimate of Curie temperature has given the value of $T_C\approx$ 10K, in
excellent agreement with experimental data of Refs. \cite{LiOH2,LiOH3}, which
reported the observation of ferromagnetic ordering of Fe in LiOH layers.
At the same time, as we mentioned above, the other experiments had cast some
doubts on this conclusion.

\subsection{FeSe monolayer}

\begin{figure}
\includegraphics[clip=true,width=0.45\textwidth]{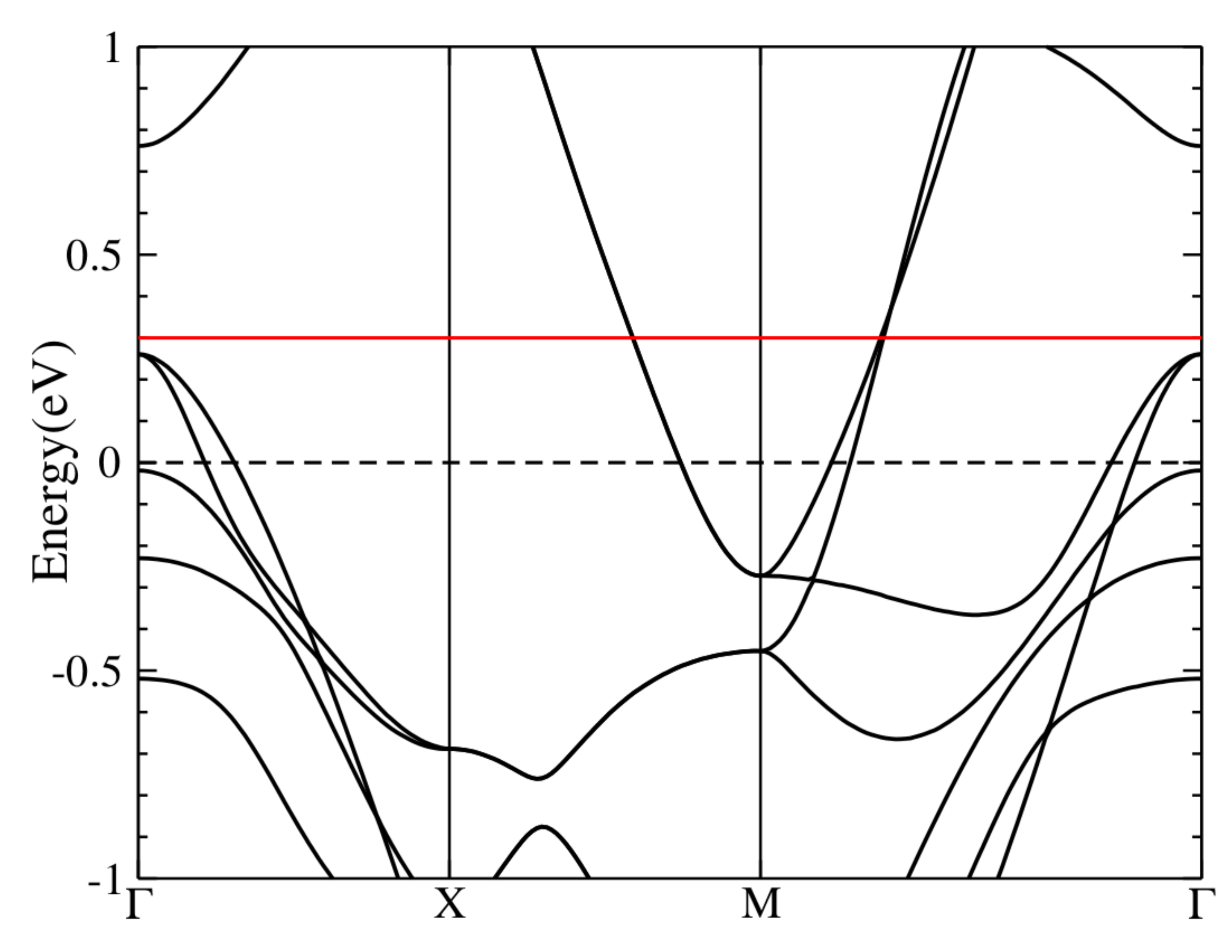}
\includegraphics[clip=true,width=0.5\textwidth]{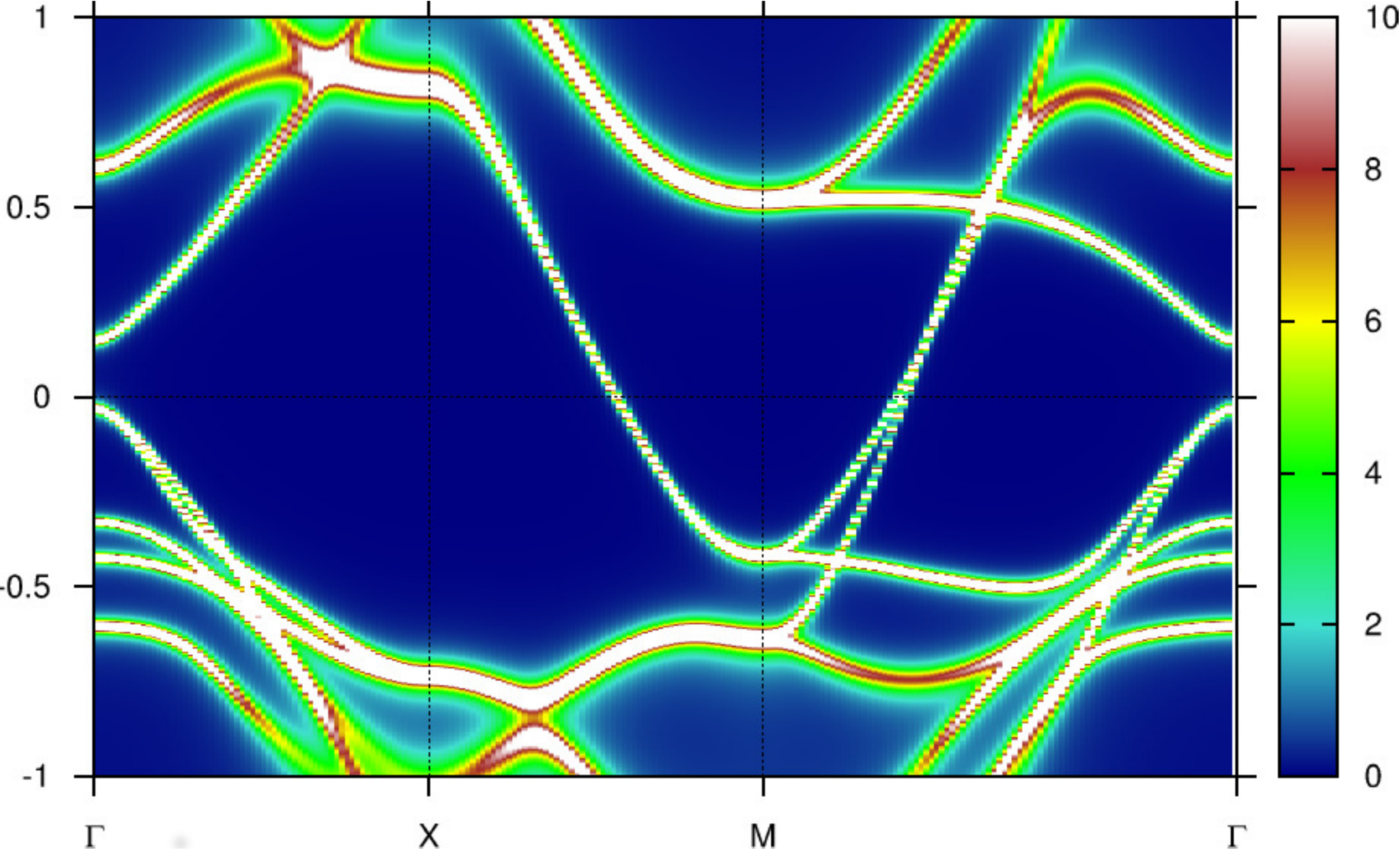}
\caption{(a) -- LDA bands of the single layer of FeSe close to the Fermi level
($E=$0). Horizontal red line denotes the approximate position of the Fermi
level, corresponding to electron doping level leading the Fermi surfaces
observed in ARPES experiments \cite{FNT_2016},
(b) -- LDA+DMFT calculated bands of single layer of FeSe close to the Fermi
level shifted by electron doping \cite{FNT_2016}.}
\label{FeSe_LDA_DMFT_bands}
\end{figure}

\begin{figure}
\includegraphics[clip=true,width=0.45\textwidth]{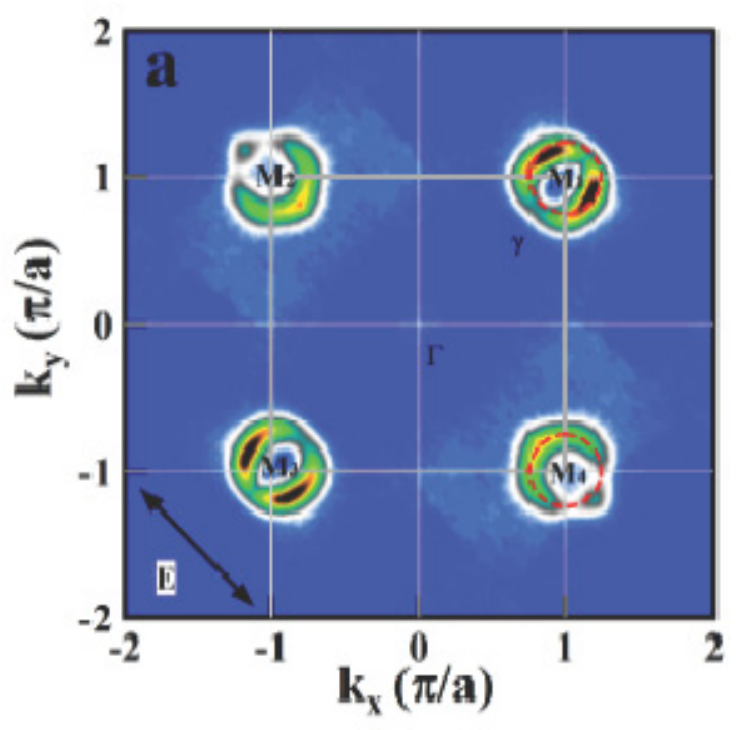}
\includegraphics[clip=true,width=0.5\textwidth]{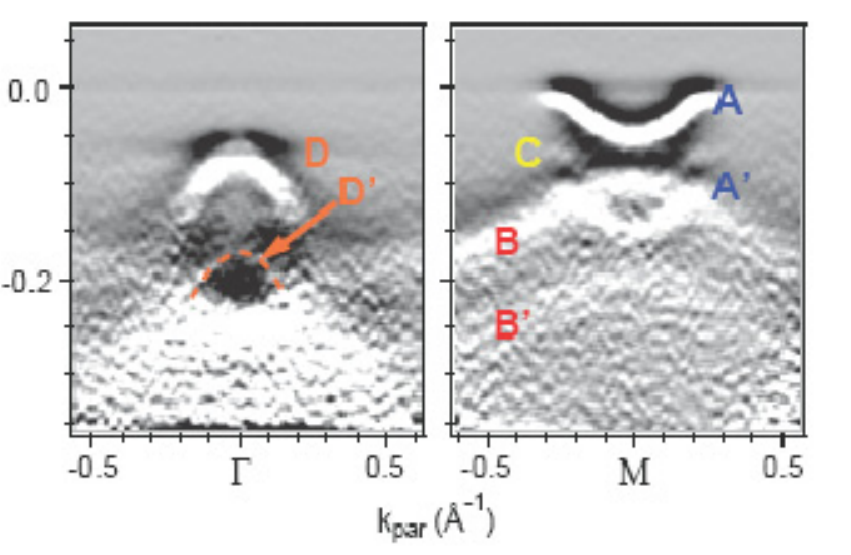}
\caption{(a) -- ARPES Fermi surface of single layer of FeSe \cite{ARP_FS_FeSe_1},
(b) -- ARPES bands of FeSe single layer close to the Fermi level
\cite{ARPES_FeSe_Nature}.}
\label{FeSe_ARPES_bands}
\end{figure}

Calculations of LDA spectra of single layer of FeSe can be done in a standard
way \cite{FNT_2016}. The results of such calculations are shown in Fig.
\ref{FeSe_LDA_DMFT_bands} (a). It is seen that the spectrum looks like the
typical for FeAs systems and the bulk FeSe, which was discussed in detail above.
However, the ARPES experiments \cite{ARP_FS_FeSe_1,ARPES_FeSe_Nature,ARP_FS_FeSe_2}
had shown convincingly that this is not so. In monolayer of FeSe on STO only
electron -- like Fermi surfaces are observed around points $M$ in Brillouin zone,
while hole -- like sheets around $\Gamma$ point (at the zone center) are just
absent. An example of this type of data is shown in Fig. \ref{FeSe_ARPES_bands}
(a) \cite{ARP_FS_FeSe_1}. Thus, similarly to the case of intercalated FeSe
systems, any kind of ``nesting'' properties are absent here. The apparent
contradiction with the results of LDA calculations has a simple qualitative
explanation --- the observed Fermi surfaces can be easily obtained assuming
that the system is electron doped, so that the Fermi level moves upward in
energy by $\sim$ 0.2 -- 0.25 eV, as shown by the red horizontal line in
Fig. \ref{FeSe_LDA_DMFT_bands} (a). This corresponds to doping level of the
order of 0.15 -- 0.2 per Fe ion.

Strictly speaking, the origin of this doping remains unclear, but there is a
general consensus that it is related to formation of oxygen vacancies in
SrTiO$_3$ substrate (in TiO$_2$ layer), appearing during different technological
operations (like annealing, etching etc.) used during the growth of the films
under study . It should be noted that the formation of electron gas at the
interface with SrTiO$_3$ is well known and was studied for rather long time
\cite{Shkl_16}. However, for FeSe/STO system of interest to us, this
problem was not studied in any detail (cf. though Refs. \cite{Mills,Agter}).

Electronic correlations influence of the spectrum of single layer of FeSe is
relatively weak. In Fig. \ref{FeSe_LDA_DMFT_bands} (b) we show the results of
LDA+DMFT calculations for the case of appropriately shifted (by electron doping)
Fermi level \cite{FNT_2016}. DMFT calculations were performed for the values
of Coulomb and exchange (Hund -- like) interactions strength in 3$d$ shell of
Fe, taken as $U=3.5$~eV and $J=0.85$~eV. As {\em impurity solver} we have used
here the continuous -- time  quantum Monte -- Carlo (CT--QMC), and dimensionless
inverse temperature was taken to be $\beta$=40. We can see that the spectrum
is only weakly renormalized by correlations and conserves LDA -- like form with
rather low bandwidth compression factor $\sim$ 1.3.

Electronic dispersions in FeSe monolayer films were measured by ARPES in a
number of works, e.g. in Refs. \cite{FeSe_BTO,ARPES_FeSe_Nature}. Results of
Ref.  \cite{ARPES_FeSe_Nature} are presented in Fig. \ref{FeSe_ARPES_bands} (b).
These are in agreement with data obtained in other papers and are, in general,
analogous to the similar data obtained for intercalated FeSe systems
(cf. e.g. Fig. \ref{LiOHFeSe_spectr_FS} (c)). In general, these data are also
qualitatively similar to the results of LDA+DMFT, but the quantitative agreement
is absent. In particular, ARPES experiments clearly demonstrate the presence
of unusually ``shallow'' electron -- like band at the $M$ point, with Fermi
energy $\sim$ 0.05 eV, while in theoretical calculations this band is almost an
order of magnitude ``deeper''.

It should also be noted that in Ref.  \cite{ARPES_FeSe_Nature} it was observed
for the first time, that a ``shadow'' electron -- like band exists at
$M$ point, which is about 100 meV below the main band and is a kind of a
``replica'' of this band. This band is clearly seen in Fig.
\ref{FeSe_ARPES_bands} (b). Such a ``shadow'' band is absent in band structure
calculations. The nature of this band and its possible significance for
high -- temperature superconductivity in monolayers of FeSe on STO will be
discussed in some detail below, in connection with possible mechanisms of
enhancement of $T_c$.

\begin{figure}
\includegraphics[clip=true,width=0.75\textwidth]{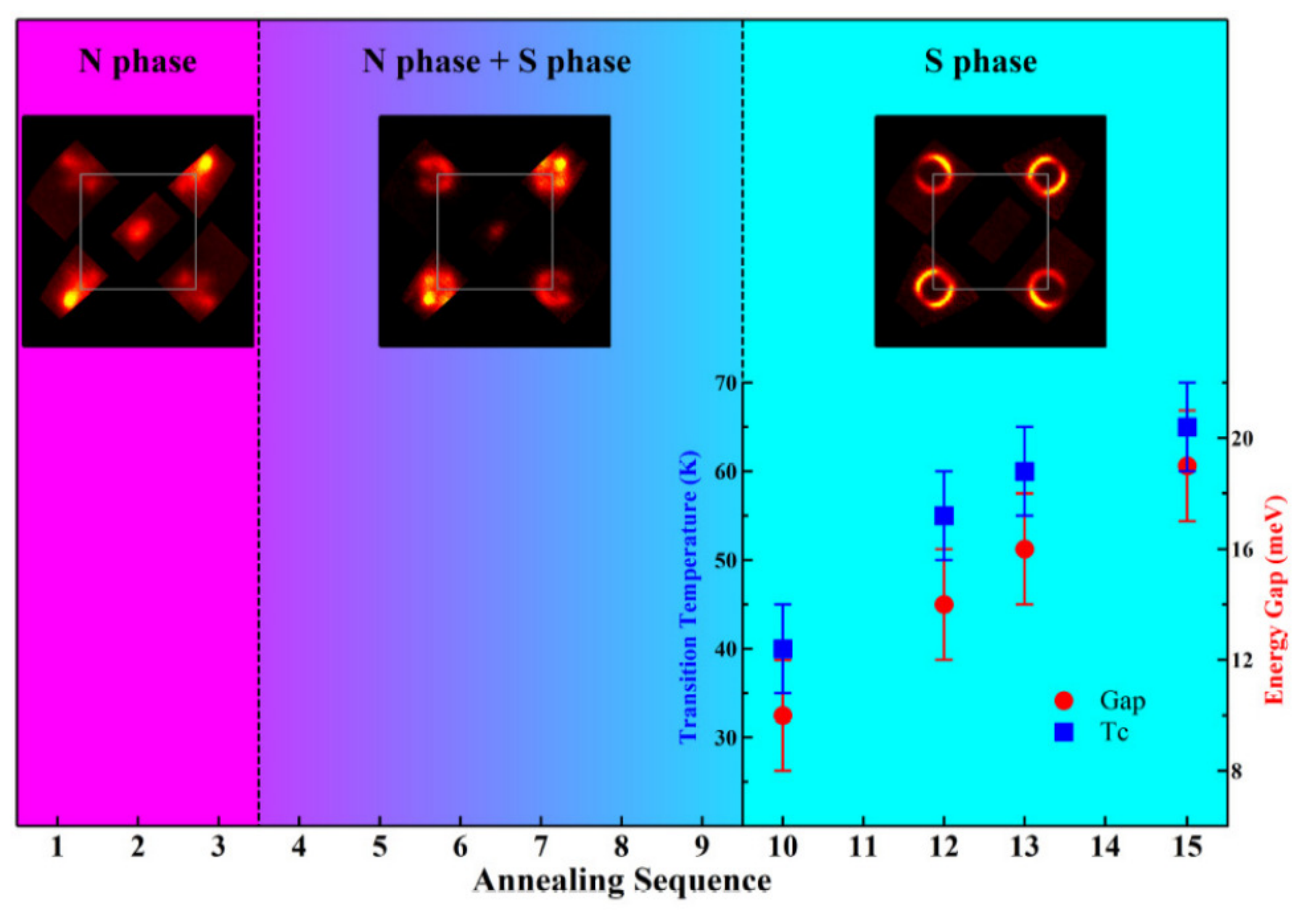}
\includegraphics[clip=true,width=0.5\textwidth]{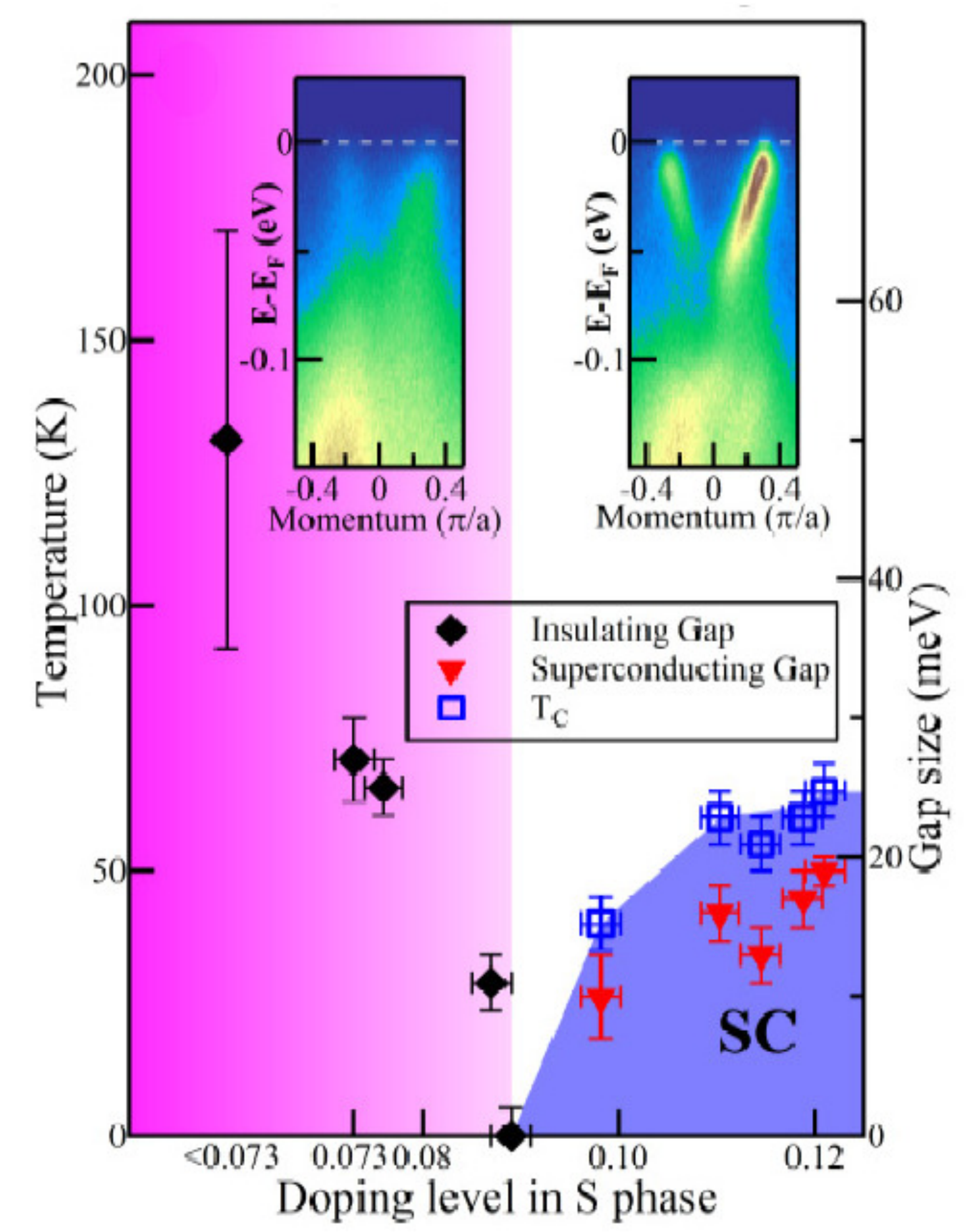}
\caption{Phase diagram of FeSe single layer on SrTiO$_3$:
(a) -- schematic phase diagram obtained in Ref. \cite{FeSe1UC_Ph_Dia1} on the
series of samples with different electron dopings. Also shown are corresponding
ARPES Fermi surfaces and in superconducting phase ARPES measured values of
superconducting gap and $T_c$.
(b) -- phase diagram obtained from ARPES measurements in Ref.
\cite{FeSe1UC_Ph_Dia2} and demonstrating the existence of insulating and
superconducting phases. The values of superconducting and insulating gaps
were also obtained from ARPES measurements.}
\label{FeSe_PD}
\end{figure}

As we noted above electron doping level of FeSe monolayers on STO is rather
poorly controlled parameter. However, in a number of papers, using different
procedures of film annealing {\em in situ}, the authors successfully made
ARPES experiments on samples with different doping levels
\cite{FeSe1UC_Ph_Dia1,FeSe1UC_Ph_Dia2}. These experiments allowed to obtain
some kind of phase diagrams of FeSe/STO system. In particular, in
Ref. \cite{FeSe1UC_Ph_Dia1} a series of samples was demonstrated with
consequent transitions from the topology of the Fermi surface typical of FeAs
systems and bulk FeSe (with Fermi surface sheets around $\Gamma$ point in the
center of Brillouin zone) to topology of Fermi surface sheets around $M$ point.
It was shown, that superconductivity with high $T_c$ appears only in samples
without central Fermi surface sheets, while the samples with typical Fermi
surface topology remain in the normal phase. Schematically these results are
shown in Fig. \ref{FeSe_PD} (a). The presence of superconductivity was determined
from ARPES measurements of the energy gap at the Fermi level, and $T_c$ was
derived from the temperature dependence of the gap.

In Ref. \cite{FeSe1UC_Ph_Dia2} similar measurements were done with electron
concentration controlled by by measurements of the area of electron pockets of
the Fermi surface around $M$ points. The obtained phase diagram is shown in
Fig. \ref{FeSe_PD} (b), where we see insulating (antiferromagnetic?) phase at
low doping levels and superconducting phase at dopings exceeding the critical
value $\sim$ 0.09, corresponding to the quantum critical point. These
conclusions are also based on ARPES measurements of superconducting and
insulating energy gaps in the spectrum and their temperature dependencies.

It is obvious that the results of Refs. \cite{FeSe1UC_Ph_Dia1,FeSe1UC_Ph_Dia2}
are in some contradiction with each other, in particular, the nature of
insulating phase observed in \cite{FeSe1UC_Ph_Dia2} remains unclear.

\section{Possible mechanisms of $T_c$ enhancement in iron -- selenium monolayers}

\subsection{Correlation of $T_c$ and the density of states}

Let us now start discussing the mechanisms of high -- temperature 
superconductivity in systems under consideration. Concerning FeAs based
superconductors there is general consensus in the literature.
Electron -- phonon mechanism of Cooper pairing is considered to be insufficient
to explain the high values of $T_c$ in these systems \cite{Sad_08} and the
preferable mechanism is assumed to be the pairing due to exchange of
antiferromagnetic fluctuations. The repulsive nature of this
interaction leads to the picture of $s^{\pm}$--pairing with different signs of
superconducting order parameter (gap) $\Delta$ on hole -- like
(around the $\Gamma$ point at the center of Brillouin zone) and hole -- like
(around $M$ points in the corners of the zone) sheets of the Fermi surface 
\cite{MazKor}. However, when we consider the systems based on monolayers of FeSe,
this picture obviously becomes inconsistent --- the observed topology of Fermi
surfaces with complete absence of any ``nesting'' electron -- like and hole -- 
like sheets or even with total absence of hole -- like Fermi surfaces clearly
contradicts this picture. There is simply no obvious way to form 
well developed spin (antiferromagnetic) fluctuations. Thus, we shall start
with the elementary analysis based on simple BCS model.
 
\begin{figure}[ht]
\includegraphics[clip=true,width=0.75\textwidth]{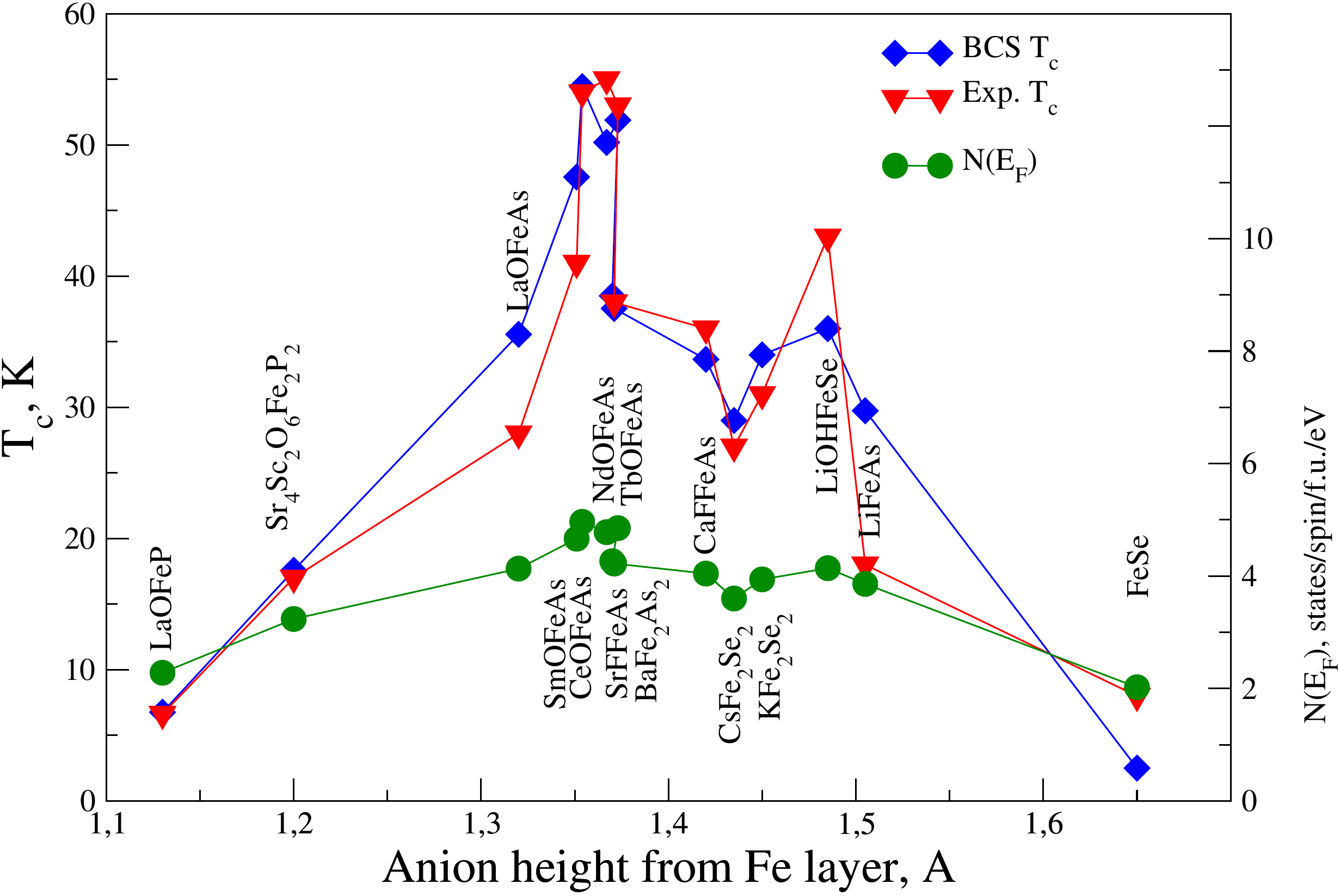}
\caption{LDA calculated values of the density of states at the Fermi level
$N(E_F)$ (circles, right axis) and superconducting critical temperature $T_c$
(left axis), obtained from elementary BCS -- like estimates (stars) and
experimental values of $T_c$ (triangles) as functions of anion height
$\Delta z_a$ over Fe layer for different iron based superconductors.}
\label{z_Tc}
\end{figure}

In Ref. \cite{hPn2} an interesting empirical dependence was discovered between
the temperature of superconducting transition T$_c$ in FeAs and FeSe systems and
the height of the anion (As or Se) $\Delta z_a$ above the Fe plane (layer) (cf.
Fig. \ref{StrucFePn}). A sharp maximum of $T_c$ was observed for systems with
$\Delta z_a\sim$1.37\AA. In Refs. \cite{JMMM,Kucinskii10} we presented the results
of systematic LDA calculations of the total density of states at the Fermi level
$N(E_F)$ for a wide choice of (stoichiometric) FeAs and FeSe based systems with
different values of $\Delta z_a$ (cf. Table \ref{tab1}). The obtained
non -- monotonous dependence of the density of states on $\Delta z_a$, shown in
Fig. \ref{z_Tc} (circles), which is determined by hybridization effects, in
principle, is sufficient to explain the corresponding dependence of $T_c$.

\begin{table}[htb]
\center
\footnotesize
\caption{Total LDA calculated density of states N(E$_F$) and the values of  
T$_c$ for iron based superconductors.}
\label{tab1}

\begin{tabular}{|l|c|c|c|c|}
\hline
Система                        &$\Delta z_a$, \AA & N(E$_F$),      & T$^{BCS}_c$, K &  T$_c^{exp}$, K\\
                              &                  & states/cell/eV &                &                \\
\hline
LaOFeP                        & 1.130            & 2.28           &    3.2         &     6.6        \\
Sr$_4$Sc$_2$O$_6$Fe$_2$P$_2$  & 1.200            & 3.24           &    19          &     17         \\
LaOFeAs                       & 1.320            & 4.13           &    36          &     28         \\
SmOFeAs                       & 1.354            & 4.96           &    54          &     54         \\
CeOFeAs                       & 1.351            & 4.66           &    48          &     41         \\
NdOFeAs                       & 1.367            & 4.78           &    50          &     53         \\
TbOFeAs                       & 1.373            & 4.85           &    52          &     54         \\
SrFFeAs                       & 1.370            & 4.26           &    38          &     36         \\
BaFe$_2$As$_2$                & 1.371            & 4.22           &    38          &     38         \\
CaFFeAs                       & 1.420            & 4.04           &    34          &     36         \\
CsFe$_2$Se$_2$                & 1.435            & 3.6            &    29          &     27         \\
KFe$_2$Se$_2$                 & 1.45             & 3.94           &    34          &     31         \\
LiOHFeSe                      & 1.485             & 4.14           &    36          &     43         \\
LiFeAs                        & 1.505            & 3.86           &    31          &     18         \\
FeSe                          & 1.650            & 2.02           &     3          &     8         \\

\hline
\end{tabular}
\end{table}
\normalsize

Corresponding dependence of $T_c$ on $\Delta z_a$ can be easily estimated along
the lines of elementary BCS model, using the usual expression
$T_c=1.14\omega_D e^{-1/\lambda}$, taking into account that $N(E_F)$ directly
enters dimensionless pairing interaction constant $\lambda=gN(E_F)$ (where $g$
is corresponding dimensional coupling constant). Taking, rather arbitrary value
$\omega_D$=350 K (which may be related to characteristic value of phonon
frequencies in FeAs systems \cite{Sad_08}) we can determine the value of
$g$ fitting experimental value of $T_c$, e.g. for Ba122 system ($\sim$ 38 K),
which gives $\lambda$=0.43. Fixing this value of $g$, we can easily
recalculate the values of $T_c$ for all other systems, just taking appropriate
values of the density of states from LDA calculations (cf. Fig. \ref{z_Tc}).
Corresponding values of $T_c$ given in Table \ref{tab1} and shown in Fig.
\ref{z_Tc} (stars) are in very reasonable agreement with experimental values,
shown in the same Figure  (triangles), which are also given in Table \ref{tab1}.

FeSe systems in general just fit this dependence. This can be seen from the data
of Table \ref{tab1} and Fig. \ref{z_Tc}.  For example for [Li$_{1-x}$Fe$_x$OH]FeSe
system the calculated value of the density of states for stoichiometric
composition LiOHFeSe is $N(E_F)$=4.14 states/cell/eV and elementary estimate of
$T_c$ yields T$_c$=36K, which is somehow lower than the experimental value
$T_c$=43K. However, introduction of Fe into LiOH layers shifts the Fermi level,
so that it moves to a higher value of $N(E_F)$=4.55 states/cell/eV ,
leading to the appropriate growth of что $T_c$ up to 45K, which is very close to
experimental value \cite{LiOHFeSe_NS}.

It should be stressed that the rough estimates given above does not necessarily
mean that we assume electron -- phonon pairing mechanism for these systems, and
$\omega_D$ in BCS expression can be considered just as a characteristic frequency
of any kind of Boson excitations responsible for pairing (e.g. magnetic
fluctuations). These results show that there is an obvious correlation between
experimental values of $T_c$ and the value of the total density of states
at the Fermi level, obtained via band structure calculations for stoichiometric
(!) compositions of FeAs and FeSe based compounds. Similar results can be
obtained using more complicated expressions for $T_c$ like McMillan or
Allen -- Dynes formulas \cite{Kucinskii10}.

At the same time, for the single layer of FeSe LDA calculations produce the
value $N(E_F)\approx$ 2 states/cell/eV, which is practically the same as for
the bulk FeSe and is weakly changing with electron doping (Fermi level shift)
\cite{FNT_2016}. Corresponding elementary estimate of $T_c$ does not produce
the values higher than 8K, so that the appearance of high values of $T_c$
in this case can not be explained from similar simple considerations.

However, there is a number of experimental papers, where the significant
increase of $T_c$ were reported up to the values of the order of 40K in
bulk crystals and multilayer films of FeSe under electron doping, achieved by
the coverage of the surface of FeSe by alkali metal atoms (sodium)
\cite{K_dop_1,K_dop_2,K_dop_3}. It is possible that this treatment has lead to
intercalation of FeSe layers by alkali metal, so that these systems were
transformed into an analogue of intercalated FeSe systems, similar to those
discussed above, and the growth of $T_c$ was related to the growth of $N(E_F)$.
This point of view is confirmed by calculations presented in Ref. \cite{K_dop_band}.
However, the growth of $T_c$ up to the values $>$ 40K in a number of papers was
achieved by doping of FeSe induced by strong electric field (at the gate) in the
field -- effects transistor structures \cite{Field_1,Field_2,Field_3}, where
similar explanation seems less probable.

\subsection{Multiple bands picture of superconductivity}

The basic feature of electronic spectrum of iron pnictide and chalcogenide
superconductors is its multiple band character --- in general case the Fermi
level is crossed by several bands, formed by $d$ -- states of Fe, so that there
appear several sheets (pockets) of the Fermi surface (electron and hole -- like)
\cite{Sad_08,MazKor,Kord_12}. In superconducting state the energy gap can open
on each of these sheets and the values of these gaps can be quite different from
each other \cite{Sad_08,Kord_12}. Thus, the elementary description of
superconductivity based on the single -- band BCS model used in the previous
section is in fact oversimplified. Below, following mainly Refs.
\cite{BGrk,KucSad08} we shall briefly describe the multiple -- band formulation
of BCS model with application to Fe based superconductors.

Consider the simplified version of electronic structure (Fermi surfaces) of
the square lattice of Fe, shown in Fig. \ref{BZFS} (b), with two hole -- like
pockets around the $\Gamma$  point and two electron -- like pockets around
$X$ and $Y$  points (in Brillouin zone for the square lattice with one Fe ion
per unit cell). Let $\Delta_i$ denotes superconducting order parameter (energy
gap) on $i$-th sheet (pocket) of the Fermi surface (in Fig. \ref{BZFS} (b)
$i$=1,2,3,4). The value of $\Delta_i$ is determined by self -- consistency
equation for corresponding anomalous Green's function in Gorkov's system of
equations \cite{BGrk}.

Pairing interaction in multiple -- band BCS model can be written in the matrix
form:
\begin{equation}
\hat V=\left(\begin{array}{cccc}
u & w & t & t \\
w & u' & t' & t' \\
t & t' & \lambda & \mu \\
t & t' & \mu & \lambda
\end{array}\right).
\label{mmatr}
\end{equation}
where matrix elements $V^{i,j}$ determine intraband and interband coupling
constants. For example, $\lambda = V^{eX,eX} = V^{eY,eY}$ determines the
pairing interaction on the same electron -- like pocket at $X$ or $Y$ points,
while $\mu = V^{eX,eY}$ connects electrons on different pockets at  $X$ and $Y$.
Constants $u = V^{h1,h1}$, $u'=V^{h2,h2}$ and $w=V^{h1,h2}$ characterize BCS
interaction on hole -- like pockets --- the smaller one ($h1$) and larger one
($h2$), and between them, while $t = V^{h,eX}=V^{h,eY}$ pair electrons at points
$X$ and $\Gamma$.

For the temperature of superconducting transition the standard BCS -- like
expression appears:
\begin{equation}
T_c = \frac{2 \gamma \omega_c}{\pi}\exp\left(- \frac{1}{g_{eff}}\right),\
\gamma\approx 1.78
\label{TC1}
\end{equation}
whereгe $\omega_c$ --- is the usual cut--off parameter in Cooper channel (for
simplicity we assume, that this parameter is the same for all pairing
interactions, while the generalization for say two characteristic cut--off
frequencies is rather direct \cite{Hirshf}), and $g_{eff}$ represents an
{\em effective} pairing constant, determined from solubility condition for the
system of linearized gap equations:
\begin{equation}
g_{eff}\Delta_{i}  =   \sum_{j}g_{ij}\Delta_{j} \,,
\label{gapeqlin}
\end{equation}
where
\begin{equation}
g_{ij} \equiv  - V^{i,j} \nu_{j},\quad g_{eff}^{-1}=
\ln \frac{2\gamma}{\pi}\frac{\omega_c}{T_c}.
\label{intmatr}
\end{equation}
is the matrix of dimensionless pairing constants $g_{ij}$ is determined by the
products of matrix elements (\ref{mmatr}) and {\em partial} densities of states
on different Fermi surface pockets -- $\nu_j$ denotes the density of states per
one spin projection on $j$-th pocket (cylinder).

From symmetry it is clear that $\nu_3=\nu_4$, so that system of
Eqs. (\ref{gapeqlin}) can produce two types of solutions \cite{BGrk}:

\begin{enumerate}

\item{Solution, corresponding to $d_{x^2 - y^2}$ pairing, when the gaps on
different sheets at points $X$ and $Y$ differ by sign, while gaps on hole --
pockets are equal to zero:

\begin{equation}
\Delta_1 = \Delta_2 = 0, \ \
\Delta_3 = - \Delta_4 = \Delta,
\end{equation}
or, as a special case, when corresponding pockets are just absent.}

\item{Solutions, corresponding to the so called $s^{\pm}$ pairing, when gaps at
points $X$ and $Y$ are equal: $\Delta_3 =  \Delta_4$, while gaps on Fermi
surface pockets surrounding the point $\Gamma$ have the different sign in case
of {\em repulsive} interaction between electron and hole pockets -- $t>0$,
and usual $s$ -- wave pairing, when gaps on electron and hole pockets have the
same sign in the case attraction --- $t<0$.}

\end{enumerate}

\begin{figure}[ht]
\includegraphics[clip=true,width=0.65\textwidth]{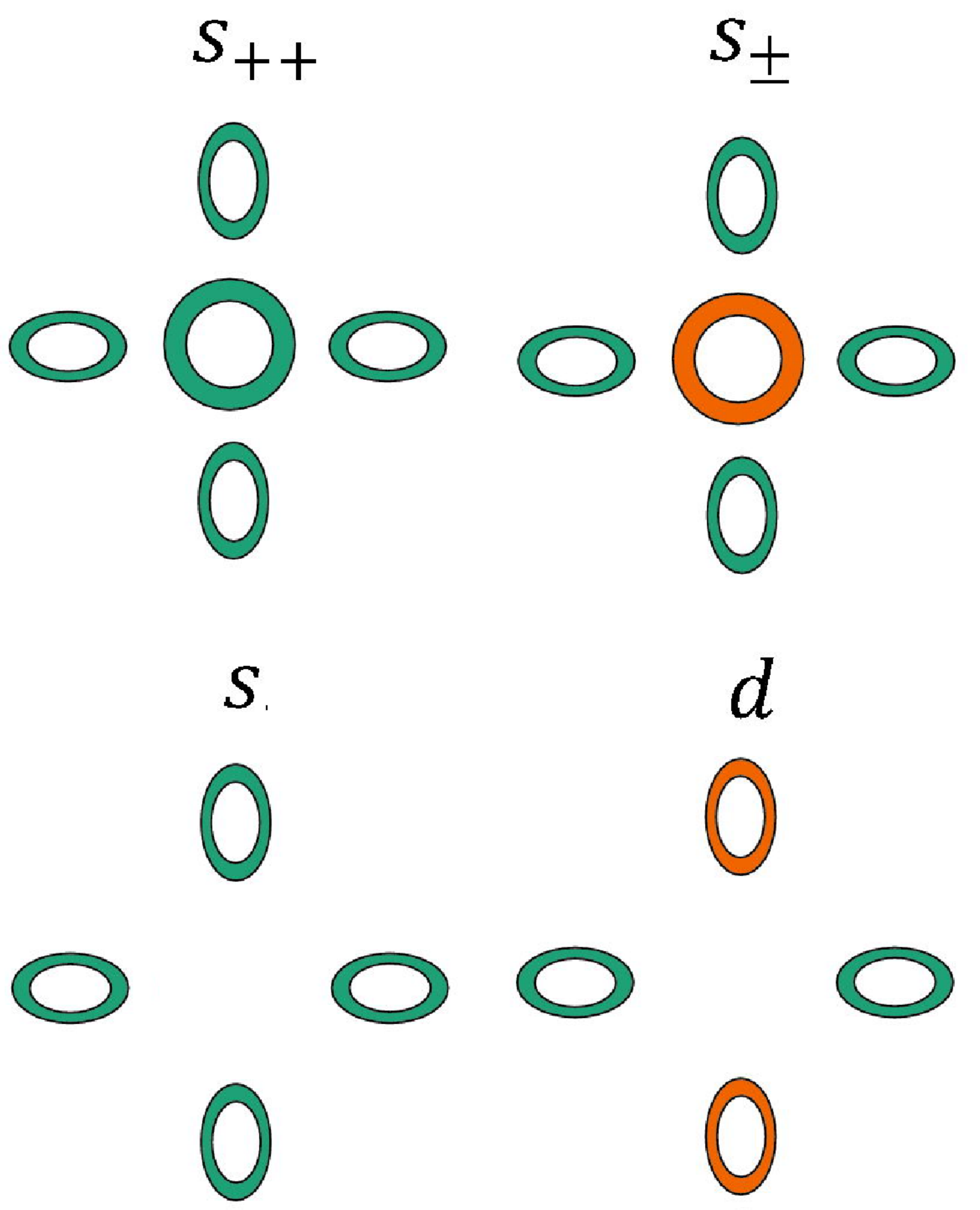}
\caption{Main type of pairing in multiple -- band scheme for superconductivity
in FeAs and FeSe systems. Different colors represent different signs of
superconducting gaps.}
\label{symm_pair}
\end{figure}

\noindent All these variants are shown qualitatively in Fig. \ref{symm_pair}.

\noindent In the first case we obtain for the effective pairing constant:
\begin{equation}
g_{eff} = (\mu - \lambda) \nu_3.
\label{dwve}
\end{equation}

\noindent In second case we have  $\Delta_3 =  \Delta_4$ and $\nu_3=\nu_4$,
so that two equations in (\ref{gapeqlin}) just coincide and instead of
(\ref{mmatr}), (\ref{intmatr}) appears the coupling matrix $3\times 3$ of the
following form:
\begin{equation}
-\hat g=\left(\begin{array}{ccc}
u\nu_1 & w\nu_2 & 2t\nu_3 \\
w\nu_1 & u'\nu_2 & 2t'\nu_3 \\
t\nu_1 & t'\nu_2 & 2\bar \lambda \nu_3
\end{array}\right),
\label{effmatr}
\end{equation}
where $\bar\lambda =\frac{\lambda +\mu }{2}$ and solution of system of
Eqs. (\ref{gapeqlin}) reduces to the standard procedure of finding the
eigenvalues (and eigenvectors) for the matrix of dimensionless coupling
constants $g_{ij}$ (\ref{effmatr}), which are determined by the cubic secular
equation:
\begin{equation}
Det(g_{ij}-g_{eff}\delta _{ij})=0
\label{solvgeff}
\end{equation}
Physical solution is determined by the maximal positive value of $g_{eff}$,
which gives the maximal value of $T_c$. Eigenvectors of the problem determine
here the ratios of the gaps $\Delta_i$ on different sheets of the Fermi surface
for $T\to T_c$. Temperature dependencies of gaps for $T<T_c$ can be found by
solving the system of generalized BCS equations:
\begin{equation}
\Delta_i =  \sum_{j} g_{ij}\Delta_j \int_{0}^{\omega_c}d\xi
\frac{th\frac{\sqrt{\xi^2+\Delta_j^2}}{2T}}{\sqrt{\xi^2+\Delta_j^2}}.
\label{gapeq_T}
\end{equation}
For $T\to 0$ these equations reduce to:
\begin{equation}
\Delta_i =  \sum_{j} g_{ij}\Delta_jF\left( \frac{\Delta_j}{\omega_c}\right),\
F(x)=ln\left( \frac{1+\sqrt{1+x^2}}{|x|}\right)
\label{gapeq_T0}
\end{equation}
This analysis makes it clear that the value of $T_c$ (effective pairing constant)
in multiple bands system is determined, in general case, not only by the value
of the total density of states at the Fermi level (multiplied by the single
dimensional coupling constant), but by rather complicated combination of
several coupling constants, multiplied by partial densities of states for
different bands. Now it becomes obvious that the multiple -- band structure of
the spectrum can lead to the growth of $T_c$ by itself, reasonably enhancing
the effective pairing constant in Eq. (\ref{TC1}) \cite{KucSad08}.
To understand the essence of this effect it is useful to analyze simple
limiting cases.

Let the matrix of dimensionless coupling constants be diagonal (i.e. there are
only intraband pairing interactions):
\begin{equation}
\hat g=\left(\begin{array}{cccc}
g_1 & 0 & 0 & 0 \\
0 & g_2 & 0 & 0 \\
0 & 0 & g_3 & 0 \\
0 & 0 & 0 & g_3
\end{array}\right).
\label{diag_g}
\end{equation}
Then obviously что $g_{eff}=Max\{{g_i}\}$ and $T_c$ is determined by the
density of states and pairing interaction of the single (and in this sense
dominating) pocket of the Fermi surface.

Let us consider in some sense opposite case, when all intraband and interband
interactions in (\ref{mmatr}) are the same and also all partial densities of
states are just equal. Then we can introduce $g_0=-u\nu$ and the matrix of
dimensionless pairing constants takes the following form:
\begin{equation}
\hat g=g_0\left(\begin{array}{cccc}
1 & 1 & 1 & 1 \\
1 & 1 & 1 & 1 \\
1 & 1 & 1 & 1 \\
1 & 1 & 1 & 1
\end{array}\right).
\label{g_unity}
\end{equation}
In  this case we obtain $g_{eff}=4g_0$, i.e. the real quadrupling of the
effective pairing coupling constant, as compared to the single -- band model
(or the model without interband pairing couplings). The generalization for the
case of $n\times n$ matrices is obvious.

In Refs. \cite{KucSad08,KucSad10} it was shown that a certain choice of
coupling constants in this model (with the account of LDA calculated values of
partial densities of states) allows, in principle rather easily, to explain the
observed (by ARPES) values of gap ratios on different pockets of the Fermi
surface for a number of FeAs based superconductors.

\begin{figure}[ht]
\includegraphics[clip=true,width=0.45\textwidth]{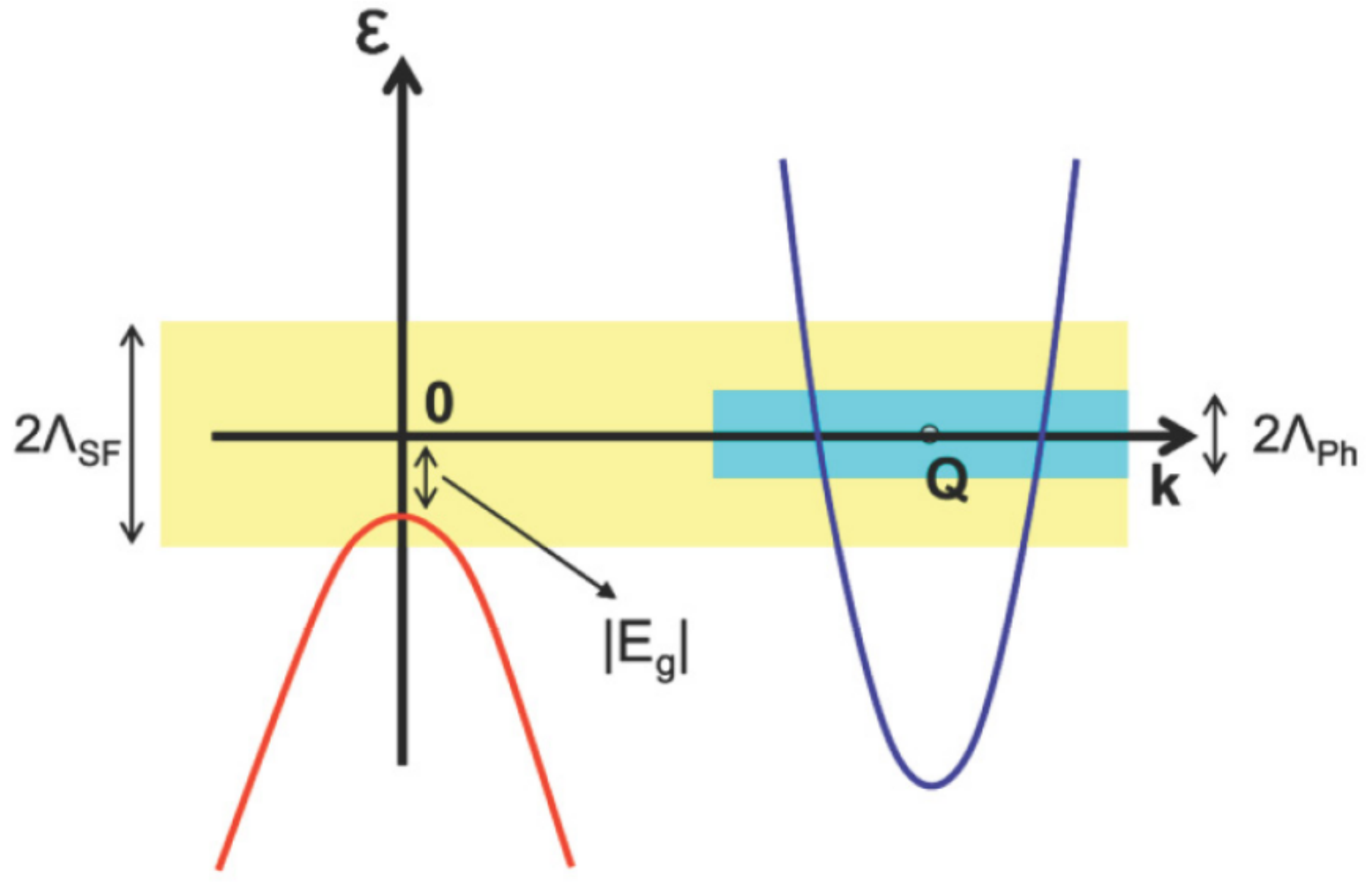}
\includegraphics[clip=true,width=0.5\textwidth]{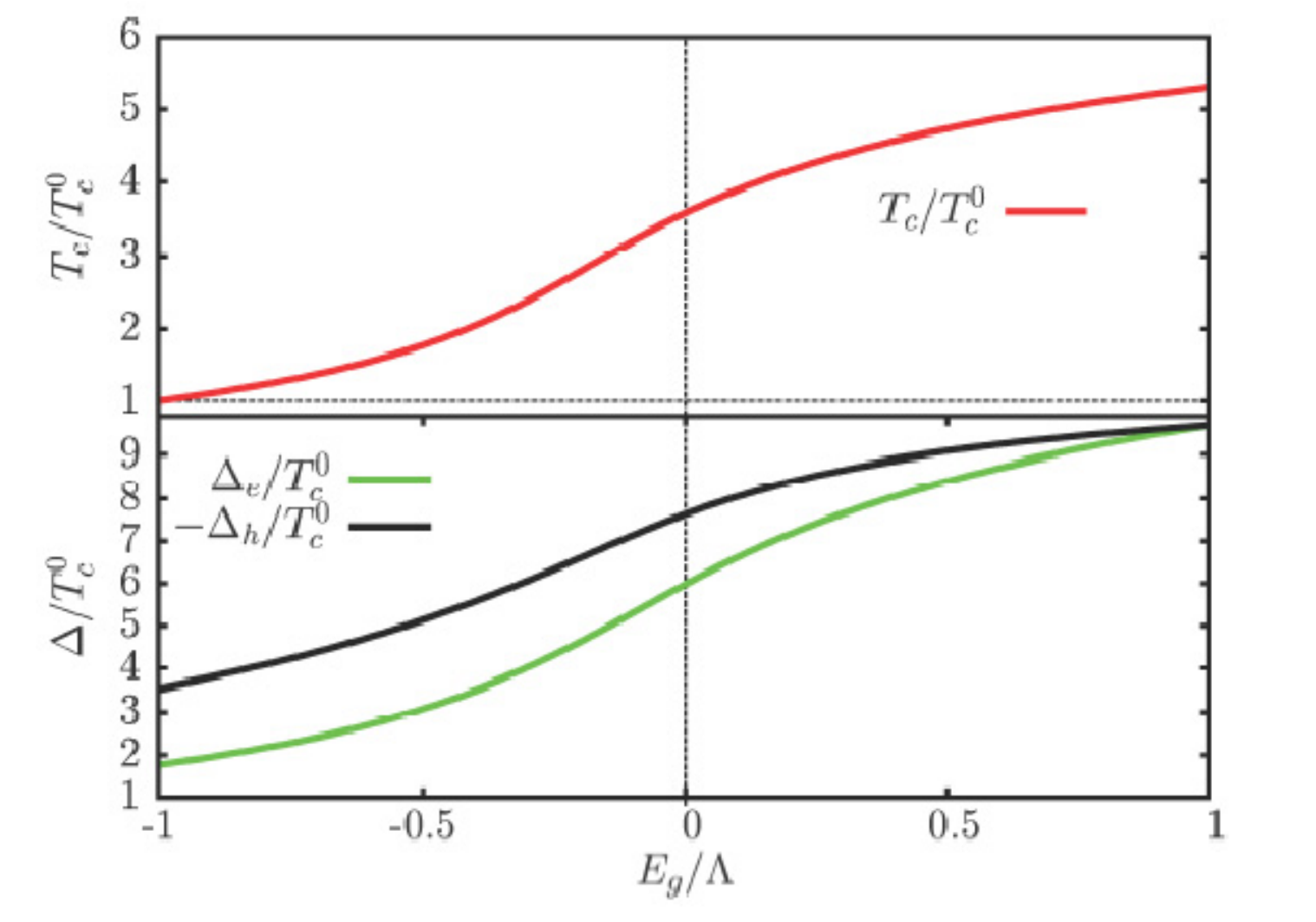}
\caption{(a) -- typical band structure of FeAs and FeSe based superconductors
with hole -- like band approaching to the Fermi level from below,
crossed areas denote the energy 
regions around the Fermi level, where two
pairing interactions operate, e.g. electron -- phonon ($\Lambda_{ph}$) and
spin -- fluctuation ($\Lambda_{sf}$).
(b) -- $T_c$ and energy gaps (at $T=0$ and on different sheets of the Fermi
surface) behavior during the crossing of the Fermi -- level by hole -- like
band. $T_{c0}$ is the temperature of superconducting transition in the absence
of hole -- like band. All energies are in units of $\Lambda_{sf}$ \cite{Hirshf}.}
\label{Lifsh_trans}
\end{figure}

In Ref. \cite{Hirshf} the similar analysis was done for a number of typical
situations of electronic spectrum evolution, which can be realized in FeAs and
FeSe based systems. It was explicitly shown that e.g. in the case of hole -- like
band approaching from below in energy (at the $\Gamma$ point) and crossing
the Fermi level (Lifshits transition) $T_c$ and the values of energy gaps on
hole -- like and electron -- like pockets of the Fermi surface actually {\em grow}.
In Fig. \ref{Lifsh_trans} we show the results of calculations of Ref.
\cite{Hirshf} for one of the typical cases, which may be realized in systems
under consideration. We can see that as the distance of hole -- like band
from the Fermi level $E_g$ diminish and change its sign (at Lifshits transition)
there is a significant growth of $T_c$ and the gap values $\Delta_i$ (at $T$=0).
Specific values of parameters used in this calculations can be found in
Ref. \cite{Hirshf}.

The basic conclusion from this elementary analysis is that the multiple -- band
structure, in general, facilitates the growth of effective pairing coupling
constant and the growth of $T_c$. It is also clear that the opening of new
pockets of the Fermi surface (during the Lifshits transition) also leads to the
growth of $T_c$, while closing of such pockets leads to the drop of $T_c$.
A number of experiments on FeAs systems under strong enough electron or hole
doping evidently confirm these conclusions \cite{dop_1,dop_2}.

At the same time, the general picture of electron spectrum evolution during the
transition from typical FeAs systems to intercalated FeSe systems, as well as
all the data obtained for single -- layer FeSe/STO, are in drastic contradiction
with this conclusion --- the high values of $T_c$ are achieved in these systems
after the disappearance of hole -- like pockets around the $\Gamma$ point and
only electron -- like pockets remain around $M$ points.

\begin{figure}[ht]
\includegraphics[clip=true,width=0.95\textwidth]{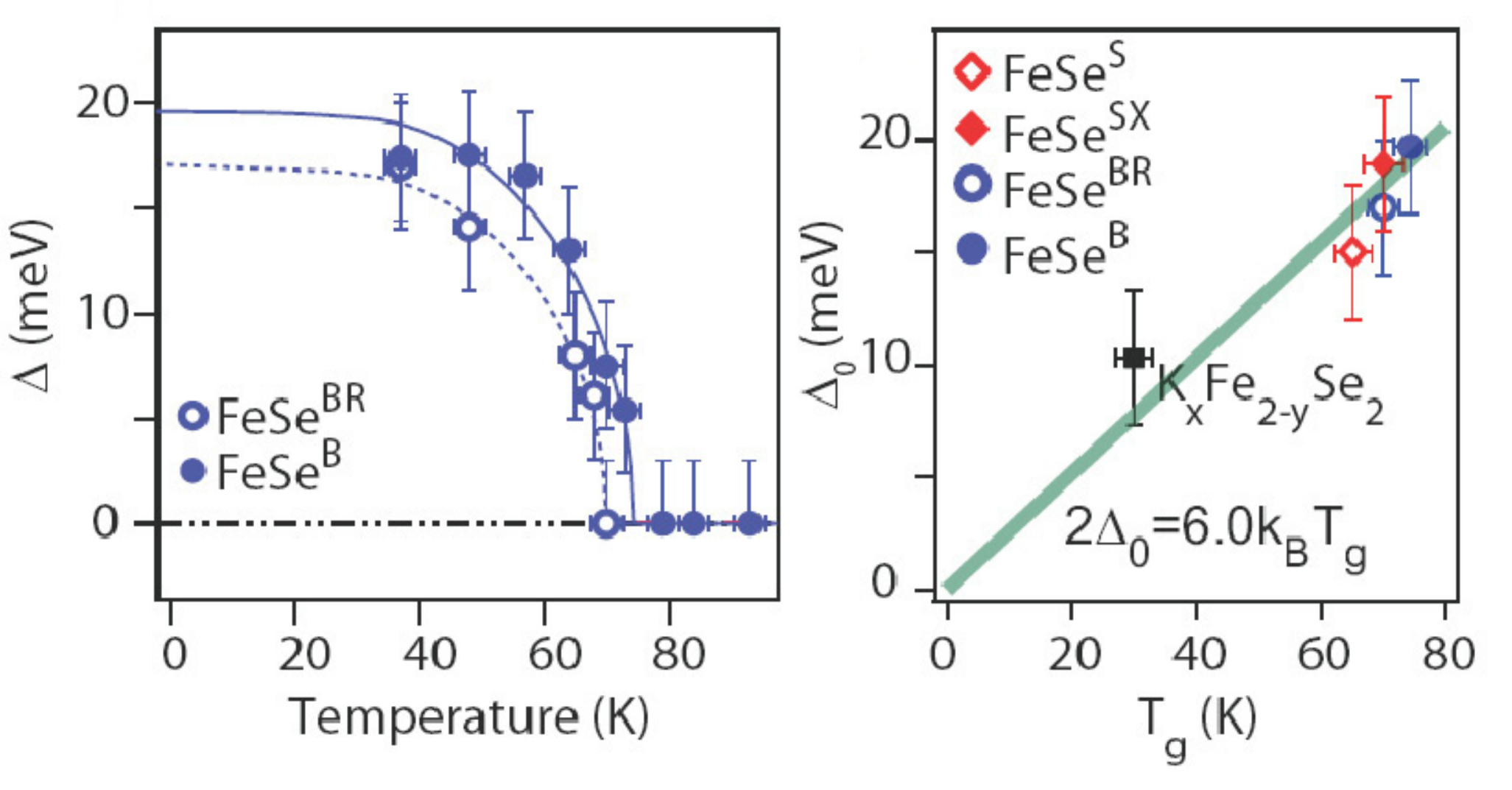}
\includegraphics[clip=true,width=0.45\textwidth]{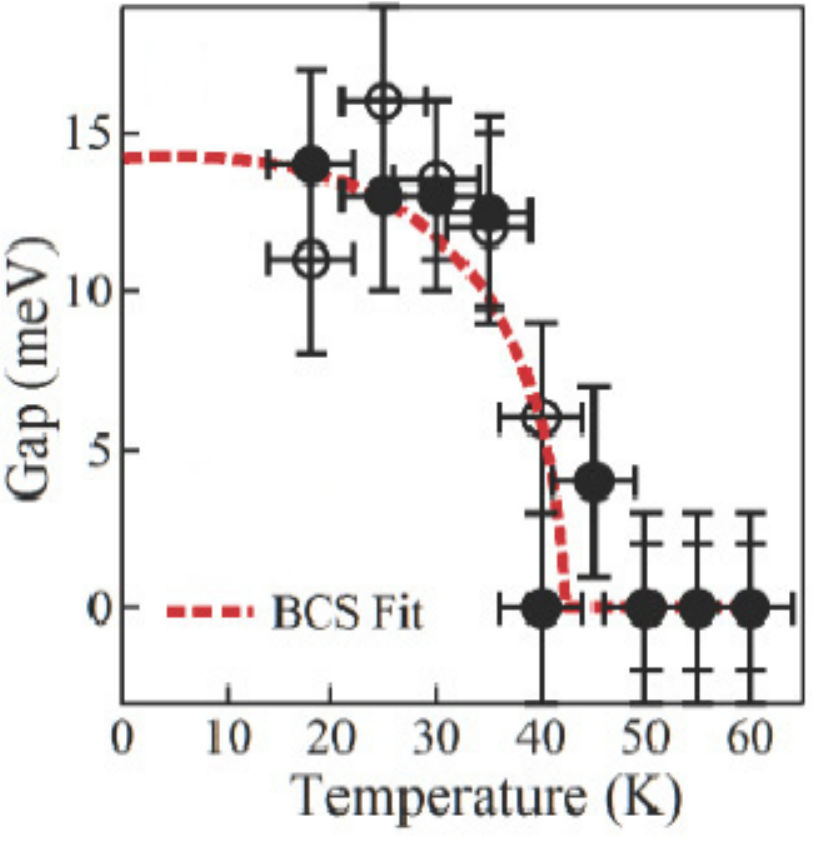}
\includegraphics[clip=true,width=0.45\textwidth]{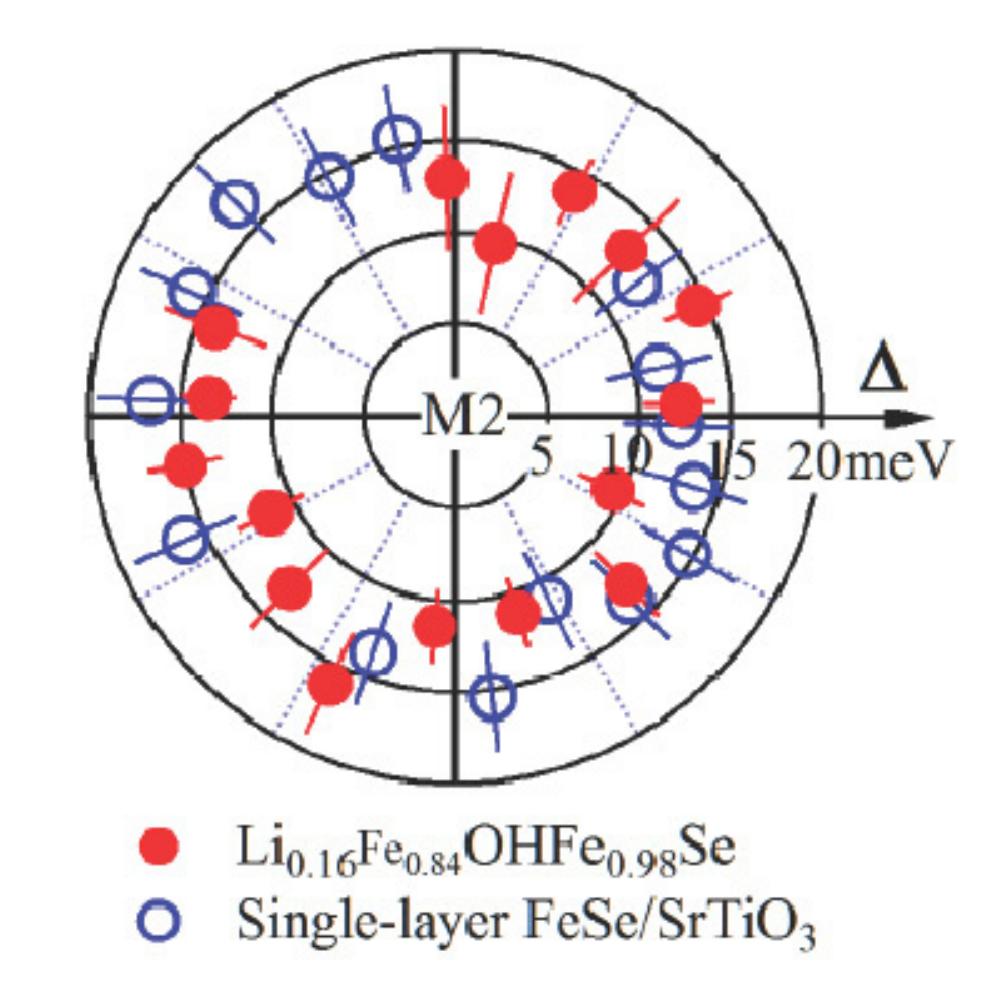}
\caption{(a) -- temperature dependence of energy gap for two FeSe/BTO films
\cite{FeSe_BTO}, (b) -- the value of the gap for K$_x$Fe$_{2-y}$Se$_2$ and monolayers
of FeSe \cite{FeSe_BTO}, (c) -- temperature dependence of the energy gaps in
Li$_{0.16}$Fe$_{0.84}$OHFe$_{0.98}$Se \cite{ARP_FS_FeSe_2},
(d) -- angular dependence of energy gap in Li$_{0.16}$Fe$_{0.84}$OHFe$_{0.98}$Se
and monolayer FeSe/STO \cite{ARP_FS_FeSe_2}. }
\label{Gaps_FeSe}
\end{figure}

The energy gaps, appearing on these pockets are reliably measured in ARPES
experiments and are practically isotropic \cite{FeSe_BTO,ARP_FS_FeSe_2}.
The relevant experimental data are shown in Fig. \ref{Gaps_FeSe}.

These data give rather convincing evidence of either $d$-wave pairing (case 1
above) or the usual $s$-wave pairing in systems under discussion. Pairing of
$s^{\pm}$-type can not be realized in these systems due to the absence (or
smallness) of Fermi surface pockets around the $\Gamma$ point. The absence of
``nesting'' of electron -- like and hole -- like pockets of the Fermi surface
also indicates the absence of well developed spin fluctuations, which can
be responsible for repulsive interaction, leading to the picture of $s^{\pm}$
pairing.

Apparently, the most probable in these systems is the scenario of $s$-wave
pairing, when the usual isotropic gap opens on electronic pockets. The variant
of $d$-wave pairing (as in case 1) seems less probable. First of all, no
microscopic mechanism (like spin fluctuations) was ever proposed for
realization of repulsive interaction on characteristic inverse lattice vectors
connecting  electronic pockets at points $M$ (or $X$ and $Y$ points in
Brillouin zone of Fig. \ref{BZFS} (b)). This picture also contradicts direct
experiments on the influence of magnetic and non -- magnetic adatoms on
superconductivity in single -- layer FeSe/STO films. It was shown in
Ref. \cite{mag_imp}, that magnetic adatoms suppress superconductivity, while
non -- magnetic adatom practically do not influence it at all. This obviously
corresponds to the picture of $s$-wave pairing.

\subsection{Models of $T_c$ enhancement in FeSe monolayer due to
interaction with elementary excitations in the substrate}

From the previous discussion it is clear that the values of $T_c\sim$40 K
in intercalated FeSe layers can be achieved, in principle, by increasing
the density of states at the Fermi level, as compared with its value for bulk
FeSe, which may be connected with the evolution of the band structure and
doping. At the same time, it is also clear that the enhancement of $T_c$ up to
the values exceeding 65 K, observed in FeSe monolayers on STO(BTO), can not be
explained along these lines. It is natural to assume that this enhancement is
somehow related to the nature of STO(BTO) substrate, e.g. with additional
pairing interaction of carriers in FeSe layer, appearing due to their
interaction with some kind of elementary excitations in the substrate, in the
spirit of ``excitonic'' mechanism, as was initially proposed by Ginzburg
\cite{VLG,HTSC77}.

It is well known that SrTiO$_3$ is a semiconductor with indirect gap equal to
3.25 eV \cite{STO_el_str}. At room temperature it is paraelectric with very
high dielectric constant, reaching the values of $\sim$10$^4$ at low temperatures,
remaining in paraelectric state \cite{STO_parael}. It is interesting to note that
under electron doping, in concentration interval from 6.9 10$^{18}$cm$^{-3}$ to
5.5 10$^{20}$cm$^{-3}$  SrTiO$_3$ becomes superconductor with maximal value
of $T_c\sim$0.25K at electronic concentration of the order of 9 10$^{19}$cm$^{-3}$
\cite{KCSHP,XLin}. The origin of superconductivity at such low concentrations
(and general form of corresponding phase diagram) is by itself the interesting
separate problem.

\subsubsection{Excitonic mechanism of Allender, Bray and Bardeen}

\begin{figure}[ht]
\includegraphics[clip=true,width=0.75\textwidth]{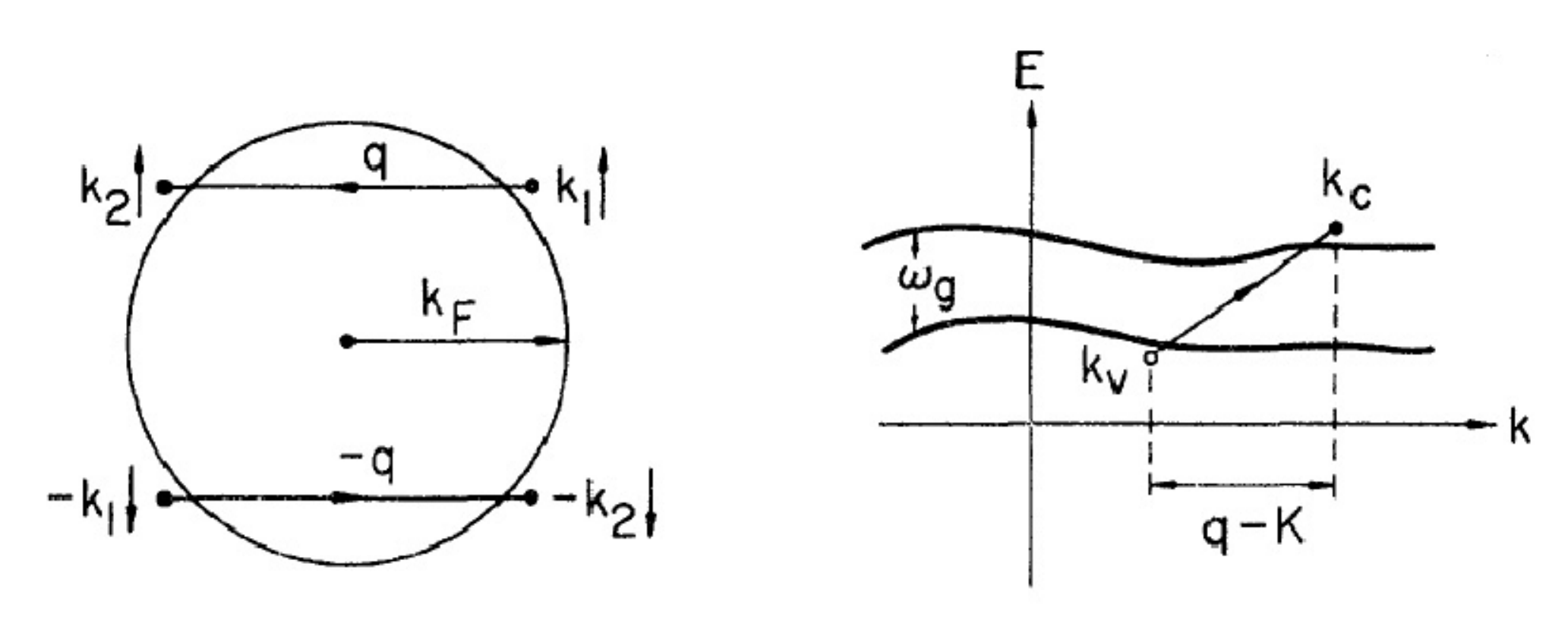}
\caption{ABB excitonic mechanism.}
\label{ABB}
\end{figure}

The structure of FeSe films on SrTiO$_3$, shown Fig. \ref{FeSeSTO}, represents
the typical Ginzburg's ``sandwich'' \cite{VLG}, which indicates the possibility
of realization of excitonic mechanism of superconductivity. Let us consider the
widely known version of this mechanism, as proposed for such a system long ago
by Allender, Bray and Bardeen (ABB) \cite{ABB}. Schematically this mechanism is
shown in Fig. \ref{ABB}. Electron from metal with momentum ${\bf k}_{1\uparrow}$
(arrow denotes spin direction) is transferred into the state ${\bf k}_{2\uparrow}$,
due to excitation of interband transition in semiconductor from valence band
state ${\bf k}_v$ into ${\bf k}_c$ state in conduction band, creating the
virtual exciton. The second electron of Cooper pair, which is initially in
$-{\bf k}_{1\downarrow}$ state, absorbs this exciton and goes into
$-{\bf k}_{2\downarrow}$ state. The momentum conservation law holds:
${\bf q}={\bf k}_2-{\bf k}_1={\bf k}_v-{\bf k}_c+{\bf K}$, where
{\bf K} is an arbitrary inverse lattice vector. As a result we obtain electron
attraction within the pair, which is conceptually identical to that appearing 
due to phonon exchange.

In Ref. \cite{ABB} a rough estimate of the corresponding attraction coupling
constant was obtained as:
\begin{equation}
\lambda_{ex}=ba\mu\frac{\omega_p^2}{\omega_g^2}
\label{l_ex}
\end{equation}
where $\mu$ is the dimensionless Coulomb potential, $\omega_p$ -- the plasma
frequency in semiconductor, while $\omega_g$ is the width of the energy gap in
semiconductor, which plays the role of exciton energy. Dimensionless constant
$b\sim$ 0.2 defines the fraction of time the metallic electron spends inside the
semiconductor, and the constant $a\sim$ 0.2-0.3 is related to screening of
Coulomb interaction within metal. This estimate was criticized in Refs.
\cite{IA_73,UspZhar} as an overestimate, additional arguments in favor of it
were given in Ref. \cite{ABB_2}. Without returning to this discussion, we
further use the estimate of Eq. (\ref{l_ex}) as obviously too optimistic.

To estimate $T_c$ due to two mechanisms of attraction (phonon and exciton)
Ref. \cite{ABB} proposed to use the following simple expression, which gives
(as was shown in \cite{ABB}) a good approximation to numerical solution of
appropriate Eliashberg equations:
\begin{equation}
T_c=\frac{\omega_D}{1.45}\exp\left(-\frac{1}{g_{eff}}\right),
\label{Tc_ABB}
\end{equation}
where
\begin{eqnarray}
g_{eff}=\lambda_{ph}^{\star}+\frac{\lambda_{ex}^{\star}-\mu^{\star}}
{1-(\lambda_{ex}^{\star}-\mu^{\star})\ln\left(\frac{\omega_g}{\omega_D}\right)}
\label{g_ABB}
\\
\mu^{\star}=\frac{1}{1+\mu\ln\left(\frac{E_F}{\omega_g}\right)}
\label{m_ABB}
\end{eqnarray}
and the constants of electron -- phonon and exciton attraction are taken here in
renormalized form:
\begin{equation}
\lambda_{ph}^{\star}=\frac{\lambda_{ph}}{1+\lambda_{ph}},\
\lambda_{ex}^{\star}=\frac{\lambda_{ex}}{1+\lambda_{ex}},\
\label{l_eff}
\end{equation}
which takes into account qualitatively the effects of strong coupling.
$E_F$ is the Fermi energy of metallic film.

\begin{figure}[ht]
\includegraphics[clip=true,width=0.48\textwidth]{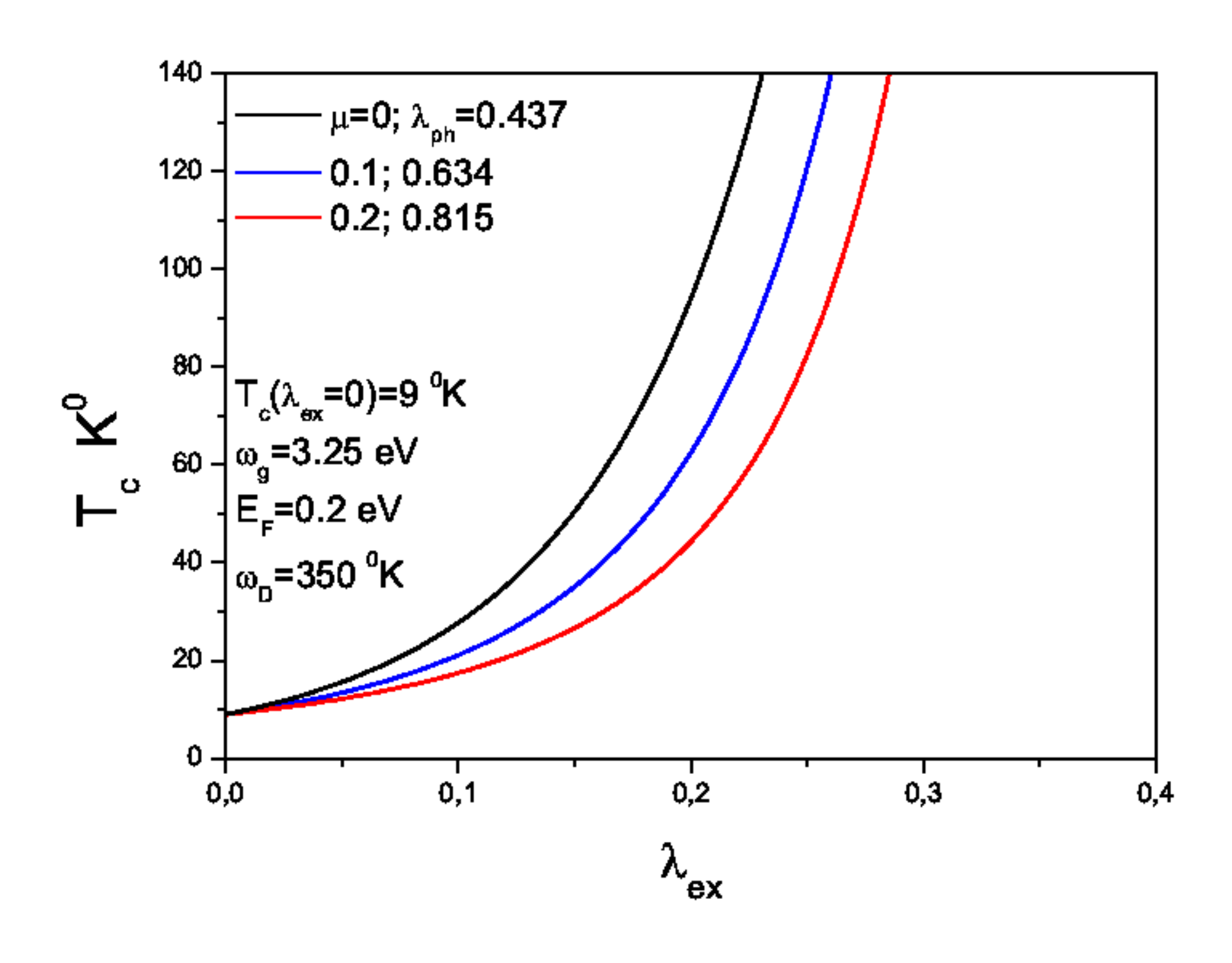}
\includegraphics[clip=true,width=0.48\textwidth]{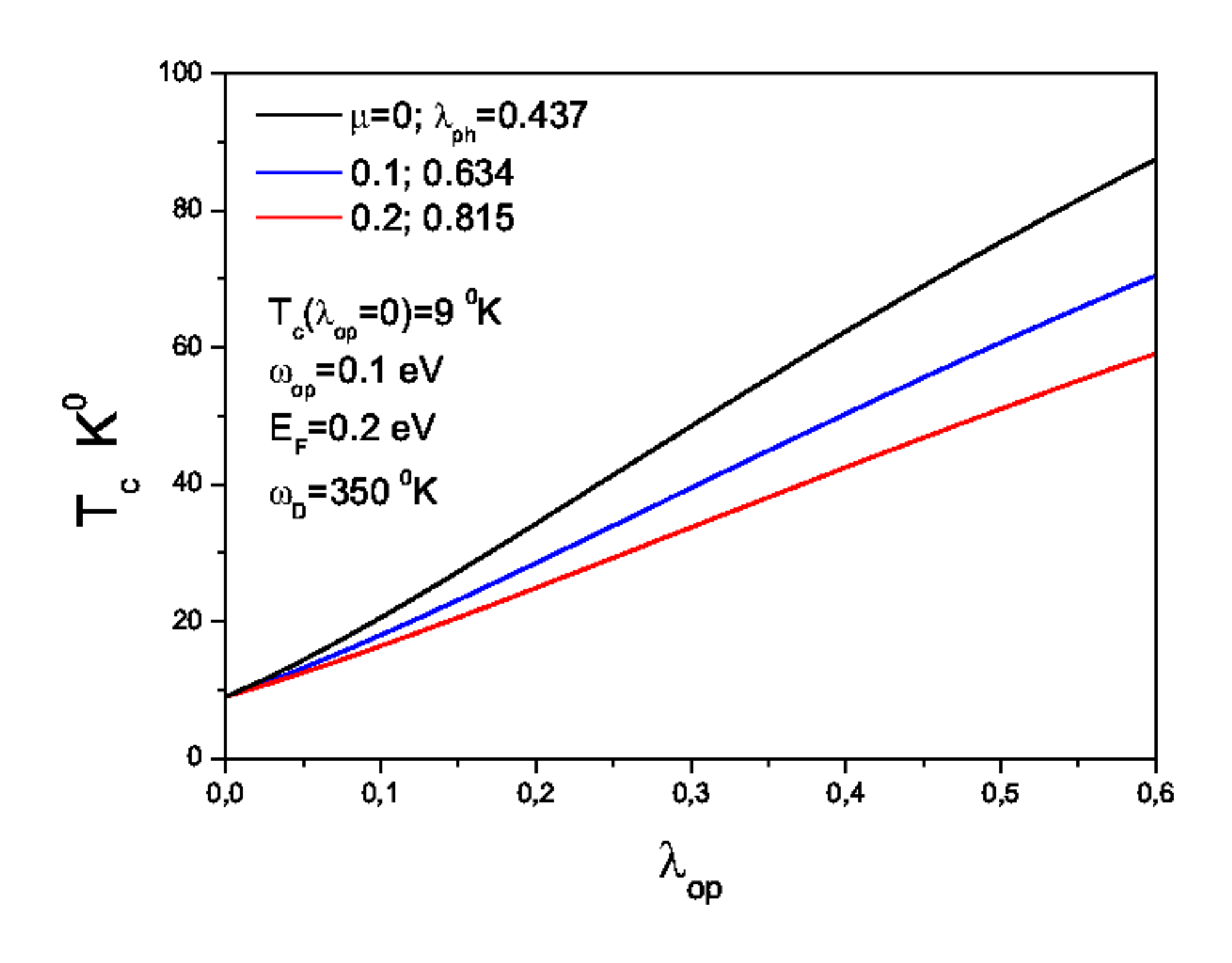}
\caption{Dependence of $T_c$ for FeSe/STO on coupling constant with ABB exciton
(a) and optical phonon in STO (b).}
\label{ABB_Tc}
\end{figure}

Is we consider $\lambda_{ex}$ a free parameter, we can easily estimate the
possible extent of $T_c$ enhancement due to excitonic mechanism. Corresponding
dependencies, calculated from Eqs. (\ref{Tc_ABB}),(\ref{g_ABB}),(\ref{m_ABB}),
(\ref{l_eff}) for typical values of Coulomb potential $\mu$, are shown Fig.
\ref{ABB_Tc} (a). The value of $\omega_D$ was taken to be 350 K, while
$\lambda_{ph}=$ 0.437, to reproduce the value of $T_c=$ 9 K, typical for bulk
FeSe, while $E_F=$ 0.2 eV was taken to be in agreement with LDA calculations of
FeSe monolayer. From Fig. \ref{ABB_Tc} (a) we can see that for large enough
values of $\lambda_{ex}$ very high values of $T_c$ can be easily obtained
(as it was predicted in Ref. \cite{ABB}).
The problem, however, is that even using the very optimistic estimate of
$\lambda_{ex}$ in Eq. (\ref{l_ex}), taking characteristic values of
$\omega_p=$ 10 eV, $\omega_g=$ 3.25 eV, for typical $\mu\sim$ 0.1-0.2, we obtain
the values of $\lambda_{ex}\sim$ 0.04-0.13. Correspondingly, as we can see from
Fig. \ref{ABB_Tc} (a), even for these over optimistic estimates, we obtain quite
modest enhancement of $T_c$ and it is very far from the desirable values of
$\sim$ 65-75 K. These estimates convincingly demonstrate ineffectiveness of
ABB excitonic mechanism in FeSe/STO monolayers.

\subsubsection{Interaction with optical phonons in STO}

\begin{figure}[ht]
\includegraphics[clip=true,width=0.45\textwidth]{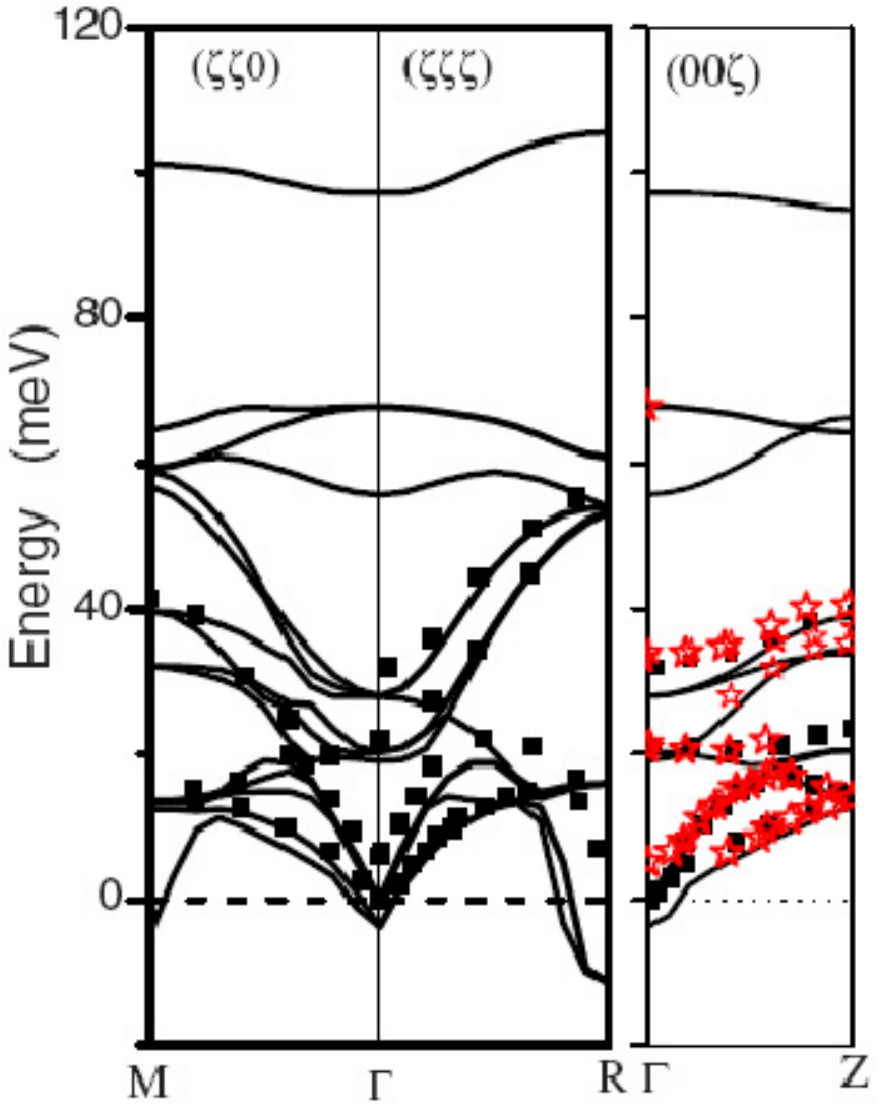}
\includegraphics[clip=true,width=0.45\textwidth]{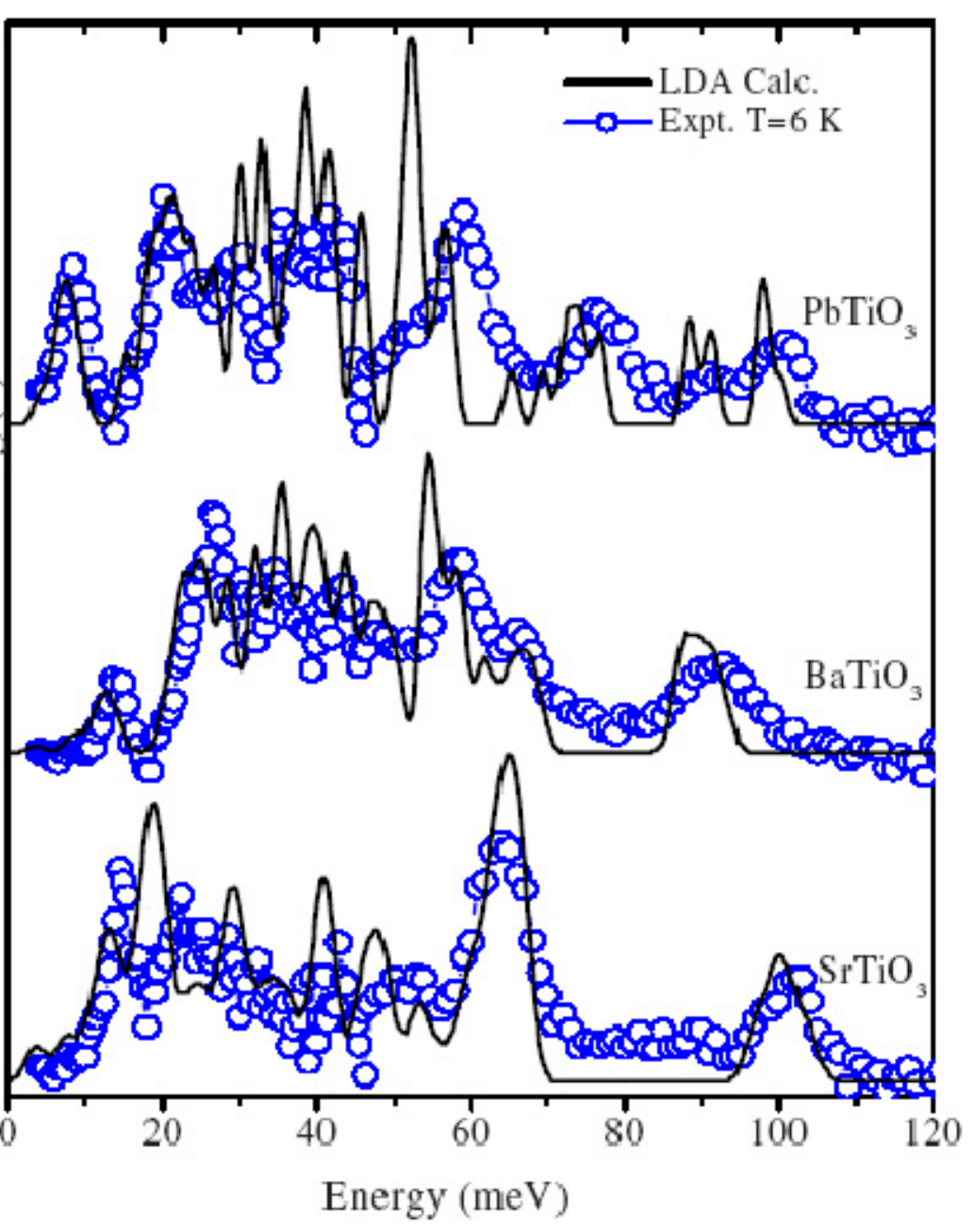}
\caption{Phonons in SrTiO$_3$ and similar compounds: (a) ---
phonon dispersions in SrTiO$_3$, both calculated and measured by inelastic
neutron scattering, (b) --- phonon density of states in в SrTiO$_3$,
BaTiO$_3$ и SrTiO$_3$ from neutron scattering and calculations \cite{CWK}.}
\label{STO_phonons}
\end{figure}

The initial Ginzburg's guess to enhance $T_c$ in ``sandwich'' type structures
\cite{VLG} was based on the idea of electron in metallic film interaction
with more or less high -- energy  excitations of electronic nature (``excitons'')
within semiconducting substrate. However, this idea can be understood in a
wider context --- interaction of electrons of metallic film with some arbitrary
Boson excitations in substrate (e.g. with phonons) can lead to the enhancement
of $T_c$. As we shall see, precisely this scenario is probably realized in
FeSe monolayers on STO(BTO). The thing is that in SrTiO$_3$ or BaTiO$_3$ like
systems almost dispersionless optical phonons exist with unusually high
excitation energy of the order of $\sim$ 100 meV \cite{CWK}. Examples of phonon
dispersions and densities of states in these systems (both calculated and
measured by neutron scattering) are shown in Fig. \ref{STO_phonons}.
To estimate the prospects of $T_c$ enhancement due to interaction with such
phonons we can again use the expressions (\ref{Tc_ABB}),(\ref{g_ABB}),
(\ref{m_ABB}),(\ref{l_eff}), with simple replacements of
$\omega_g\to\omega_{op}$ and $\lambda_{ex}\to\lambda_{op}$, where $\omega_{op}$
is characteristic frequency of optical phonon, $\lambda_{op}$ is dimensionless
coupling constant for such phonon with electrons within metallic film.
Results of such calculations of $T_c$ versus $\lambda_{op}$ (similar to those
shown in Fig. \ref{ABB_Tc} (a) for ABB excitonic mechanism) with the choice of
$\omega_{op}=$ 0.1 eV are shown in Fig. \ref{ABB_Tc} (b). It can be seen that
for large enough values of $\lambda_{op}\sim$ 0.5-0.6 and not very large $\mu$
we can easily achieve values of $T_c\sim$ 60-80 K, corresponding to experiments
on FeSe/STO(BTO), even if we start from relatively low initial $T_c\sim$ 9 K
for FeSe in the absence of additional pairing interaction. Corresponding values
of $\lambda_{op}$ seem to be realistic enough and below we shall present concrete
evidence that interaction with optical phonons in these structures can be
strong enough.

The idea that interaction with optical phonons in STO can play a significant
role in physics of FeSe/STO monolayers was first proposed in Ref.
\cite{ARPES_FeSe_Nature} in connection with ARPES measurements done in this work,
which demonstrated the formation of a ``shadow'' band at the $M$  point in
Brillouin zone, as shown in Fig. \ref{FeSe_ARPES_bands} (b). This band is
situated approximately 100 meV below the main conduction electronic band and
practically replicates its dispersion. Formation of such band can be linked with
interaction of FeSe electrons with optical phonon of appropriate energy in STO.
To understand this situation we have to consider a realistic enough picture of
FeSe monolayer electrons interacting with optical phonons in STO, which was
proposed in Refs. \cite{ARPES_FeSe_Nature} and will be briefly described below
(cf. also \cite{DHL}).

\begin{figure}[ht]
\includegraphics[clip=true,width=0.45\textwidth]{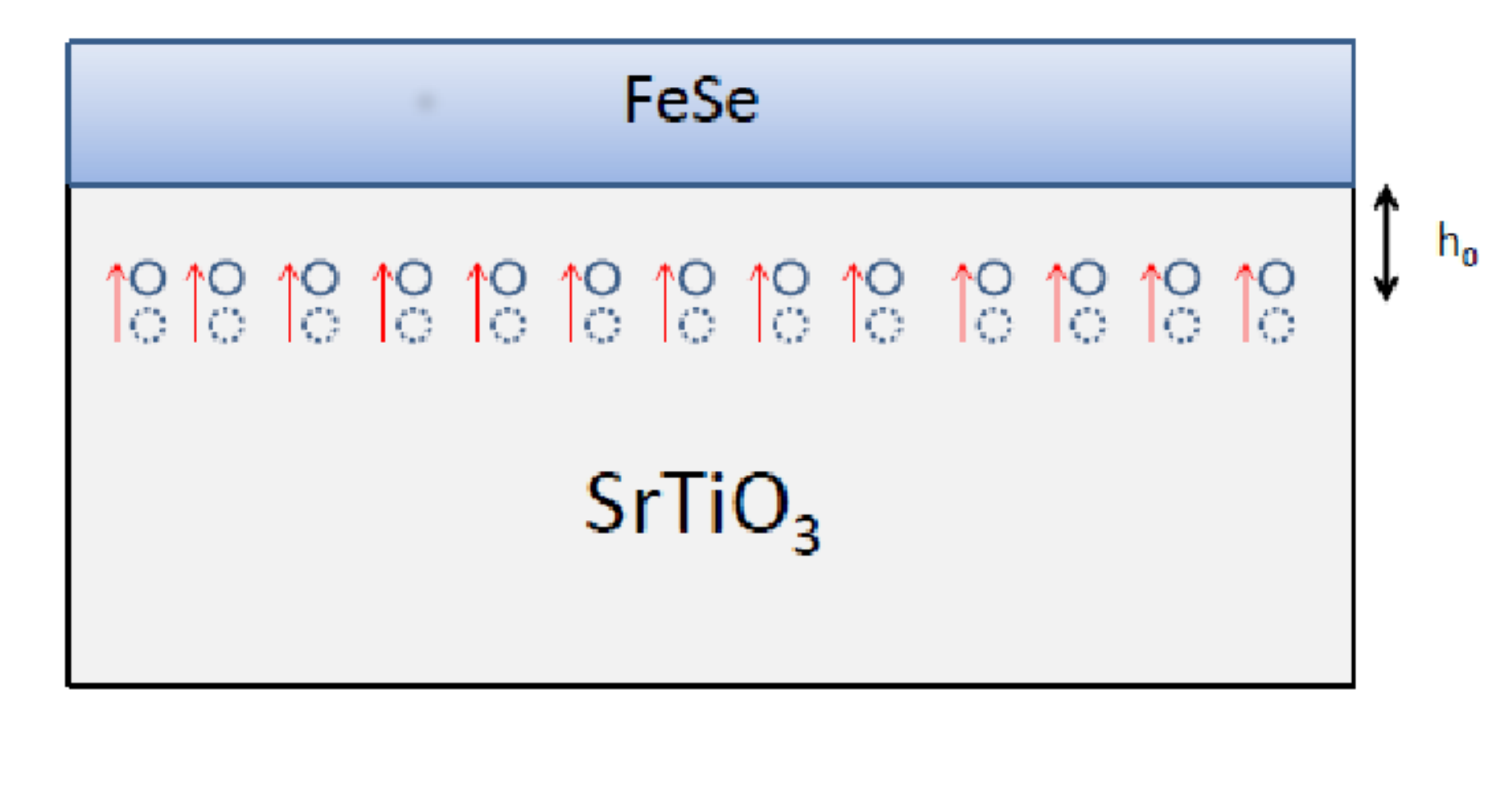}
\includegraphics[clip=true,width=0.45\textwidth]{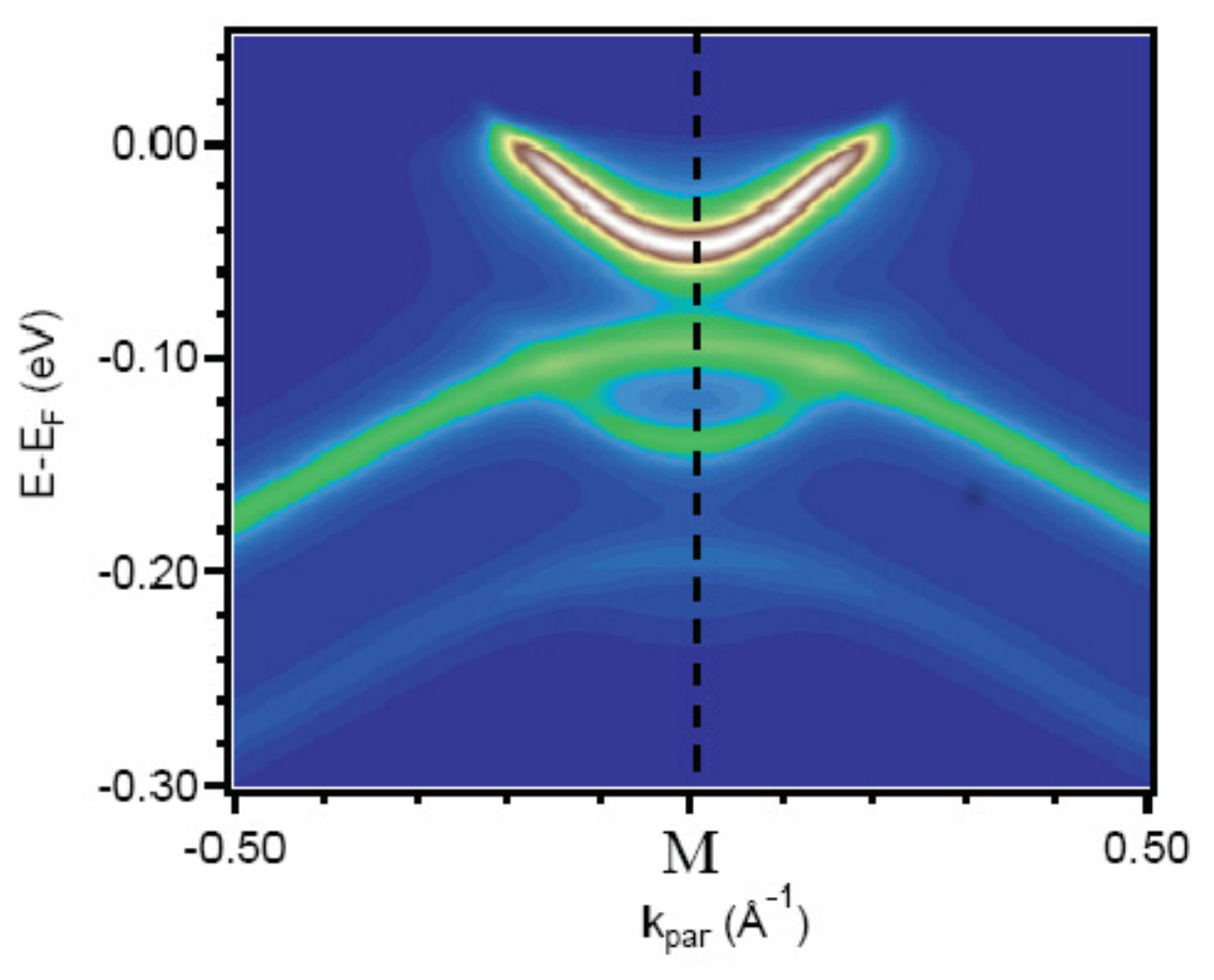}
\caption{(a) -- schematic picture of dipole excitations close to FeSe/STO
interfaceс, (b) -- calculated electron spectral density in в FeSe/STO in the
model with dominating forward scattering \cite{ARPES_FeSe_Nature}.}
\label{Sh_band}
\end{figure}

As STO is in almost ferroelectric state it is natural to expect that charge
transfer at the interface can induce the appearance of the layer of ordered
dipoles. Free carriers in STO, appearing for example due to oxygen vacancies
(or Nb doping) will screen the electric field far from the interface. Then, the
dipole layer will be localized close to the interface. The appearance of dipoles
is connected with the displacement of Ti cations relative to oxygen anions, so
that oscillations of these anions will lead to modulation of dipole potential
along the FeSe layer. Schematically, this situation is shown in Fig.
\ref{Sh_band} (a).

Let $\delta P_z$ denote the change of dipole moment due to displacement of
oxygen anions in the direction perpendicular to interface:
\begin{equation}
\delta P_z(x,y,-h_0)=q_{eff}\delta h(x,y,-h_0). 
\end{equation}
Here $x,y$ are coordinates in the plane parallel to interface and the origin of
$z$ - axis is chosen in Fe plane, $q_{eff}$ is dipole charge. With respect to Fe plane the
dipole layer is at $z=-h_0$. The induced change of dipole potential in Fe plane
connected with the ``frozen'' displacement of oxygens is given by the
following expression:
\begin{equation}
\Phi(x,y,0)=\frac{{\epsilon}_\parallel^{1/2} q_{eff} h_0} {\epsilon_\perp^{3/2}}
n_d\int dx' dy'\frac{\delta h(x',y',-h_0)} {\left(\frac{\epsilon_\parallel}
{\epsilon_\perp} h_0^2 +(x-x')^2+(y-y')^2\right)^{3/2}}.
\end{equation}
Performing Fourier transformation over $x,y$, we get:
\begin{equation}
\Phi({\bf q}_\parallel,0)=
\frac{2\pi q_{eff}n_d}{\epsilon_\perp} \exp\left({-|{\bf q}_\parallel| h_0
\sqrt{\epsilon_\parallel/\epsilon_\perp}}\right)\delta h({\bf q}_\parallel,-h_0).
\label{eph}
\end{equation}
Here ${\bf q}_\parallel$ is the wave-vector parallel to the interface and
$\epsilon_\parallel,\epsilon_\perp$ are dielectric constants parallel and
perpendicular to the interface, $n_d$ is density of dipole per unit square of
the interface. As electrons in FeSe move parallel to interface,
they contribute only to $\epsilon_\parallel$. As to carriers in STO, besides
their role in screening which we have mentioned above, they give approximately
equal contributions (STO has cubic structure) both to $\epsilon_{\parallel}$ and
$\epsilon_\perp$. Thus, we can expect that the total dielectric constant
$\epsilon_\parallel$ is much greater than $\epsilon_\perp$.

From Eq. (\ref{eph}) it becomes clear that the value of the matrix element of
electron -- phonon interaction has an important dependence on
${\bf q}_{\parallel}$, so that it can be written as:
\begin{eqnarray} 
\Gamma(p_\parallel,q_\parallel)=
\frac{2\pi q_{eff}n_d}{\epsilon_\perp} 
\exp\left({-|{\bf q}_\parallel|/q_0}\right),\\
q_0^{-1}=h_0\sqrt{\epsilon_\parallel/\epsilon_\perp}.
\label{epc}
\end{eqnarray}
The fact that $\epsilon_\parallel \gg \epsilon_\perp$ leads to 
$q_0$ suppression by the factor of 
$1/\sqrt{\epsilon_\parallel/\epsilon_\perp}$, which in turn leads to a sharp 
enough peak in electron -- phonon interaction at ${\bf q}_\parallel = 0$.
Such dominating role of forward scattering explains the appearance of the
``shadow'' band in electronic spectrum, which replicates the dispersion of the
main band. In the case of electron -- phonon interaction acting in the wide
range of transferred momenta, it will lead to a superposition of many bands,
each being moved by its own scattering vector, which will lead to a general
smearing of the ``shadow'' band.

The standard numerical calculation of second -- order electron self -- energy
due to electron -- phonon interaction was performed in Ref. \cite{ARPES_FeSe_Nature}
with coupling constant written as $g({\bf q})=g_{0}\exp(-|{\bf q}|/q_0)$, with
$g_0=$ 0.04 eV, $q_0=0.3/a$ (a=3.9\AA), optical phonon frequency $\Omega_0=$ 80
meV, and the bare spectrum of electrons and holes (one-dimensional --- along
$\Gamma - M$ direction) close to the $M$ point written as
$\epsilon_{e,h}(k)=-2t_{e.h}\cos(k/a)-\mu_{e,h}$
with $t_{e}=$ 125 meV, $t_{h}=$ 30 meV, $\mu_{e}=$ - 185 meV and
$\mu_{h}=$ 175 meV, where all numerical parameters were taken from fitting the
ARPES experiment. Results of such calculation for electron spectral density
(imaginary part of Green's function) are shown in Fig. \ref{Sh_band} (b).
We can see that these calculations are in excellent agreement with ARPES data
of Fig. \ref{FeSe_ARPES_bands} (b). The standard dimensionless electron -- 
phonon coupling constant can be estimated numerically using the same values of 
all parameters giving ($N$ is the number of lattice sites) 
\cite{ARPES_FeSe_Nature}: 
\begin{equation}
\lambda=\frac{2}{N\Omega_0}\frac{\sum_{{\bf k},{\bf q}}|g({\bf q})|^2
\delta{(\epsilon_{e}{(\bf k)})}\delta(\epsilon_{e}{(\bf k-q)})}
{\sum_{\bf k}\delta(\epsilon_{e}({\bf k}))} = 0.5
\label{e-ph-const}
\end{equation}
which is (as noted above) quite sufficient for 
significant enhancement of $T_c$ in FeSe/STO monolayer. As we shall see below, 
the peculiarities of the model of electron -- phonon interaction with dominating 
forward scattering lead also to some other, even more important, effects  
enhancing $T_c$.

\subsubsection{Cooper pairing in the model with dominating forward
scattering}

Dominating forward scattering in electron -- phonon interaction was for a long
time considered as a special cause for $T_c$ enhancement due to specific
dependencies differing from the standard BCS, which appear in this model
\cite{DaDoKuO,Kulic}. These papers analyzed  the possible role of such
interactions in cuprates. An application of these ideas to FeSe/STO was
considered  recently in \cite{Rade_1,Rade_2}.

In weak coupling approximation, for the case of $s$-wave pairing, the gap
equation in Eliashberg theory reduces to ($\varepsilon_n=(2n+1)\pi T$ -- is Fermion
Matsubara frequency):
\begin{eqnarray}
\Delta (i \varepsilon_n) &=&
-\frac{T}{N}\sum_{{\bf q},m} |g({\bf q})|^2 D({\bf q},i\varepsilon_n-i\varepsilon_m)
\frac{\Delta (i \varepsilon_m)}{(\varepsilon_m)^2 + \xi_{{\bf k} + {\bf q}}^2 +
\Delta^2 (i \varepsilon_m)}.
\label{Elias_1}
\end{eqnarray}
where  $D({\bf q},i\varepsilon_n-i\varepsilon_m)=
-\frac{2\Omega_{\bf q}}{(\varepsilon_n-\varepsilon_m)^2+\Omega_{\bf q}^2}$ is
Matsubara Green's function of an optical phonon with frequency
$\Omega_{\bf q}$, $\xi_{\bf k}=v_F(|{\bf k}|-p_F)$ is the electronic spectrum
close to the Fermi level ($v_F$, $p_F$ are Fermi velocity and momentum).

Before going to the results of numerical solution of this equation, let us
consider the elementary model of exactly forward scattering by phonons, when all
calculations can be done analytically. For this purpose we introduce
$|g({\bf q})|^2 = g^2_0N\delta_{\bf q}=(2\pi)^2\delta({\bf q})$. Then the gap
equation (\ref{Elias_1}) at the Fermi surface is easily transformed to:
\begin{equation}
\Delta(i\varepsilon_n) = \lambda_m \Omega^2_0 T_c\sum_m
\frac{\Delta(i\varepsilon_m)}{\varepsilon_m^2 + \Delta^2(i\varepsilon_m)}
\frac{2\Omega_0}{\Omega^2_0 + (\varepsilon_n - \varepsilon_m)^2},
\label{Eq_gap}
\end{equation}
where we have introduced the dimensionless coupling constant
\begin{equation}
\lambda_m=g^2_0/\Omega^2_0. 
\label{lamb_m}
\end{equation}
Note that this definition is somehow different from the standard definition of
electron -- phonon coupling constant (\ref{e-ph-const}).

To find the critical temperature $T_c$ the authors of Ref. \cite{Rade_1} used the 
following {\em Ansatz} for the gap function:
\begin{equation}
\Delta(i\varepsilon_n) = \Delta_0/[1+(\varepsilon_n/\Omega_0)^2]
\label{Ans_Delta}
\end{equation}
Then, linearizing the gap equation we can obtain the following equation for
$T_c$ \cite{Rade_1}:
\begin{eqnarray}
1 = \lambda_m\Omega^2_0 T_c\sum_m \frac{2\Omega_0}
{\varepsilon^2_m(1 + \varepsilon^2_m/\Omega^2_0)(\Omega^2_0 + \varepsilon_m^2)}.
\label{Tc_1}
\end{eqnarray}
The sum over Matsubara frequencies is calculated directly and we obtain:
\begin{eqnarray}
1 = \frac{\lambda_m}{2T_c}\frac{2\Omega_0 + \Omega_0\cosh(\Omega_0/T_c)
-(3 T_c)\sinh(\Omega_0/T_c)}{1+\cosh(\Omega_0/T_c)}.
\label{Tc_eq_fin}
\end{eqnarray}
For FeSe/STO $T_c \ll \Omega_0$, so that we can use the asymptotics of
hyperbolic functions and in the leading approximation the critical temperature
becomes the quasi -- linear  function of the coupling constant
(for its small values):
\begin{equation}
T_c = \frac{\lambda_m}{2+3\lambda_m}\Omega_0.
\label{T_c_forward}
\end{equation}
Similar result was previously obtained in the context of cuprates physics
\cite{DaDoKuO,Kulic}. For $\lambda_m=$ 0.16 and $\Omega_0=$ 100 meV we get
$T_c = 75$ K, which is rather unexpected for such a small value of $\lambda_m$.

This value of $T_c$ can be compared with the standard expression of BCS theory
where the linearized equation for $T_c$ takes the following form:
\begin{eqnarray}
1 = \pi T_c \lambda_m \sum_{|\varepsilon_m|<\omega_D}\frac{1}{|\varepsilon_m|}
= \lambda_m \left[\ln\left(\frac{\omega_D}{2\pi T_c}\right) -
\psi\left(\frac{1}{2}\right)\right],
\label{BCS_Tc}
\end{eqnarray}
where we have used the asymptotics of large $\omega_D/T_c$. This leads to the
usual BCS expression:  $T_c = 1.13\omega_{D}\exp(-1/\lambda_m)$, so that for
$\lambda_m=0.16$ and $\omega_{D}=$ 100 meV we obtain  $T_c = $ 2.5K.

Comparing these results for $T_c$ we can conclude that the significant $T_c$
enhancement obtained above appears due to effective exclusion of momentum
integration in Eliashberg equation, which is related to the strong interaction
peak at ${\bf q} = 0$. In BCS model we integrate over the whole of the Fermi
surface and all momenta enter with the same weight, which leads to the
appearance of $\sum_m 1/|\varepsilon_m|$ term in the equation for $T_c$ and
corresponding logarithmic behavior. In the case of forward scattering integration
over momenta is lifted, so that in the sum over frequencies only the
$\varepsilon_m^{-2}$ term remains, which leads to $1/T_c$ behavior. Due to this
the model with strong forward scattering leads to the effective mechanism of
$T_c$ enhancement \cite{DaDoKuO,Kulic}.

\begin{figure}[ht]
\includegraphics[clip=true,width=0.65\textwidth]{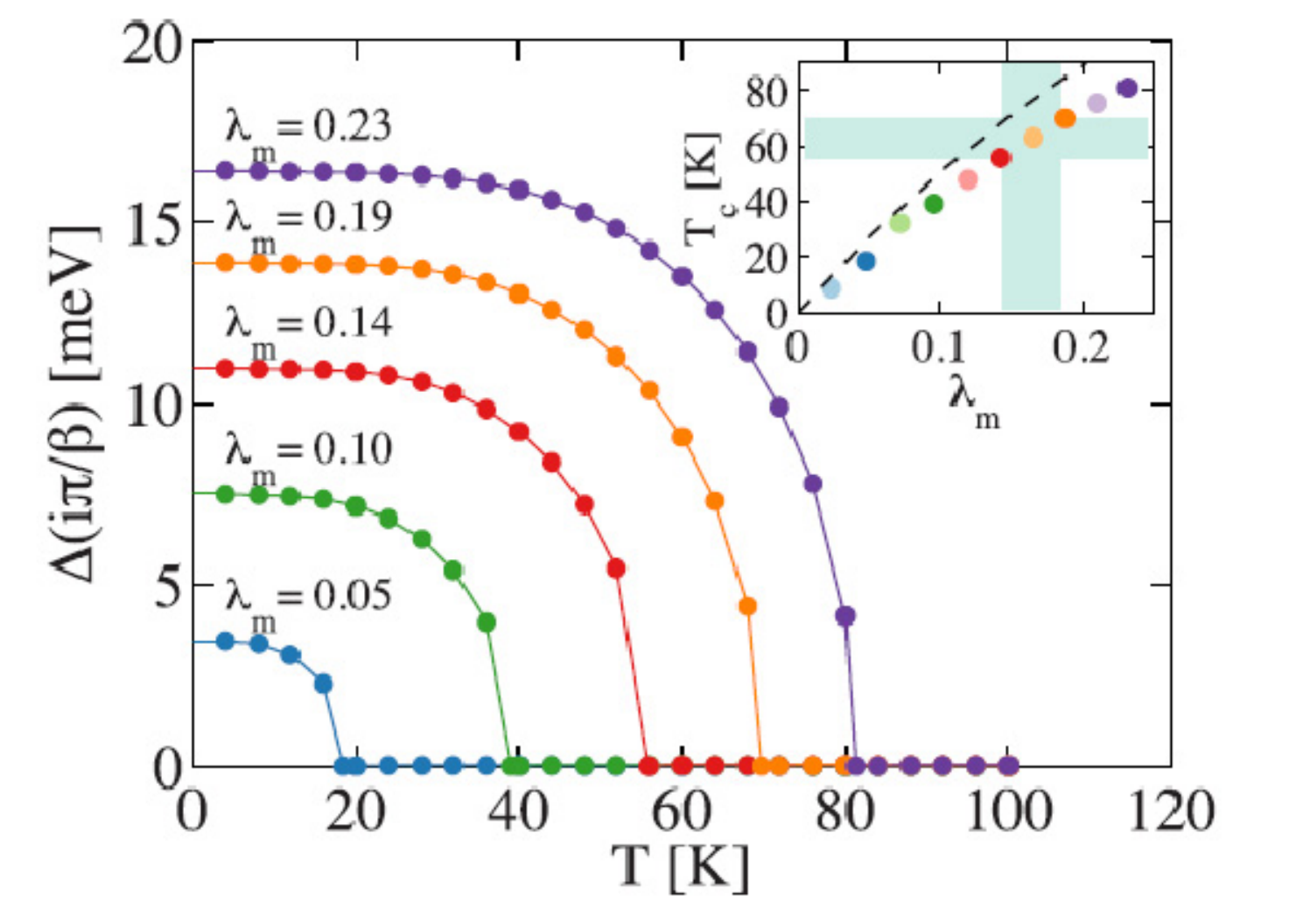}
\caption{Temperature dependence of the energy gap $\Delta(i\pi T)$
(at the smallest Matsubara frequency) in the model with dominating forward
scattering ($q_0 = 0.1/a$) for different values of the coupling constant.
At the insert the dashed line shows $T_c$ coupling constant dependence in the
case of exactly forward scattering (\ref{T_c_forward}), while the dots
represent the results of numerical calculation \cite{Rade_1}. The dashed region
show the interval of $\lambda_m$ values appropriate for FeSe/STO.}
\label{Gap_Tfs}
\end{figure}

Let us now discuss the numerical results for the general case \cite{Rade_1}.
In realistic situation, the forward scattering dominates in the finite region
of momentum space, with the size determined by the parameter $q_0$.
Numerical solution of Eliashberg equations with the coupling constant of the
form $g({\bf q}) = g_0\exp(-|{\bf q}|/q_0)$, gives the temperature behavior of
the superconducting gap (at the lowest Matsubara frequency)
$\Delta(i\pi T)$ shown in Fig. \ref{Gap_Tfs} (for several values
of $\lambda_m$ and $q_0 = 0.1/a$). We can see that $T_c$ is high enough
already for modest enough values of $\lambda_m$ and grows approximately linearly
over $\lambda_m$, while we remain in the weak coupling region. The finiteness of
$q_0$ leads to some suppression of $T_c$ as compared with the case of exact
forward scattering (cf. the insert in Fig.~\ref{Gap_Tfs}), but in general case
the quasi -- linear dependence of $T_c$ on $\lambda_m$ can guarantee the values
of $T_c$ observed in FeSe/STO films.

In the framework of this model it is rather easy to explain the formation of
the ``shadow'' band in the vicinity of the $M$ point \cite{Rade_1,Rade_2}.

\subsubsection{Nonadiabatic superconductivity and other problems}

We have already noted above, that the characteristic feature of electronic
spectrum of superconductors containing FeSe monolayers is the formation of
unusually ``shallow'' electronic band in the vicinity of $M$ point in
Brillouin zone (cf. Fig. \ref{LiOHFeSe_spectr_FS} (d), \ref{FeSe_ARPES_bands} (b)).
The value of Fermi energy $E_F\sim$ 0.05 eV in these systems is almost an order
of magnitude less than the values obtained in LDA and LDA+DMFT calculations.
Such small value of $E_F$ creates additional difficulties for consistent
theory of superconductivity in FeSe/STO system. Gor'kov was the first to note
\cite{Gork_1} that here we are dealing with an unusual situation, when the
energy of an optical phonon in STO $\sim$ 100 meV is significantly larger than
Fermi energy $\sim$ 50 meV. Let us remind that in the great majority of
superconductors we have the opposite inequality $\omega_D\ll E_F$, which allows
us to use the adiabatic approximation to describe the effects of electron --
phonon interaction, which is based upon the inequality
$\frac{\omega_D}{E_F}\sim\sqrt\frac{m}{M}\ll 1$ ($m$ is electron mass, $M$ is
an ion mass). Then (as in the normal state) we can apply Migdal theorem and
neglect all vertex corrections in electron -- phonon interaction, limiting
ourselves to second -- order diagrams for electron self -- energy. In particular,
the standard derivation of Eliashberg equations is entirely based on adiabatic
approximation, so that the common term is ``Migdal -- Eliashberg theory''.
Breaking the relevant inequality in FeSe/STO means that the theory explaining
$T_c$ enhancement is to be developed, from the very beginning, in the
{\em antiadiabatic} approximation. An attempt to build such a theory was
undertaken in recent papers by Gor'kov \cite{Gork_1,Gork_2,Gork_3}.
In particular, Refs. \cite{Gork_1,Gork_2} were devoted to FeSe/STO system and
the general aspects of the problem, while in Ref. \cite{Gork_3} a new theory
was proposed for superconductivity in doped SrTiO$_3$, which, as was noted above,
is by itself quite unusual superconductor \cite{KCSHP,XLin}.

Obviously, the nature of our review does not allow us to go deeply inside the
discussion of rather complicated theoretical problems, so that we shall limit
ourselves only to qualitative presentation of the results of Refs.
\cite{Gork_1,Gork_2}, which are directly relevant to superconductivity in
FeSe/STO. The only approximation, which can be apparently used here is the
{\em weak coupling} approximation, when the smallness of electron -- phonon 
coupling constant by itself allows to sum the usual (ladder) series of Feynman 
diagrams in Cooper channel. It is natural that in antiadiabatic approximation 
the cut-off of logarithmic divergence in Cooper channel takes place not on 
phonon frequencies, but at energies of the order of Fermi energy $E_F$ 
(or the bandwidth) \cite{Gork_2}, so that we can expect 
$T_c\sim E_F\exp(-1/\lambda)$, where $\lambda$ is determined by the details of 
pairing interaction.

The interaction of FeSe electrons with longitudinal surface phonons at the
STO interface can be introduced \cite{Gork_1} via interactions with polarization 
induced by these phonons:
\begin{equation}
{\bf P}=F_C{\bf u},
\label{Polrz}
\end{equation}
where ${\bf u}$ is atomic displacement and coefficient $F_C$ is determined by
the model of electron interaction with optical surface (SLO) phonons at
the surface of an insulator \cite{Mahan}:
\begin{equation}
F_{C,i}=\left[4\pi e^2\frac{\omega^i_{SLO}}{2}\left(\frac{1}{\epsilon_{\infty}+1}
-\frac{1}{\epsilon_0}\right)\right]^{1/2}
\end{equation}
where $i$ enumerates phonon branches, while $\epsilon_0$ and $\epsilon_{\infty}$
are static and optical dielectric constants of the bulk insulator, and
$\omega^i_{SLO}$ is the frequency of $i$-th SLO phonon.

Then the matrix element for two -- electron scattering due to exchange of
surface phonon takes the form:
\begin{equation}
M_i({\bf q},\varepsilon_n-\varepsilon_m)=
-\frac{4\pi e^2}{|{\bf q}|}\left(\frac{1}{\epsilon_{\infty}+1}
-\frac{1}{\epsilon_0}\right)D^i_{SLO}({\bf q},\varepsilon_n-\varepsilon_m)
\label{M_ep}
\end{equation}
where $D^i_{SLO}({\bf q})$ is Green's function of STO phonon:
\begin{equation}
D^i_{SLO}({\bf q},\varepsilon_n-\varepsilon_m)=\frac{(\omega^i_{SLO})^2}
{(\omega^i_{SLO})^2+(\varepsilon_n-\varepsilon_m)^2},
\label{D_STO}
\end{equation}
where ${\bf q}={\bf p}-{\bf k}$ and $\varepsilon_n-\varepsilon_m$ are momentum
and (Matsubara) frequency exchanged between electrons.

In the bulk insulator the well known Lyddane -- Sachs -- Teller relation holds
between the frequencies of longitudinal (LO) and transverse (TO) optical
phonons: $\omega^2_{LO}/\omega^2_{TO}=\epsilon_0/\epsilon_{\infty}$. According
to Ref. \cite{Mahan}, the frequency of longitudinal surface phonon is given
by the following expression:
$\omega^2_{SLO}/\omega^2_{TO}=\epsilon_0+1/\epsilon_{\infty}+1$.
It should be stressed that the values of $\epsilon_0$ and $\epsilon_{\infty}$
are considered here as model parameters, depending on the details of STO
surface preparation in the process of creation of FeSe/STO structures
\cite{Gork_1} (e.g. SrTiO$_3$ doping by Nb) \cite{Gork_1}.

Finally, for the matrix element of two -- electron scattering due to the exchange
of surface LO phonons and (two -- dimensional) Coulomb repulsion, dropping some
irrelevant at the moment factors \cite{Gork_1}, we obtain:
\begin{equation}
M_{tot}({\bf p},\varepsilon_n|{\bf k},\varepsilon_m)=
\frac{4\pi e^2}{(\epsilon_{\infty}+1)q}-\sum_{i}\frac{4\pi e^2}{(\epsilon_{\infty}+1)q}
D^i_{SLO}(\varepsilon_n-\varepsilon_m)
\label{M_ep_C}
\end{equation}
Here the summation is performed over three IR -- active phonons at $\Gamma$ 
point of the bulk SrTiO$_3$, with frequencies satisfying the inequality
$\omega^i_{LO}>T_c$ \cite{CWK}. In fact, in SrTiO$_3$ we have the single LO
mode, which has a very large gap, as compared to the frequencies of all TO
phonons, and  which is of principal importance here compensating the Coulomb
repulsion in Eq. (\ref{M_ep_C}) for $|\varepsilon_n - \varepsilon_m|\ll \omega_{LO}$.
The remaining LO phonons, as usual, provide the additional contribution to
attraction, As in SrTiO$_3$ we have $\epsilon_0\gg \epsilon_{\infty}$, in
Eq. (\ref{M_ep_C}) we have left only the terms with $\epsilon_{\infty}+1$.

In extremely antiadiabatic limit, when $\omega_{SLO}\gg E_F$, we can neglect
$(\varepsilon_n-\varepsilon_m)^2$ terms in the denominator of phonon Green's
function, so that the matrix element of two -- electron interaction can be
written as:
\begin{equation}
M_{tot}({\bf p},\varepsilon_n|{\bf k},\varepsilon_m)=M({\bf p}-{\bf k})\approx
-2\alpha^2\frac{4\pi e^2}{|{\bf p}-{\bf k}|(\epsilon_{\infty}+1)} < 0.
\label{M_tot}
\end{equation}
Here $\alpha^2 < 1$ are some numerical correction factors \cite{Gork_1}.

Now we also have to take into account the screening of Coulomb interaction by
two -- dimensional electron gas of FeSe. Then, in RPA approximation we get
\cite{Gork_1}:
\begin{equation}
M_{scr}({\bf p}-{\bf k})\approx -2\alpha^2\frac{4\pi e^2}{\epsilon_{\infty}}
\frac{1}{|{\bf p}-{\bf k}|+4e^2m/(\epsilon_{\infty}+1)}.
\label{M_tot_scr}
\end{equation}
In experimental situation typical for FeSe/STO the inverse screening length
$q_0$ is small as compared with Fermi momentum $p_F$, so that the following
inequality always holds:
\begin{equation}
p_F/q_0=p_F(\epsilon_{\infty}+1)/e^2m\gg 1.
\label{scr_rad}
\end{equation}
Introducing the effective Bohr radius $a_B=(\epsilon_{\infty}+1)/e^2m$ this
inequality can be rewritten as $p_Fa_B\gg 1$.

In weak coupling approximation the linearized gap equation can be written as
\cite{Gork_1}:
\begin{equation}
\Delta({\bf p})=-T\sum_{m}\int\frac{d^2k}{(2\pi^2)}M_{scr}({\bf p}-{\bf k})
G(-{\bf k})G({\bf k})\Delta({\bf k}),
\label{Gap_eq_anti}
\end{equation}
where the product of two Green's functions
$G(-{\bf k})G({\bf k})=[\varepsilon_{m}^2+\xi^2_{\bf k}]^{-1}$.

\begin{figure}[ht]
\includegraphics[clip=true,width=0.65\textwidth]{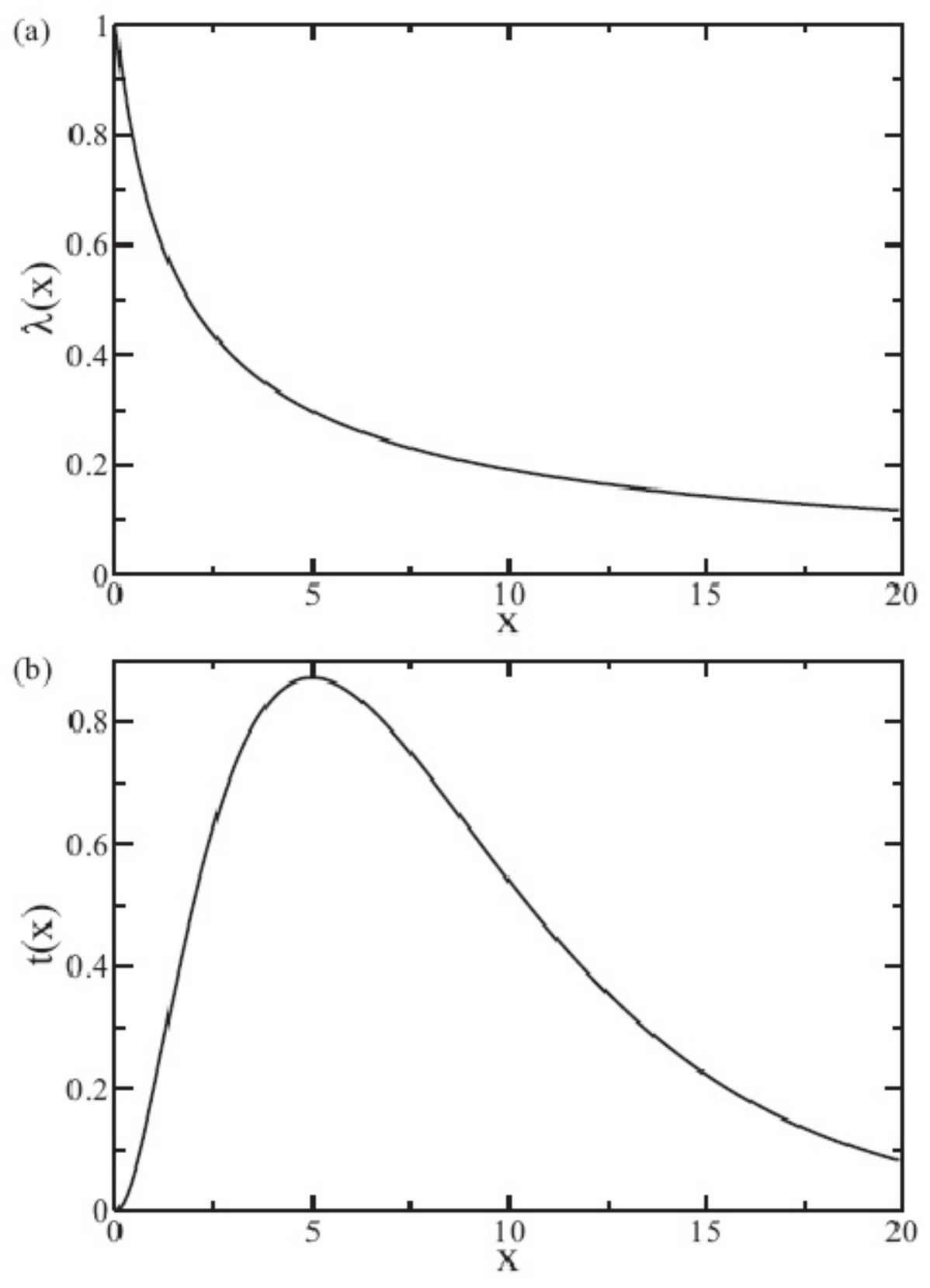}
\caption{Dimensionless functions determining $T_c$ in Gor'kov's model:
(a) $\lambda(x)$ function, (b) $t(x)=x^2\exp[-1/\lambda(x)]$ function.}
\label{Gorkov_curve}
\end{figure}

Then, after some a little bit cumbersome, though direct, analysis we can obtain
the following result for the critical temperature $T_c$:
\begin{eqnarray}
T_c(x)\sim\frac{p_F^2}{2m}\exp\left[-\frac{1}{\alpha^2\lambda(x)}\right]=
\nonumber\\
=\frac{2}{ma_B^2}x^2\exp\left[-\frac{1}{\alpha^2\lambda(x)}\right]
\label{Gork_Tc}
\end{eqnarray}
where we have introduced the dimensionless parameter $x=(p_Fa_B)/2$ and
\begin{equation}
\lambda(x)=\frac{2}{\pi}\int_{0}^{\pi/2}\frac{du}{x\sin u + 1}.
\label{Gork_lambda}
\end{equation}
For our estimates we can just put $\alpha^2=1$. Two dimensionless functions
$\lambda(x)$ and $t(x)=x^2\exp[-1/\lambda(x)]$ are shown in Fig. \ref{Gorkov_curve}.
The maximum in $t(x)$ appears due to two competing factors: for the given value
of $a_B$ the critical temperature first grows with growth of electron
concentration and then the increased screening suppresses the effective
coupling constant.

Direct calculations \cite{Gork_1} show, that this model also reproduces the
``shadow'' band in the electronic spectrum in the vicinity of $M$ point.
This is essentially due to the fact, that from the form of pairing interaction
(\ref{M_tot_scr}) it becomes clear, that Gor'kov's model produces the
significant growth of interaction at small transferred momenta. Effective
interaction is concentrated in momentum region inside the inverse screening
length  $q_0$, satisfying inequality (\ref{scr_rad}), so that $q_0\ll 1/a$,
in accordance with the estimates given above for the model with dominating
forward scattering.

Let us make the simplest estimate of the maximal value of $T_c$, which can be
achieved in this model. We take $E_F=$ 60 meV, which approximately corresponds
to ARPES experiments. Maximum of $t(x)$ as can be seen from Fig. \ref{Gorkov_curve}
(b) is close to $x=5$, which corresponds to $\lambda(5)=$ 0.3
(cf. Fig. \ref{Gorkov_curve}(a)). Then we get $T_c\approx$ 0.03$\times$60 meV
$\approx$ 20 K. Thus, this mechanism by itself can not explain the values of
$T_c >$ 60 K, observed in experiments on FeSe/STO. However, in combination with
some additional pairing mechanism, responsible for the initial value of
$T_c\sim$ 8 K in bulk FeSe (either due to the usual electron -- phonon mechanism
or pairing due exchange of antiferromagnetic fluctuations) we can obtain
significantly higher values of $T_c$ \cite{Gork_1}. For example, if for
spin -- fluctuation mechanism we use the estimate
$T_c\sim E_F\exp(-1/\lambda_{sf})$, then for $E_F=$ 60 meV the initial value of
$T_c$ is obtained for $\lambda_{sf}=0.23$. Then the combined pairing constant
$\lambda=\lambda_{sf}+\lambda(5)=$ 0.48 (assuming the same cut-off in Cooper
channel the coupling constants are just summed) leading to
$T_c\approx$ 0.15$\times$60 meV$\approx$ 90 K. In the case of combination with
the usual electron -- phonon mechanism, we can estimate $T_c$ using the upper
curve (corresponding to $\mu$=0) in Fig. \ref{ABB_Tc} (b). Then, taking
$\lambda_{op}=\lambda(5)=$ 0.3 we immediately obtain $T_c\approx$ 50 K.

The situation with nonadiabatic effects in the model with dominating forward
scattering was recently analyzed in Ref. \cite{Rade_2} by direct calculations
of vertex corrections to electron -- phonon interaction with coupling constant
$|g({\bf q})|^2=g_0^2N\delta_{\bf q}$. It was shown, that in this model Migdal
theorem is invalid for any values of $\Omega_0/E_F$ ratio, which does not appear
at all in vertex corrections. However, vertex corrections remain small for small
values of the parameter $\lambda_m=g_0^2/\Omega_0^2$, and we have seen above,
that to explain the current experiments on FeSe/STO it is sufficient to take
the values $\lambda_m\sim$ 0.15-0.2.

The small values of Fermi energy $E_F$ in electron band at $M$ point,
observed in intercalated FeSe systems and FeSe/STO(BTO), lead to one more
important consequence. Typical values of superconducting gap at low temperatures,
observed in ARPES measurements on these systems, are $\Delta\sim$ 15-20 meV
(cf. Fig. \ref{Gaps_FeSe}). Correspondingly, here we have unusually large values
of $\Delta/E_F\sim$ 0.25-0.3, which unambiguously show, that these systems belong
to the region of BCS -- Bose crossover \cite{NozSR,Rand}, when the size of
Cooper pairs, determined by coherence length $\xi$, becomes small and approaches
interelectron spacing, when $p_F\xi\sim\xi/a\sim$ 1. The picture of superconducting
transition and all estimates for the physical characteristics like $T_c$ in this
region are different from those for the weak coupling BCS theory and are closer to the
picture of Bose -- Einstein condensation of compact Cooper pairs \cite{NozSR,Rand}.

The development of such situation was earlier noted in connection with some
experiments on FeSe$_x$Te$_{1-x}$ system \cite{FeSeTe_BEC}, and also for the
bulkа FeSe in external magnetic field \cite{FeSe_BEC}.

From theoretical point of view we need here a special treatment \cite{NozSR,Rand}.
Unfortunately, for multiple -- band systems like FeSe theoretical description
of BCS -- Bose crossover remains, up to now, almost undeveloped. We can quote
only the recent Ref. \cite{ChErEfr}, but the detailed discussion of different
possibilities appearing here is outside the scope of the current review.

\section{Conclusion}

Basic conclusions from our discussion can be formulated as follows. The number of
aspects of the physics of systems under investigation is more or less clear:

\begin{itemize}

\item{Electronic spectrum of intercalated FeSe systems and FeSe/STO(BTO) is
significantly different from the spectrum of the systems  based on FeAs and
the bulk FeSe. Here we have only electron -- like Fermi surfaces, surrounding
the $M$  points in Brillouin zone. Hole -- like Fermi surfaces ``sink'' under
the Fermi level. There are no ``nesting'' properties of Fermi surfaces at all;}

\item{The values of superconducting critical temperature $T_c$ in intercalated
systems are well correlated with the value of the total density of states at the
Fermi level, obtained by LDA calculations, independently of the microscopic
nature of pairing;}

\item{Cooper pairing is most probably the usual $s$-wave pairing, there is no
possibility for $s^{\pm}$-pairing, because of the absence of hole -- like
Fermi surfaces, while $d$-wave pairing also seems less probable;}

\item{The record values of $T_c$, observed in FeSe monolayers on STO(BTO),
are related to the additional pairing mechanism, due to interaction with
high -- energy optical phonons of STO(BTO) in the geometry of Ginzburg
``sandwich''. In this sense here we may speak of the realization o
``pseudoexcitonic'' pairing mechanism.}

\end{itemize}

At the same time many questions remain to be resolved:

\begin{itemize}

\item{Until now the observation of the values of $T_c\sim$ 100K, reported
in Ref. \cite{FeSe_STO2}, remain unconfirmed;}

\item{The origin of unusually ``shallow'' electronic bands with extremely small
values of the Fermi energy in the vicinity of $M$ points remains unclear.
Probably, this is related to our poor understanding of the role of electron
correlations;}

\item{The data on possible magnetically ordered phases in intercalated FeSe
systems remain rather indeterminate. Practically nothing is known on the
possible types of magnetic ordering in FeSe/STO(BTO) films;}

\item{From the theoretical point of view it is unclear why the disappearance
of some of the Fermi surfaces in FeSe systems is followed by the significant
{\em increase} of $T_c$, in contradiction with general expectations, based on
the multiple -- band BCS model;}

\item{Practically no serious theoretical developments are known concerning
the possible manifestations of BCS -- Bose crossover effects in these systems,
as well as its experimental consequences and the role of these effects in the
formation of high values of $T_c$.}

\end{itemize}

\begin{figure}[ht]
\includegraphics[clip=true,width=0.95\textwidth]{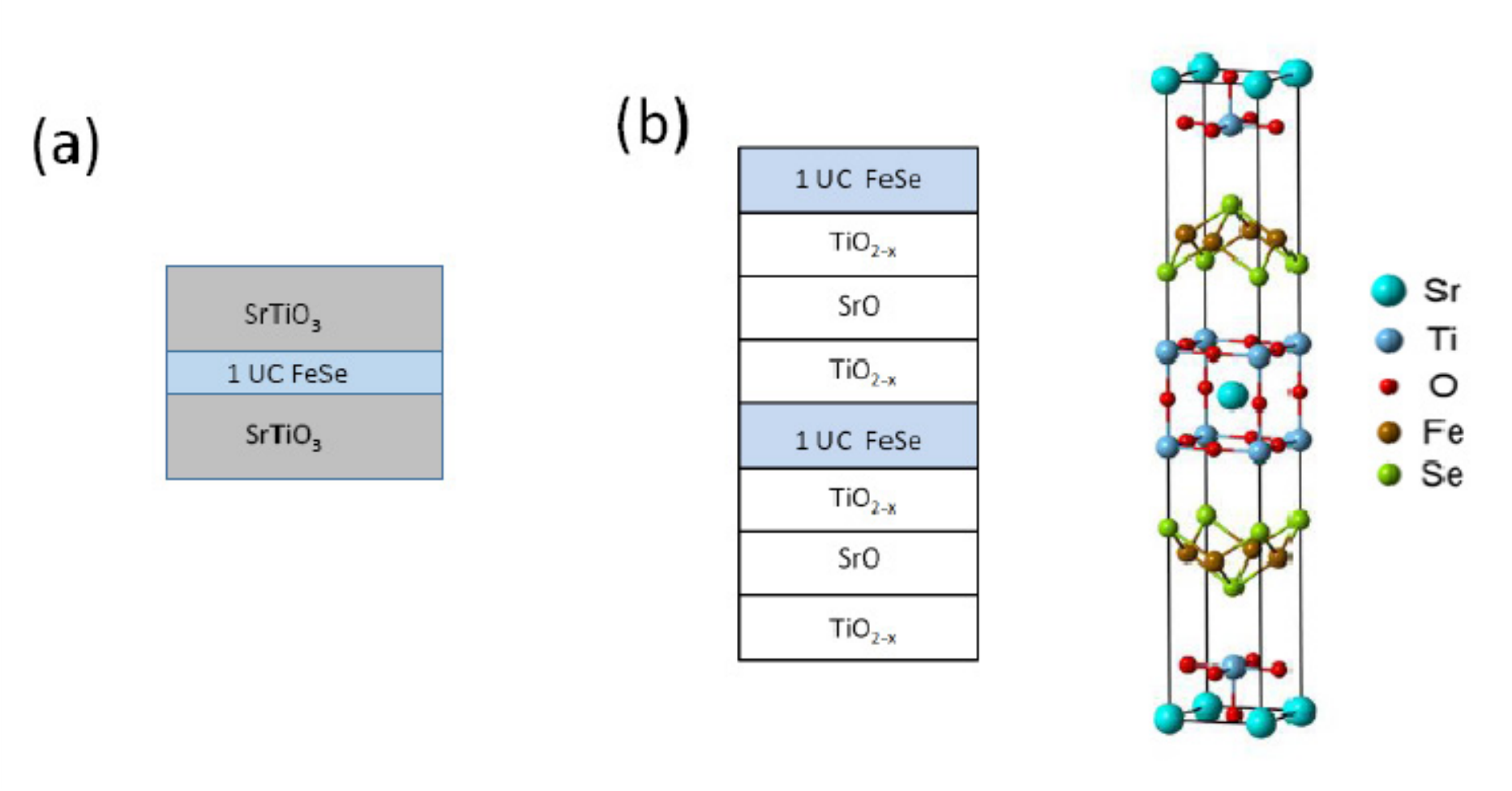}
\caption{Possible FeSe/STO based superstructures, where the enhancement of $T_c$
can be expected \cite{DHL}: (a) -- Ginzburg's ``sandwich'' with two STO layers,
(b) -- multiple layers superstructure.}
\label{super_struct}
\end{figure}

Finally, let us discuss several proposals for possible ways of further increase
of $T_c$ in FeSe monolayers on STO (or BTO) . If we accept the picture of
decisive role of interactions with elementary excitations in the substrate
(most probably with optical phonons), the natural idea appears of creation
of multiple layer films and superstructures, like those shown in
Fig. \ref{super_struct} \cite{DHL}. In particular, the structure shown in
Fig. {\ref{super_struct}} (a), is the direct realization of Ginzburg ``sandwich'',
precisely as was proposed in his original works \cite{VLG}. It seems obvious,
that the presence of the second SrTiO$_3$ layer (or the similar BaTiO$_3$ layer)
will lead to the effective enhancement of the pairing constant due to interaction
with optical phonons in the second STO layer. Obviously, the presence of the
second STO layer will also serve as a good protection of FeSe layer from
external environment. Similarly, very promising seems to be the attempts to create
the bulk superstructures (compounds), like that shown in Fig. \ref{super_struct}
(b). Despite all technical problems appearing on the way to create such
structures (or their analogues), this way seems to be very perspective.
There is no doubt that the last word in the studies of high -- temperature
superconductivity in FeSe monolayers and other similar systems is yet to be
heard.

The author is grateful to E.Z. Kuchinskii and I.A. Nekrasov for discussions of
the number of problems dealed with in this review, as well as for their help
in some of numerical calculations.

This review was supported by RSF grant 14-12-00502. Calculations of electronic
spectra of FeSe systems and comparative analysys of mechanisms of Cooper 
pairing were performed under FASO State contract No.
0389-2014-0001 with partial support by RFBR grant 14-02-00065.


\newpage








{\bf Notes added in proof:}

\footnotesize

During the time after the submission of this review, a number of new 
experimental and theoretical works dealing with systems under discussion have 
appeared in the literature. Below we quote some of these with brief comments.

In Ref. \cite{Shi} single -- layer films of FeSe on STO were studied for
different doping levels, which were achieved by surface deposition of potassium
{\em in situ}. The sharp growth of $T_c$ from 60K to 75K was observed accompanied
by Lifshits transition with formation of a small {\em electron -- like} pocket of
the Fermi surface around $\Gamma$ -- point, which was confirmed by ARPES
measurements. Note that such $T_c$ behavior is in complete accordance with
qualitative conclusions of multiple -- bands superconductivity theory,
discussed in our review.

Important results were obtained in Ref. \cite{Shen}, where the high values
of $T_c\sim$65K were obtained for monolayers of FeSe on 100 plane of rutile 
TiO$_2$. These results show that ferroelectric properties of SrTiO$_3$
(absent for TiO$_2$) are irrelevant for $T_c$ growth in systems under
disussionв and almost unambiguously confirm the important role of interactions
with longitudinal optical phonons in the substrate, which in TiO$_2$ are
practically the same as in STO. Electronic spectrum of FeSe films on  TiO$_2$ 
measured by ARPES was observed to be practically the same as in FeSe/STO, 
with ``replica'' band well observed approximately 100 meV below electronic
band at $M$ -- point, similar to that observed in Ref. [47] in FeSe/STO.

Experiments on high resolution electron energy loss spectroscopy (HREELS) 
performed in Ref. \cite{Zh} confirmed the presence of strong electron --
phonon interaction at FeSe/STO interface, giving the experimental  estimate
of coupling constant with 92 meV optical phonon in STO $\sim$ 1.0.

Theoretical results of Refs. [47,79] and [82,83] were critically reconsidered
in Ref. \cite{Dolg}. However, the qualitative conclusion on important role of
dominating forward scattering of electron in FeSe monolayer by the optical
phonons of SrTiO$_3$ for the increase of $T_c$ in FeSe/STO was essentially 
confirmed.

In Ref. \cite{John} the ``first principles'' calculations of electron --
phonon coupling if FeSe/STO system were performed, confirming the significant
enhancement of this interaction in the region of small transferred momenta.
However, the numerical values of corresponding coupling constant were too
low to explain the experimentally observed high values of $T_c$. At the same
time, it should be noted, that calculations of electronic spectrum for
FeSe/STO sytem, performed in this work, were made neglecting the possible
role of electron correlations, and spectra obtained were quite different from
those observed in ARPES experiments (absence of the ''shallow`` band).
Thus, the conclusions on the value of the coupling constant made in this work,
may be rather approximate.

Gor'kov's approach to explanation of superconductivity in SrTiO$_3$ [86] 
was criticized in Refs. \cite{Lee,vdM}. In principle, this criticism can 
can be extended to Refs. [84,85] dealing with superconductivity in FeSe/STO. 
At the same time, the use of phenomenological values of dielectric 
permeabilities at  FeSe/STO interface in [84,85] makes the arguments of Refs.
\cite{Lee,vdM} only partly relevant for this case.

Finally, we can mention the recent rather detailed review of experiments on
intercalated A$_x$Fe$_2$Se$_2$ systems \cite{Maziopa}.



\begin{thebibliography} {100}
\label{bibl}


\bibitem{VLG}Ginzburg V L. Usp. Fiz. Nauk {\bf 95} 91 (1968); {\bf 101} 185 
(1970); {\bf 118} 316 (1976)
[Contemporary Physics {\bf 9} 355 (1968); 
Physics Uspekhi {\bf 13} 535 (1970); {\bf 19} 174 (1976)]

\bibitem{HTSC77}Problema visokotemperaturnoi sverkhprovodimosti. Ed. by 
V.L. Ginzburg and D.A. Kirzhits. GRFML ``Nauka'', Moscow, 1977
[High -- Temperature Superconductivity. Ed. by V.L. Ginzburg and D.A. Kirzhnits,
Consultants Bureau, NY, 1982]  

\bibitem{Sad_08}Sadovskii M V. Usp. Fiz. Nauk {\bf 178} 1243 (2008)
[Physics Uspekhi {\bf 51} 1201 (2008)]

\bibitem{Hoso_09}Ishida K, Nakai Y, Hosono H.
J. Phys. Soc. Jpn. {\bf 78} 062001 (2009)

\bibitem{John}Johnson D C, Adv. Phys. {\bf 59}, 83 (2010)

\bibitem{MazKor}Hirshfeld P J, Korshunov M M, Mazin I I.
Rep. Prog. Phys. {\bf 74}, 124508 (2011)

\bibitem{Stew}Stewart G R, Rev. Mod. Phys. {\bf 83}, 1589 (2011)

\bibitem{Kord_12}Kordyuk A A, Fizika Nizkikh Temperatur {\bf 38} 1119 (2012)
[Low Temperature Physics {\ 38} 888 (2012)]

\bibitem{FeSe}Mizugushi Y, Takano Y. J. Phys. Soc. Jpn. {\bf 79} 102001 (2010)

\bibitem{JMMM}Sadovskii M V, Kuchinskii E Z, Nekrasov N A. JMMM {\bf 324}
3481 (2012)

\bibitem{JTLRev}Nekrasov I A, Sadovskii M V. Pis'ma Zh. Eksp. Teor. Fiz.
{\bf 99} 687 (2014) [JETP Letters{\bf 99} 598 (2014)]

\bibitem{Boz2014}Bozovic I, Ahn C. Nature Physics {\bf 10} 892 (2014)

\bibitem{VivRod}Vivanco H K, Rodriguez E E, ArXiv:1603.02334

\bibitem{AFeSe}Jiangang Guo, Shifeng Jin, Shunchong Wang, Kaixing Zhu,
Tingting Zhou, Meng He, Xialong Chen. Phys. Rev. B {\bf 82} 180520 (2010)

\bibitem{AFeSe2}Yan Y J, Wang A F, Ying J J, Li Z Y, Qin W, Luo J Q, Hu J,
Chen X H. Chin. Sci. Rep. {\bf 2} 212 (2012)

\bibitem{intCH}Hatakeda T, Noji T, Kawamata T, Kato M, Koike Y. J. Phys. Soc.
Jpn. {\bf 82} 123705 (2013)

\bibitem{intNH}Burrad-Lucas M, Free D G, Sedlmaier S J, Wright J D, Cassidy S J,
Hara Y, Corkett A J, Lancaster T, Baker P J, Blundell S J, Clarke S J.
Nature Materials {\bf 12} 15 (2013)

\bibitem{LiOH1}Lu X F, Wang N Z, Wu H, Wu Y P, Zhao D, Zeng X Z, Luo X G,
Wu T, Bao W, Zhang G H, Huang F Q, Huang Q Z, Chen X H. Nature Materials
{\bf 14} 325 (2015)

\bibitem{LiOH2}Pachmayr U, Nitsche F, Luetkens H, Kamusella S, Br\"uckner F,
Sarkar R, Klauss H.-H, Johrendt D. Angew. Chem. Int. Ed. {\bf 54} 293 (2015)

\bibitem{LiOH3}Lynn J W, Zhou X, Borg C K H, Saha S R, Paglione J, Rodriguez E E.
Phys. Rev. B {\bf 92}, 0605510(R) (2015)

\bibitem{LiOH4}Nejasattari F, Stadnik Z M. J. Alloys Compounds {\bf 652}, 470 (2015)

\bibitem{FeSe_STO1}Wang Qing-Yan, Li Zhi, Zhang Wen-Hao, Zhang Zuo-Cheng,
Zhang Jin-Song, Li Wei, Ding Hao, Ou Yun-Bo, Deng Peng, Ghang Kai, Wen Jing,
Song Can-Li, He Ke, Jia Jin-Feng, Ji Shuai-Hua, Wang Ya-Yu, Wang Li-Li, Chan Xi,
Ma Xu-Cun, Xue Qi-Kun. Chin. Phys. Lett. {\bf 29} 037402 (2012)

\bibitem{FeSe_STO2}Jian-Feng Ge, Zhi-Long Liu, Chun-Lei Gao, Dong Qian,
Qi-Kun Xue, Ying Liu, Jin-Feng Jia. Nature Materials {\bf 14} 285 (2015)

\bibitem{FeSe13UCK}Miyata Y, Nakayama K, Sugawara K, Sato T, Takahashi T.
Nature Materials {\bf 14} 775 (2015)

\bibitem{FeSe_STO_FeTe}Guanyu Zhou, Ding Zhang, Chong Liu, Chenjia Tang,
Xiaoxiao Wang, Zheng Li, Canli Song, Shuaihua Ji, Ke He, Lili Wang, Xucun Ma,
Qi-Kun Xue. ArXiv:1512.01948

\bibitem{FeSe_BTO}Peng R, Xu H C, Tan S Y, Xia M, Shan X P, Huang Z C,
Wen C H P, Song Q, Zhang T, Xie B P, Feng D L. Nature Communications {\bf 5}
5044 (2014)

\bibitem{FeSe_Anatas}Hao Ding, Yan-Feng Lv, Kun Zhao, Wen-Lin Wang, Lili Wang,
Can-Li Song, Xi Chen, Xu-Cun Ma, Qi-Kun Xue. ArXiv:1603.00999

\bibitem{FeSeGraph}Can-Li Song, Yi-Lin Wang, Ye-Ping Jiang, Zhi Li, Lili Wang,
Ke He, Xi Chen, Xu-Cun Ma, Qi-Kun Xue. Phys. Rev. B {\bf 84} 020503(R) (2011)

\bibitem{FeSe1UC_rev}Xu Liu, Lin Zhao, Shaolong He, Junfeng He, Defa Liu,
Daixiang Mou, Bing Shen, Yong Hu, Jianwei Huang, Zhou X J. J. Phys. Cond. Mat.
{\bf 27} 183201 (2015)

\bibitem{KuchSad10}Kuchinskii E Z, Sadovskii M V. Pis'ma Zh. Eksp. Teor. Fiz.
{\bf 91} 729 (2010) [JETP Letters {\bf 91} 660 (2010)]

\bibitem{DMFT1}Skornyakov S L, Efremov A V, Skorikov N A, Korotin M A,
Izyumov Yu A, Anisimov V I, Kozhevnikov A V, Vollhardt D. 
Phys. Rev. B {\bf 80} 092501 (2009)

\bibitem{DMFT2}Nekrasov I A, Pavlov N S, Sadovskii M V. Pis'ma Zh. Eksp. Teor.
Fiz. {\bf 102} 30 (2015) [JETP Letters {\bf 102} 26 (2015)]

\bibitem{KFe2Se2}Nekrasov I A, Sadovskii M V. Pis'ma Zh. Eksp. Teor. Fiz.
{\bf 93} 182 (2011) [JETP Letters {\bf 93} 166 (2011)]

\bibitem{KFe2Se2SI}Shein I R, Ivanovskii A L. Phys. Lett. A {\bf 375} 1028 (2011)

\bibitem{Ba122}Nekrasov I A, Pchelkina Z V, Sadovskii M V. Pis'ma Zh. Eksp. Teor.
Fiz. {\bf 88} 155 (2008) [JETP Letters {\bf 88} 144 (2008)]

\bibitem{KFe2Se2ARPES}Zhao L, Mou D, Liu S, He J, Peng Y, Yu L, Liu X,
Liu G, He S, Dong X, Zhang J, He J B, Wang D M, Chen G F, Guo J G, Chen X L, 
Wang X, Peng Q, Wang Z, Zhang S, Yang F, Xu Z, Chen C, Zhou X J.
Phys. Rev. B {\bf 83}, 140508(R) (2011)

\bibitem{KFeSeLDADMFT1}Nekrasov I A, Pavlov N S, Sadovskii M V.
Pis'ma Zh. Eksp. Teor. Fiz. {\bf 97} 18 (2013) [JETP Letters {\bf 97} 15 (2013)]

\bibitem{KFeSeLDADMFT2}Nekrasov I A, Pavlov N S, Sadovskii M V.
Zh. Eksp. Teor. Fiz. {\bf 144} 1061 (2013) [JETP {\bf 117} 926 (2013)]

\bibitem{CLDA}Nekrasov I A, Pavlov N S, Sadovskii M V,
Pis'ma Zh. Eksp. Teor. Fiz. {\bf 95} 659 (2012)
[JETP Letters {\bf 95} 581 (2012)]

\bibitem{CLDA_long}Nekrasov I A, Pavlov N S, Sadovskii M V, Zh. Eksp. Teor. Fiz.
{\bf 143} 713 (2013)  [JETP {\bf 116} 620 (2013)]

\bibitem{KFe2Se2_ARPES}Yi M, Lu D H, Yu R, Riggs S C, Chu J H, Lv B, Liu Z K,
Lu M, Cui Y T, Hashimoto M, Mo S K, Hussain Z, Chu C W, Fisher I R, Si Q,
Shen Z X, Phys. Rev. Lett. {\bf 110} 067003 (2013)

\bibitem{KFe2Se2_ARPES_2}Niu X H, Chen S D, Jiang J, Ye Z R, Yu T L, Xu D F,
Xu M, Feng Y, Yan Y J, Xie B P, Zhao J, Gu D C, Sun L L, Mao Q, Wang H, Fang M,
Zhang C J, Hu J P, Sun Z, Feng D L. Phys. Rev. B {\bf 93} 054516 (2016)

\bibitem{LiOHFeSe_NS} Nekrasov I A, Sadovskii M V. Pis'ma Zh. Eksp. Teor. Fiz.
{\bf 101} 50 (2015)

\bibitem{LiOHFeSe_ARP}Niu X H, Peng R, Xu H C, Yan Y J, Jiang J, Xu D F, Yu T L,
Song Q, Huang Z C, Wang Y X, Xie B P, Lu X F, Wang N Z, Chen X H, Sun Z,
Feng D L. Phys. Rev. B {\bf 92} 060504(R) (2015)

\bibitem{FNT_2016}Nekrasov I A, Pavlov N S, Sadovskii M V, Slobodchikov A.A.
ArXiv:1605.02404

\bibitem{ARP_FS_FeSe_1}Defa Liu, Wenhao Zhang, Daixiang Mou, Junfeng He,
Yun-Bo Ou, Qing-Yan Wang, Zhi Li, Lili Wang, Lin Zhao, Shaolong He,
Yingying Peng, Xu Liu, Chaoyu Chen, Li Yu, Guodong Liu, Xiaoli Dong, Jun Zhang,
Chuangtian Chen, Zuyan Xu, Jiangping Hu, Xi Chen, Xucun Ma, Qikun Xue, X. J. Zhou.
Nature Communications {\bf 3}, 931 (2012)

\bibitem{ARPES_FeSe_Nature}Lee J J, Schmitt F T, Moore R G, Johnston S, Cui Y T,
Li W, Liu Z K, Hashimoto M, Zhang Y, Lu D H, Devereaux T P, Lee D H, Shen Z X.
Nature {\bf 515}, 245 (2014)

\bibitem{ARP_FS_FeSe_2}Lin Zhao, Aiji Liang, Dongna Yuan, Yong Hu, Defa Liu,
Jianwei Huang, Shaolong He, Bing Shen, Yu Xu, Xu Liu, Li Yu, Guodong Liu,
Huaxue Zhou, Yulong Huang, Xiaoli Dong, Fang Zhou, Zhongxian Zhao,
Chuangtian Chen, Zuyan Xu, Zhou X J. Nature Communications {\bf 7}, 10608 (2016)

\bibitem{Shkl_16}Han Fu, Reich K V, Shklovskii B I. Zh. Eksp. Teor. Fiz.
{\bf 130} 530 (2016) [JETP {\bf 122} No.3 (2016)]

\bibitem{Mills}Yuanjun Zhou, Mills A J. ArXiv:1603.02728

\bibitem{Agter}Chen M X, Agterberg D F, Weinert M. ArXiv:1603.03841

\bibitem{FeSe1UC_Ph_Dia1}Shaolong He, Junfeng He, Wenhao Zhang, Lin Zhao,
Defa Liu, Xu Liu, Daixiang Mou, Yun-Bo Ou, Qing-Yan Wang, Zhi Li, Lili Wang,
Yingying Peng, Yan Liu, Chaoyu Chen, Li Yu, Guodong Liu, Xiaoli Dong, Jun Zhang,
Chuangtian Chen, Zuyan Xu, Xi Chen, Xucun Ma, Qikun Xue, Zhou X J.
Nature Materials {\bf 12} 605 (2013)

\bibitem{FeSe1UC_Ph_Dia2}Junfeng He, Xu Liu, Wenhao Zhang, Lin Zhao, Defa Liu,
Shaolong He, Daixiang Mou, Fansen Li, Chenjia Tang, Zhi Li, Lili Wang,
Yingying Peng, Yan Liu, Chaoyu Chen, Li Yu, Guodong Liu, Xiaoli Dong, Jun Zhang,
Chuangtian Chen, Zuyan Xu, Xi Chen, Xucun Ma, Qikun Xue, Zhou X J,
PNAS {\bf 111} 18501 (2014)

\bibitem{hPn2} Mizuguhci Y, Hara Y, Deguchi K, Tsuda S, Yamaguchi T,
Takeda K, Kotegawa H, Tou H, Takano Y, Supercond. Sci. Technol. {\bf 23}
054013 (2010)

\bibitem{Kucinskii10}Kuchinskii E Z, Nekrasov I A, Sadovskii M V,
Pis'ma Zh. Eksp. Teor. Fiz. {\bf 91} 567 (2010)
[JETP Letters {\bf 91} 518 (2010)]

\bibitem{K_dop_1}Miyata Y, Nakayama K, Sugawara K, Sato T, Takahashi T.
Nature Materials {\bf 14} 775 (2015)

\bibitem{K_dop_2}Wen C H P, Xu H C, Chen C, Huang Z C, Pu Y J, Song Q, Xie B P,
Abdel-Hafez M, Chareev D A, Vasiliev A N, Peng R, Feng D L.
Nature Communications {\bf 7} 10840 (2016)

\bibitem{K_dop_3}Ye Z R, Zhang C F, Ning H L, Li W, Chen L, Jia T, Hashimoto M,
Lu D H, Shen Z X, Zhang Y. ArXiv:1512.02526

\bibitem{K_dop_band}Fawei Zheng, Li-Li Wang, Qi-Kun Xue, Ping Zhang.
Phys. Rev. B {\bf 93} 075428

\bibitem{Field_1}Shiogai J, Ito Y, Mitsuhashi T, Nojima T, Tsukazaki A.
Nature Physics {\bf 12} 42 (2016)

\bibitem{Field_2}Hanzawa K, Sato H, Hiramatsu T, Hosono H. ArXiv:1508.07689;
PNAS (2016); DOI:10.1073/pnas.1520810113

\bibitem{Field_3}Lei B, Chui J H, Xiang Z J, Shang C, Wang N Z, Ye G J, Luo X G,
Wu T, Sun Z, Chen X H. Phys. Rev. Lett. {\bf 116} 077002 (2016)

\bibitem{BGrk}Barzykin V, Gor'kov L P. Письма ЖЭТФ {\bf 88} 142 (2008)
[JETP Letters {\bf 88} 131 (2008)]

\bibitem{KucSad08}Kuchinskii E Z, Sadovskii M V. Письма ЖЭТФ {\bf 89} 176 (2009)
[JETP Letters {\bf 89} 156 (2009)]

\bibitem{Hirshf}Xiao Chen, Maiti S, Linschfeld A, Hirschfeld P J. Phys. Rev.
B {\bf 92} 224514 (2015)

\bibitem{KucSad10}Kuchinskii E Z, Sadovskii M V. Physica C {\bf 470} S418 (2010)

\bibitem{dop_1}Sat T, Nakayama K, Sekiba Y, Richard P, Xu Y M, Souma S,
Takahashi T, Chen G F, Luo J L, Wang N L, Ding H. Phys. Rev. Lett. {\bf 103}
047002 (2009)

\bibitem{dop_2}Sekiba Y, Sato T, Nakayama K, Terashima K, Richard P, Bowen J H,
Ding H, Xu Y M, Li L J, Gao G H, Xu Z A, Takahashu T. New J. Phys. {\bf 11}
025020 (2009)

\bibitem{mag_imp}Fan Q, Zhang W H, Liu X, Yan Y J, Ren M Q, Peng R, Xu H C,
Xie B P, Hu J P, Zhang T, Feng D L. Nature Physics {\bf 11} 946 (2015)

\bibitem{STO_el_str}van Benthem K, Els\"asser C, French R H. J. Appl. Phys.
{\bf 90} 6156 (2001)

\bibitem{STO_parael}M\"uller K A, Burkard H. Phys. Rev. B {\bf 19} 3593 (1979)

\bibitem{KCSHP}Koonce C S, Cohen M L, Schooley J F, Hosler W R, Pfeiffer E R.
Phys. Rev {\bf 163} 380 (1967)

\bibitem{XLin}Lin X, Zhu Z, Fauqu\'e, Behnia K. Phys. Rev. X {\bf } 021002 (2013)

\bibitem{ABB}Allender D, Bray J, Bardeen J. Phys. Rev. B {\bf 7} 1020 (1973)

\bibitem{IA_73}Inkson J C, Anderson P W. Phys. Rev. B {\bf 8} 4429 (1973)

\bibitem{UspZhar}Uspenskii Y A, Zharkov G F. Zh. Eksp. Teor. Fiz {\bf 65} 2511 
(1974) [JETP {\bf 38} 1254 (1974)]

\bibitem{ABB_2}Allender D, Bray J, Bardeen J. Phys. Rev. B {\bf 8} 4433 (1973)

\bibitem{CWK}Choudhury N, Walter E J, Kolesnikov A I, Chun-Keung Loong.
Phys. Rev. B {\bf 77} 134111 (2008)

\bibitem{DHL}Dung-Hai Lee. ArXiv:1508.02461

\bibitem{DaDoKuO}Danylenko O V, Dolgov O V, Kuli\'c M L, Oudovenko V.
Eur. J. Phys. B {\bf 9} 201 (1999)

\bibitem{Kulic}Kuli\'c M L, AIP Conference Proceedings {\bf 715} 75 (2004)

\bibitem{Rade_1}Rademaker L, Wang Y, Berlijn T, Johnston S. New J. Phys.
{\bf 18} 022001 (2016)

\bibitem{Rade_2}Wang Y, Nakatsukasa K, Rademaker L, Berlijn T, Johnston S.
ArXiv:1602.00656

\bibitem{Gork_1}Gor'kov L P. Phys. Rev. B {\bf 93} 060507 (2016)

\bibitem{Gork_2}Gor'kov L P. Phys. Rev. B {\bf 93} 054517 (2016)

\bibitem{Gork_3}Gor'kov L P. PNAS {\bf 113} 4646 (2016); ArXiv:1508.00529

\bibitem{Mahan}Wang S Q, Mahan G D. Phys. Rev. B {\bf 6} 4517 (1972)

\bibitem{NozSR}Nozieres P, Schmitt-Rink S. J. Low. Temp. Phys. {\bf 59} 195 (1985)

\bibitem{Rand}Randeria M. In ``Bose-Einstein Condensation'', Eds. A. Griffin,
D. W. Snoke, S. Stringari. Cambridge University Press, 1995, p. 355

\bibitem{FeSeTe_BEC}Lubashevsky Y, Lahoud E, Chashka K, Podolsky D, Kanigel A.
Nature Physics {\bf 8} 309 (2012)

\bibitem{FeSe_BEC}Kasahara S, Watashige T, Hanaguri T, Kohsaka Y, Yamashita T,
Shimoyama Y, Mizukami Y, Endo R, Ikeda H, Aoyama K, Terashima T, Uji S, Wolf T,
L\"ohneysen H v, Shibauchi T, Matsuda Y. PNAS {\bf 111} 16309 (2014)

\bibitem{ChErEfr}Chubukov A V, Eremin I, Efremov D V. ArXiv:1601.01678
\end{thebibliography}

\begin{thebibliography}{99}


\bibitem{Shi}Shi X, Han Z.-Q, Peng X.-L, Richard P, Qian T, Wu X.-X,
Qiu M.-W, Wang S C, Hu J P, Sun Y,-J, Ding H. ArXiv:1606.01470

\bibitem{Shen}Rebec S N, Jia T, Zhang C, Hashimoto M, Lu D.-H, Moore R G,
Shen Z.-X. ArXiv:1606.09358

\bibitem{Zh}Zhang S, Guan J, Jia X, Liu B, Wang W, Li F, Wang L, Ma X,
Xue Q, Zhang J, Plummer E W, Zhu X, Guo J. ArXiv:1605.06941

\bibitem{Dolg}Kulic M L, Dolgov O V. ArXiv:1607.00843

\bibitem{John}Wang Y, Linscheld A, Berlijn T, Johnson S. ArXiv:1602.03288

\bibitem{Lee}Ruhman J, Lee P A. ArXiv:1605.01737

\bibitem{vdM}Klimin S N, Tempere J, Devreese J T, van der Marel D.
ArXiv:1606.00644

\bibitem{Maziopa}Krzton-Maziopa A, Svitlyk V, Pomjakushina E, Puzniak R.
J. Phys. Condens. Matter {\bf 28} 293002 (2016)


\end{thebibliography}
\end{document}